\documentclass[12pt,a4paper]{article}

\usepackage{helvet}

\usepackage{ifthen} 
\newboolean{pdflatex}
\setboolean{pdflatex}{true} 

\newboolean{articletitles}
\setboolean{articletitles}{true} 

\newboolean{uprightparticles}
\setboolean{uprightparticles}{false} 

\newboolean{inbibliography}
\setboolean{inbibliography}{false} 

\def\useSections{1}
\def\paperauthors{P. Koppenburg} 
\def\paperasciititle{Statistical biases in measurements with multiple candidates} 
\def\papertitle{Statistical biases in measurements with multiple candidates} 
\def\paperkeywords{{High Energy Physics}, {Flavour Physics}} 
\def\papercopyright{\the\year\ P. Koppenburg} 

\def\paperlicenceurl{https://creativecommons.org/licenses/by/4.0/}

\usepackage[top=1in, bottom=1.25in, left=1in, right=1in]{geometry}

\usepackage{microtype}
\usepackage{lineno}  
\usepackage{xspace} 
\usepackage{caption} 

\usepackage{graphicx}  
\usepackage{color}
\usepackage{colortbl}
\graphicspath{{./figs/}} 

\usepackage{amsmath} 
\usepackage{amssymb}
\usepackage{amsfonts}
\usepackage{upgreek} 

\newcommand*\patchAmsMathEnvironmentForLineno[1]{%
\expandafter\let\csname old#1\expandafter\endcsname\csname #1\endcsname
\expandafter\let\csname oldend#1\expandafter\endcsname\csname
end#1\endcsname
 \renewenvironment{#1}%
   {\linenomath\csname old#1\endcsname}%
   {\csname oldend#1\endcsname\endlinenomath}%
}
\newcommand*\patchBothAmsMathEnvironmentsForLineno[1]{%
  \patchAmsMathEnvironmentForLineno{#1}%
  \patchAmsMathEnvironmentForLineno{#1*}%
}
\AtBeginDocument{%
\patchBothAmsMathEnvironmentsForLineno{equation}%
\patchBothAmsMathEnvironmentsForLineno{align}%
\patchBothAmsMathEnvironmentsForLineno{flalign}%
\patchBothAmsMathEnvironmentsForLineno{alignat}%
\patchBothAmsMathEnvironmentsForLineno{gather}%
\patchBothAmsMathEnvironmentsForLineno{multline}%
\patchBothAmsMathEnvironmentsForLineno{eqnarray}%
}

\usepackage{xspace} 
\usepackage{upgreek}

\def\MagUp {\mbox{\em Mag\kern -0.05em Up}\xspace}

\ifthenelse{\boolean{uprightparticles}}%
{

 \def\Pmu         {\ensuremath{\upmu}\xspace}

 \def\PDelta      {\ensuremath{\Delta}\xspace}                 
 \def\PXi         {\ensuremath{\Xi}\xspace}                 
 \def\PLambda     {\ensuremath{\Lambda}\xspace}                 
 \def\PSigma      {\ensuremath{\Sigma}\xspace}                 
 \def\POmega      {\ensuremath{\Omega}\xspace}                 
 \def\PUpsilon    {\ensuremath{\Upsilon}\xspace}

 \def\PB      {\ensuremath{\mathrm{B}}\xspace}                 
                  
 \def\PD      {\ensuremath{\mathrm{D}}\xspace}

 \def\PK      {\ensuremath{\mathrm{K}}\xspace}

 \def\PZ      {\ensuremath{\mathrm{Z}}\xspace}

 \def\Pe      {\ensuremath{\mathrm{e}}\xspace}

 \def\Pi      {\ensuremath{\mathrm{i}}\xspace}

 \def\Ps      {\ensuremath{\mathrm{s}}\xspace}

 \def\thebaroffset{0.0em}
}
{

 \def\Pmu         {\ensuremath{\mu}\xspace}

 \mathchardef\PDelta="7101
 \mathchardef\PXi="7104
 \mathchardef\PLambda="7103
 \mathchardef\PSigma="7106
 \mathchardef\POmega="710A
 \mathchardef\PUpsilon="7107
                  
 \def\PB      {\ensuremath{B}\xspace}                 
                  
 \def\PD      {\ensuremath{D}\xspace}

 \def\PK      {\ensuremath{K}\xspace}

 \def\PZ      {\ensuremath{Z}\xspace}

 \def\Pe      {\ensuremath{e}\xspace}

 \def\Pi      {\ensuremath{i}\xspace}

 \def\Ps      {\ensuremath{s}\xspace}

 \def\thebaroffset{0.18em}
}
\newcommand{\offsetoverline}[2][\thebaroffset]{\kern #1\overline{\kern -#1 #2}}%

\makeatletter
\ifcase \@ptsize \relax
  \newcommand{\miniscule}{\@setfontsize\miniscule{4}{5}}
\or
  \newcommand{\miniscule}{\@setfontsize\miniscule{5}{6}}
\or
  \newcommand{\miniscule}{\@setfontsize\miniscule{5}{6}}
\fi
\makeatother

\DeclareRobustCommand{\optbar}[1]{\shortstack{{\miniscule (\rule[.5ex]{1.25em}{.18mm})}
  \\ [-.7ex] $#1$}}

\def\epem       {{\ensuremath{\Pe^+\Pe^-}}\xspace}

\def\mup        {{\ensuremath{\Pmu^+}}\xspace}
\def\mun        {{\ensuremath{\Pmu^-}}\xspace} 

\def\Z      {{\ensuremath{\PZ}}\xspace}

\def\squark    {{\ensuremath{\Ps}}\xspace}

\def\KorKbar {\kern \thebaroffset\optbar{\kern -\thebaroffset \PK}{}\xspace}

\def\DorDbar {\kern \thebaroffset\optbar{\kern -\thebaroffset \PD}\xspace}

\def\B       {{\ensuremath{\PB}}\xspace}
\def\Bbar    {{\ensuremath{\offsetoverline{\PB}}}\xspace}

\def\BorBbar {\kern \thebaroffset\optbar{\kern -\thebaroffset \PB}\xspace}
\def\Bz      {{\ensuremath{\B^0}}\xspace}
\def\Bzb     {{\ensuremath{\Bbar{}^0}}\xspace}
\def\Bd      {{\ensuremath{\B^0}}\xspace}
\def\Bdb     {{\ensuremath{\Bbar{}^0}}\xspace}
\def\BdorBdbar {\kern \thebaroffset\optbar{\kern -\thebaroffset \Bd}\xspace}

\def\Bs      {{\ensuremath{\B^0_\squark}}\xspace}

\def\BsorBsbar {\kern \thebaroffset\optbar{\kern -\thebaroffset \Bs}\xspace}

\def\Y#1S{\ensuremath{\PUpsilon{(#1S)}}\xspace}

\def\LorLbar     {\kern \thebaroffset\optbar{\kern -\thebaroffset \PLambda}\xspace}

\newcommand{\decay}[2]{\ensuremath{#1\!\to #2}\xspace} 

\def\to                 {\ensuremath{\rightarrow}\xspace}

\def\CP                {{\ensuremath{C\!P}}\xspace}

\def\AT#1     {\ensuremath{A_{\mathrm{T}}^{#1}}\xspace}           

\def\Bsmm     {\decay{\Bs}{\mup\mun}}

\def\C#1      {\ensuremath{\mathcal{C}_{#1}}\xspace}                       
\def\Cp#1     {\ensuremath{\mathcal{C}_{#1}^{'}}\xspace}                    
\def\Ceff#1   {\ensuremath{\mathcal{C}_{#1}^{\mathrm{(eff)}}}\xspace}        
\def\Cpeff#1  {\ensuremath{\mathcal{C}_{#1}^{'\mathrm{(eff)}}}\xspace}       
\def\Ope#1    {\ensuremath{\mathcal{O}_{#1}}\xspace}                       
\def\Opep#1   {\ensuremath{\mathcal{O}_{#1}^{'}}\xspace}                    

\newcommand{\aunit}[1]{\ensuremath{\text{\,#1}}}       

\newcommand{\tev}{\aunit{Te\kern -0.1em V}\xspace}
\newcommand{\gev}{\aunit{Ge\kern -0.1em V}\xspace}
\newcommand{\mev}{\aunit{Me\kern -0.1em V}\xspace}
\newcommand{\kev}{\aunit{ke\kern -0.1em V}\xspace}
\newcommand{\ev}{\aunit{e\kern -0.1em V}\xspace}
\newcommand{\mevc}{\ensuremath{\aunit{Me\kern -0.1em V\!/}c}\xspace}
\newcommand{\gevc}{\ensuremath{\aunit{Ge\kern -0.1em V\!/}c}\xspace}
\newcommand{\mevcc}{\ensuremath{\aunit{Me\kern -0.1em V\!/}c^2}\xspace}
\newcommand{\gevcc}{\ensuremath{\aunit{Ge\kern -0.1em V\!/}c^2}\xspace}

\def\ps   {\ensuremath{\aunit{ps}}\xspace}

\def\gsim{{~\raise.15em\hbox{$>$}\kern-.85em
          \lower.35em\hbox{$\sim$}~}\xspace}
\def\lsim{{~\raise.15em\hbox{$<$}\kern-.85em
          \lower.35em\hbox{$\sim$}~}\xspace}

\def\tell1  {TELL1\xspace}
\def\ukl1   {UKL1\xspace}

\newcommand{\eg}{\mbox{\itshape e.g.}\xspace}

\usepackage{xfrac} 
\newcommand{\IF}[4]{\ifthenelse{\equal{#1}{#2}}{#3}{#4}}%
\newcommand{\withSections}[2]{\IF{\useSections}{1}{#1}{#2}\xspace}%
\newcommand{\IG}[2][1]{{\includegraphics[width=#1\textwidth]{#2}}}%
\newcommand{\eq}[2][dummy]{\ifthenelse{\equal{#1}{dummy}}%
                                      {\begin{equation*}{#2}\end{equation*}}%
                                      {\begin{equation}\label{#1}#2\end{equation}}}%
\newcommand{\eqa}[2][dummy]{\ifthenelse{\equal{#1}{dummy}}%
                                      {\begin{align*}{#2}\end{align*}}%
                                      {\begin{align}\label{#1}#2\end{align}}}%

\usepackage{symbols}

\usepackage{hyperxmp}
\usepackage[pdftex,
            pdfauthor={\paperauthors},
            pdftitle={\paperasciititle},
            pdfkeywords={\paperkeywords},
            pdfcopyright={Copyright (C) \papercopyright},
            pdflicenseurl={\paperlicenceurl}]{hyperref}
\usepackage[all]{hypcap} 

\usepackage{cite} 
\usepackage{mciteplus}

\begin{document}

\renewcommand{\thefootnote}{\fnsymbol{footnote}}
\setcounter{footnote}{1}


\begin{titlepage}
\pagenumbering{roman}

\noindent
\begin{tabular*}{\linewidth}{lc@{\extracolsep{\fill}}r@{\extracolsep{0pt}}}
 & & Nikhef-2016-059 \\  
 & & 21 August 2019 \\ 
 & & \\
\end{tabular*}

\vspace*{4.0cm}

{\normalfont\bfseries\boldmath\huge
\begin{center}
  \papertitle
\end{center}
}

\vspace*{2.0cm}

\begin{center}
P.~Koppenburg$^1$
\bigskip\\
{\normalfont\itshape\footnotesize
$ ^1$Nikhef, Amsterdam, Netherlands.
}
\end{center}

\vspace{\fill}

\begin{abstract}
  \noindent
  Many measurements at collider experiments study 
  physics candidates that are a subset of a collision event.
  The presence of multiple such candidates in a given event can cause
  raw biases which are large compared to typical statistical uncertainties.
  Selecting a single candidate is common practice
  but only helps if the likelihood of selecting the true
  candidate is very high. 
  Otherwise the precision of the measurement can be
  affected, and biases can be generated, even if none are present
  in the data sample prior to this operation. 

  This paper aims at describing the problem in a systematic way.
  It sets definitions, provides examples of potential biases using
  pseudoexperiments and gives recommendations.
  \vskip 1em
  This is the 2019 update of the original 2017 preprint.
\end{abstract}

\vspace*{2.0cm}

\vspace{\fill}

{\footnotesize 
\centerline{\copyright~P. Koppenburg, licence \href{http://creativecommons.org/licenses/by/4.0/}{CC-BY-4.0}.}}
\vspace*{2mm}

\end{titlepage}


\newpage
\setcounter{page}{2}
\mbox{~}

\tableofcontents

\cleardoublepage


\renewcommand{\thefootnote}{\arabic{footnote}}
\setcounter{footnote}{0}



\pagestyle{plain} 
\setcounter{page}{1}
\pagenumbering{arabic}


%
\withSections{\section{Introduction}\label{Sec:Intro}}{}%
In absence of new elementary particles produced
in high-energy collisions,
precision measurements provide the best way of finding signs of New Physics
at colliders.
These measurements require large data samples as well as
a very detailed understanding of potential biases.
As luminosities increase, much larger data samples are anticipated in the near future, notably at the Large Hadron
Collider (LHC) or the KEKB \epem collider.
Many decay-rate or asymmetry measurements with a precision of
order $10^{-4}$ are or will soon be available in
flavour~\cite{LHCb-PAPER-2019-006,LHCb-TDR-012,Aushev:2010bq}
physics, while electroweak and Higgs physics also enter the precision
regime~\cite{Azzi:2019yne,*Cepeda:2019klc,*Cerri:2018ypt,Abada:2019lih}.
The presence and the handling of multiple candidates per event can cause
raw biases which may be large compared to anticipated accuracies.
These biases can be corrected for, provided a proper strategy is employed,
with some loss in precision.

In the following, {\it event} stands for all signals originating from a collision process,
or a set of simultaneous processes, as in a single beam-bunch crossing. 
A {\it candidate} is the reconstructed particle of interest, including its decay chain, 
on which the measurement is being
performed --- typically a Higgs or \Z boson, a top quark, or a \B meson. The latter example
is used below, as flavour physics is presently most affected by the effects described here.

The processes most commonly studied in modern high-energy physics have a small
probability, either because the production rate is small 
(as for the Higgs boson~\cite{Aad:2012tfa,Chatrchyan:2012ufa}) or
because the decay rate is low 
(as for the decay \Bsmm~\cite{LHCb-PAPER-2017-001,Aaboud:2016ire}), 
or both. 
(Studies of processes that are not rare, and where 
several candidates per event are to be expected,
are not in the scope of this paper.) 
The same applies to background, as analysts invest considerable effort to achieve a 
background rate of the same order as that of the signal, or less. 
Yet many high-energy physics analyses are affected by a
nonnegligible fraction of events containing several candidates,
typically in the range $0.1$ to $20\%$.
The probabilities of selecting one or a second candidate in an event
are not uncorrelated. 
If they were --- given the low probability involved --- the rate of events
containing two or more candidates would be vanishingly small.
The presence of multiple candidates in an event is therefore an indication
that the event or the candidates are atypical. 
An investigation is required prior to any action being taken.


Multiple-candidate events are a nuisance for the analysis.
They contribute to the background level and thus degrade
the sensitivity of the measurement. 
Most importantly,
additional candidates can cause biases if their rate is correlated with the
observable to be measured.
These biases may be corrected using simulation or 
control samples, provided these are good representations of the data.
Otherwise, additional corrections are required.
Such situations are discussed \withSections{ in Sec.~\ref{Sec:Toy}}{below}.

In some publications 
the rate of multiple candidates and 
the procedure to handle them is described,
but less frequently are their origin discussed and
potential biases addressed. 
It is usually not possible for the reader
to assess their nature and their effect on the measurement, 
which affects the reproducibility
of the analysis. 
No standard procedure to address multiple candidates is publicly
documented (see Note~\cite{LHCbINT}), 
which forces every collaboration or analysis group to decide how to address the problem 
(or to ignore it), and often
to re-invent the wheel.
Procedures applied in previous publications tend to be replicated without thinking
about potential bias and loss of sensitivity.
While the effects described here are small,
their magnitude may be similar to the precision of the measurement.

The present paper is an attempt at describing the problem in a systematic way.
It is based on recommendations internal to the LHCb collaboration and is organised as follows: \withSections{Sec.~\ref{Sec:Defs}}{first}
gives definitions of sources of multiple candidates and techniques to address them.
A set of pseudoexperiments assessing the size of potential 
biases is described\withSections{ in Sec.~\ref{Sec:Toy}}{}.
Recommendations are given\withSections{ in Sec.~\ref{Sec:Recommendations}}{}.

\withSections{\section{Definitions}\label{Sec:Defs}}{\vskip 1em}%
Multiple candidates can be of several kinds, listed below.
\begin{description}
\item[Overlaps] share some of their reconstructed
objects (tracks, clusters, jets\dots). If the signals are fully reconstructed,
at most one of the overlapping candidates can be signal.

\item[Reflections] are candidates which share all reconstructed
objects but with different particle-identification assignments.
Depending on the analysis they can be peaking in mass (when applicable) or not.
Their treatment is analysis dependent and thus not discussed further in this paper.
\item[Genuine multiples]
do not share any of their properties.
They could in principle be both signal,
depending on the signal rate and selection
efficiency~\cite{expectation}\nocite{LHCb-PAPER-2012-041},
but are more likely to be both background (\eg caused by a high-multiplicity event) or one of each.
\item[Reconstruction features] may make physics objects appear multiple times.
They are experiment specific and therefore their effect on the measurement
is not discussed here. However,
they need to be considered when dealing with multiple candidates.
\end{description}


\subsection{Techniques}\label{Sec:techniques}
Several techniques of handling multiple candidates are reported in the literature.
\begin{description}
\item[No action:] The simplest and most frequent approach is
to take {\it no action}, i.e.\ keep all candidates for further analysis. 
The reader has to assume that this was the chosen approach when multiple candidates are 
not mentioned in a publication~\cite{Proceedings}.
In some publications this choice is explicitly 
mentioned~\cite{LHCb-PAPER-2011-023,Aad:2012oxa,LHCb-PAPER-2014-061,Lees:2015jwa,LHCb-PAPER-2015-055,LHCb-PAPER-2016-026,LHCb-PAPER-2016-029,LHCb-PAPER-2016-031,LHCb-PAPER-2016-061,LHCb-PAPER-2017-044,LHCb-PAPER-2018-005,LHCb-PAPER-2018-021,LHCb-PAPER-2018-047,Sirunyan:2018grk,Aaboud:2018hgx,LHCb-PAPER-2019-003,LHCb-PAPER-2019-013}.
It is the correct approach in production cross-section measurements~\cite{Abbott:1998sb,Acosta:2004yw,Adare:2006kf,Khachatryan:2010yr,LHCb-PAPER-2011-003,LHCb-PAPER-2011-013,Aamodt:2011gj,Aad:2011sp,Chatrchyan:2011kc,Khachatryan:2014iia}.

\item[Sample splitting:] In rare occasions, the data sample is {\it split} by candidate multiplicity~\cite{Aaltonen:2013atp}.
A different background parametrisation depending on the presence of
multiple candidates is then used.

All other techniques aim at reducing the level of multiple candidates.
\item[Arbitration] consists
in attempting to identify the true signal candidate by using a discriminating variable,
thus applying a tighter selection in events with multiple candidates compared
to the remaining events. 
Typical choices of variables are
the (transverse~\cite{Khachatryan:2016hje,Khachatryan:2014ira}) momentum~\cite{TheBABAR:2016lja} of the candidate or of a decay product, 
the $\chi^2$ of a vertex~\cite{PhysRevD.93.052016,LHCb-PAPER-2015-027,LHCb-PAPER-2012-001,LHCb-PAPER-2018-015,LHCb-PAPER-2017-021}
or of a decay tree fit~\cite{Ablikim:2015swa,LHCb-PAPER-2013-065,Sandilya:2013rhy},
an impact parameter~\cite{Abdallah:2005cx,Abdallah:2004rz},
a flight length~\cite{Barate:1998yi},
a mass of a decay product~\cite{Schumann:2005ej,Nishida:2004fk,Aad:2012tfa,Aaboud:2016zpr,Khachatryan:2015isa,Chatrchyan:2013mxa,Aad:2014aba,Aaboud:2018krd},
the particle-identification likelihood of a final-state particle~\cite{LHCb-PAPER-2018-014},
the output of a multivariate classifier~\cite{LHCb-PAPER-2018-042},
the expected rate of the final state~\cite{Aad:2014aba}, 
or, the beam-constrained energy at \epem colliders~\cite{Bevan:2014iga,Aubert:2003jq,Lees:2014lra,Ishikawa:2006fh,Ablikim:2015hih,Adam:2007pv,Choi:2015lpc}. 
Arbitration may also be part of the event reconstruction~\cite{Pattern} and the selection
of the collision point~\cite{PV}.
\item[Random picking] is a special case of arbitration where the chosen variable is a random number and 
thus not discriminating~\cite{LHCb-PAPER-2011-041,LHCb-PAPER-2013-015,LHCb-PAPER-2014-020,ATLAS:2014fka,LHCb-PAPER-2015-005,LHCb-PAPER-2015-050,LHCb-PAPER-2016-017,LHCb-PAPER-2016-030,LHCb-PAPER-2016-035,LHCb-PAPER-2016-057,LHCb-PAPER-2017-013,LHCb-PAPER-2017-018,LHCb-PAPER-2017-029,LHCb-PAPER-2017-041,LHCb-PAPER-2017-048,LHCb-PAPER-2018-001,LHCb-PAPER-2018-025,LHCb-PAPER-2019-006,LHCb-PAPER-2018-006,LHCb-PAPER-2018-020}.
A variant of random picking is {\it weighting}, in which candidates are weighted by the inverse 
of their multiplicity~\cite{Artuso:1999dd}. 
\item[Event removal,] where all events with multiple
candidates are discarded, is also occasionally 
used~\cite{Adams:2011sq,Aad:2011dm,LHCb-PAPER-2015-049,Aghasyan:2017utv,LHCb-PAPER-2018-036}.
\end{description}

The above-mentioned techniques have an associated signal efficiency, 
which contributes to the total selection efficiency. 
The latter is often determined using simulated data 
or a control channel, which may incorrectly model the
candidate multiplicity. 
This is discussed further \withSections{in Sec~\ref{Sec:Efficiencies}}{below}.


\withSections{\section{Pseudoexperiment}\label{Sec:Toy}}{\vskip 1em}%

The effect of the candidate multiplicity is tested with a set of
pseudoexperiments. 
The model employed here is typical of a \B physics experiment,
but is kept intentionally simple, so that it can be translated to any scenario.
The only needed features are that the signal is separated from the background by
a fit to a discriminating variable (here a mass). The properties
of the signal are then investigated.

The problem is formulated as follows.
A process resulting in a Gaussian-shaped signal peak (a \Bz meson mass in this example) is 
studied. The rate (which could result in the measurement of a cross-section or a branching fraction),
an asymmetry, and a property (here the lifetime) are to be measured. 
Ratios and asymmetries are of particular interest as many of the
detector- and selection-related biases cancel at first order. 
They allow for high-precision measurements which can be compared with precise predictions.
Here a \CP or production asymmetry between the \Bz and \Bzb yields is taken as an example,
but the problem can be generalised to any fraction, as for instance a polarisation.

In the present example $\NBgen=50\:000$ signal \Bz mesons are generated (including Poisson fluctuations)
on top of as many combinatorial background candidates.
The \Bz signal is represented by a Gaussian shape and the background has a uniform mass
distribution~\cite{Uniform} in a range of 25 times the resolution
and is centred on the signal.
The same is done for \Bzb mesons.
This starting point, before additional candidates are added, is referred to as the {\it clean} case and the initial candidates are called {\it original}.

Next, additional candidates (called {\it companions}) are generated as follows.
\begin{enumerate}
\item Each signal \Bz meson has a probability
\PBsig to have a companion candidate in the same event. 
  The mass distribution of these candidates is background-like~\cite{FlatMass}.
This candidate can have a swapped flavour\cite{swaps}, with probability $\PBswapB$, or have the same flavour, 
with probability $(1-\PBswapB)$.
\item The same is done for
  \Bzb mesons, introducing the probabilities \PBbsig and \PBbswapB, which do
  not need to be identical to those for \Bz mesons.
\item Each background \Bz candidate has a probability \PBbkg to have a
  companion candidate with probability of swapped flavour \PBswapBkg,
generated with uniform mass.         
\item The same is done for \Bzb background candidates with probabilities \PBbbkg and \PBbswapBkg,
      which do not need to be identical to those for \Bz background candidates.
\end{enumerate}
All symbols defined above are listed in Table~\ref{Tab:Symbols}, along with symbols defined later in this section.
\begin{table}[b]\centering
\caption{Symbols used in this document.}\label{Tab:Symbols} 
\symbolsTable 
\end{table}


In absence of flavour swaps, the fraction of events with multiple candidates is given by
\eq[eq:MCrate]{
  \Rall=\frac{\PBsig\NBgen+\PBbkg\NBKgen}{\NBgen+\NBKgen}.
}
The analogous quantity can be defined {\it mutatis mutandis} for \Bzb mesons
but is not used in the following.
This emulation is a simplified model of reality
as there are no events with three candidates.
As all candidates have passed all selection requirements, they are equally
signal-like except for their mass.

After the addition of companions, the \Bz and \Bzb background levels are increased by
amounts depending on the values of the above-defined probabilities.
\begin{figure}[t]%
\IG[0.48]{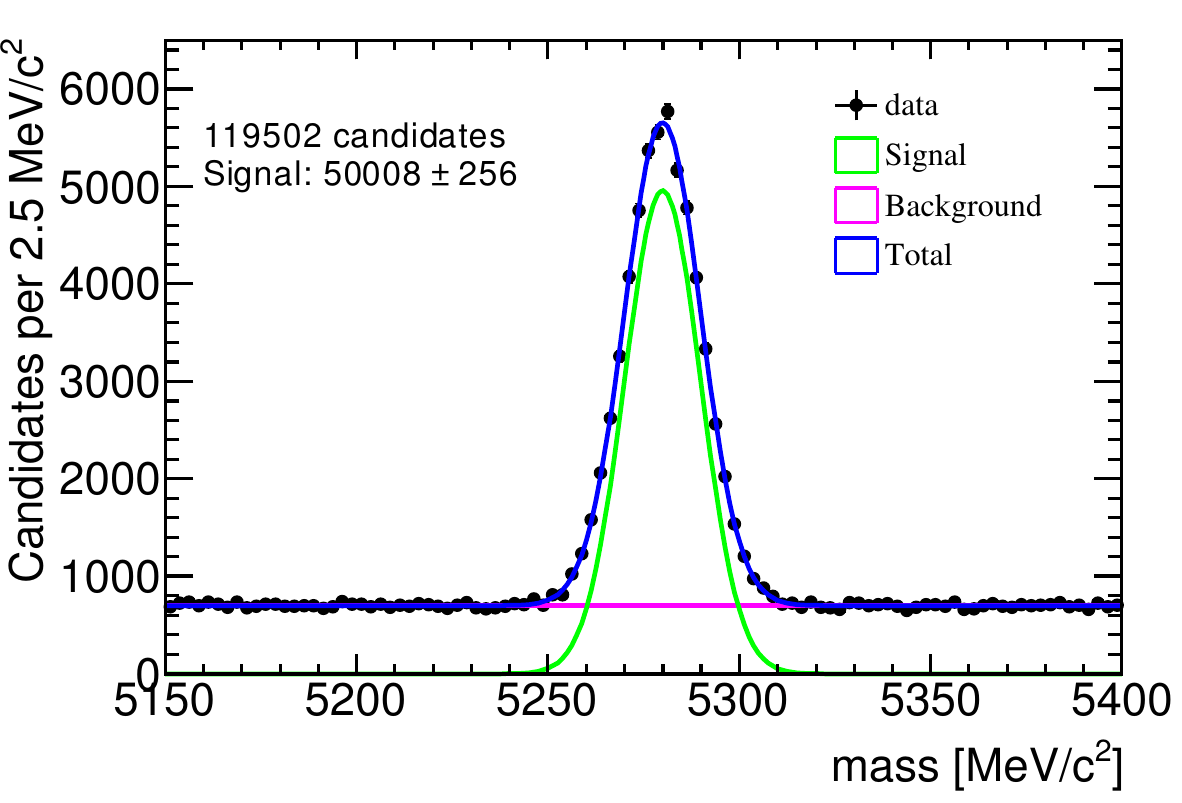}\hskip 0.04\textwidth%
\IG[0.48]{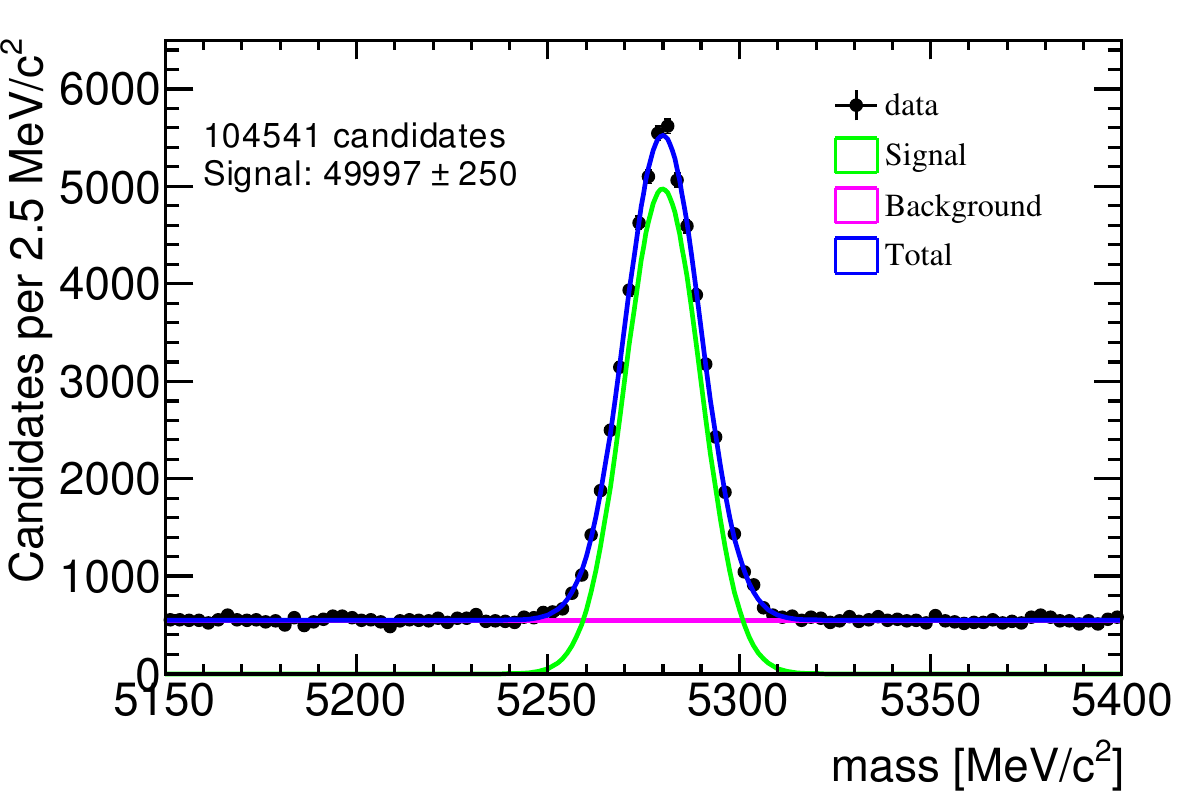}%
\caption{Typical mass distributions after companion addition %
  for (left) \Bz and (right) \Bzb candidates with %
  $\PBsig=0.4$, $\PBbsig=0.1$. 
  All other probabilities of Table~\ref{Tab:Symbols} are set to zero.
  The number of candidates and the fitted signal yield are indicated.}\label{Fig:Peaks}%
\end{figure}%
Typical mass distributions
are shown in Fig.~\ref{Fig:Peaks} for $\PBsig=0.4$ and $\PBbsig=0.1$.
Here and in the following, large probabilities are used to make
the effects easily visible in graphics.
In a real experiment, the probabilities would be smaller, but the
total yields may be much larger, making the described effects significant.

Taking no action results in keeping all candidates, and is referred to
as the {\it all} scenario. Alternatively, the analyst can choose to pick a {\it random} candidate
or perform an {\it arbitration}. 
Weighting and event removal are not tested.
Their effect can  be inferred from those of the random selection~\cite{infer}. 
After this operation the \Bz and \Bzb yields
are determined separately from unbinned maximum-likelihood fits to the mass distributions.
The probability distribution function used in the fit 
is the sum of a Gaussian and a uniform background; the same 
as is used in the generation of the sample. 
All parameters are left free in the fit. 

The uncertainty on the signal yield \NBfit depends on the background yield \NBK as
\eq[eq:Uncertainty]{
\sigma(\NBfit)=\sqrt{\NBfit+\alpha \NBK},
} 
where to a good level of approximation $\alpha$ only depends on the mass window and the signal peak resolution. 
In this case $\sigma(\NBfit)=248$ and $\alpha=0.23$.
The example in Fig.~\ref{Fig:Peaks} shows that while the probabilities of
companion candidates are different,
resulting in a higher background level for \Bz candidates, the fits return signal yields
consistent with the generated numbers. As expected, the uncertainty is larger in
the case of a higher candidate multiplicity.

Multiple-candidate handling techniques
are studied as function of the probabilities defined above
with thousand pseudoexperiments per set of probabilities. The results shown below are
averages on these pseudoexperiments.
Many of the subsequent results can be determined analytically using the probabilities
listed in Table~\ref{Tab:Symbols} and behave almost linearly with the input probabilities~\cite{supplementary}.

\withSections{\subsection[Varying companion probabilities]{Varying \boldmath\Bz companion probabilities}\label{Sec:Toy:B0}}{}
\begin{figure}[t]\centering
\IG[0.48]{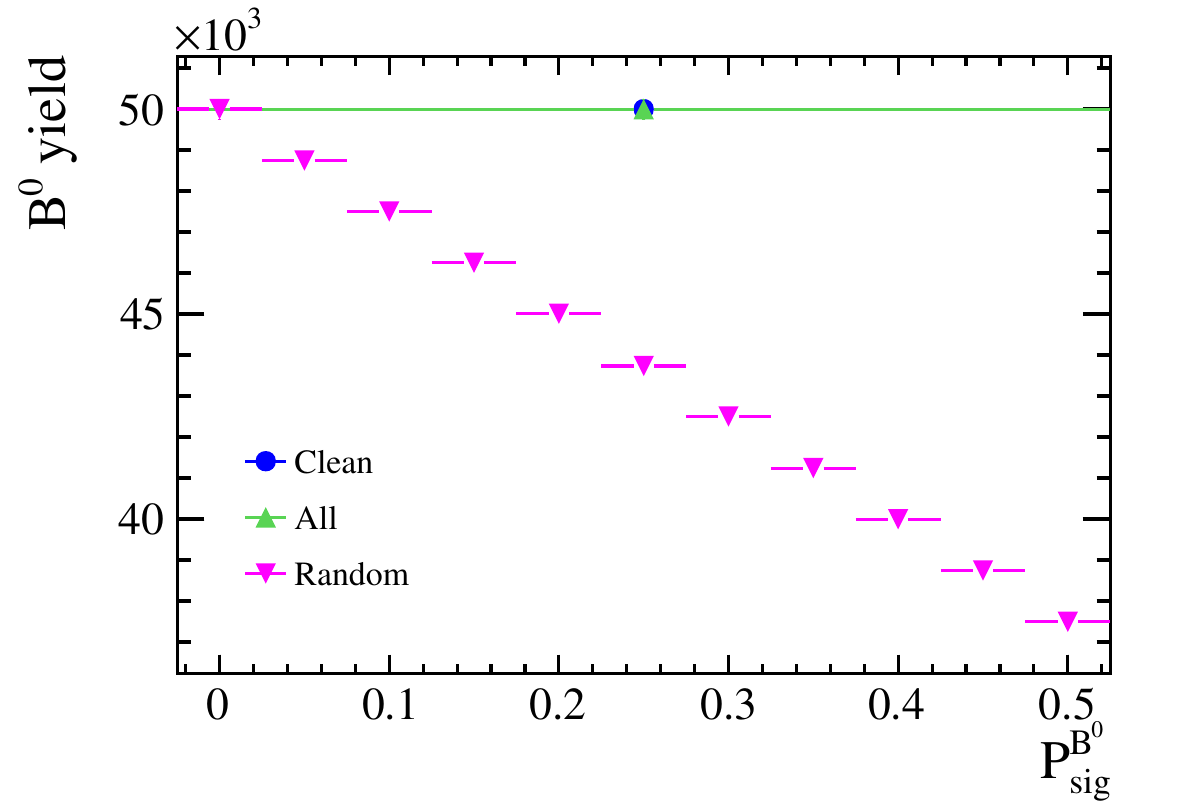}\hskip 0.02\textwidth
\IG[0.48]{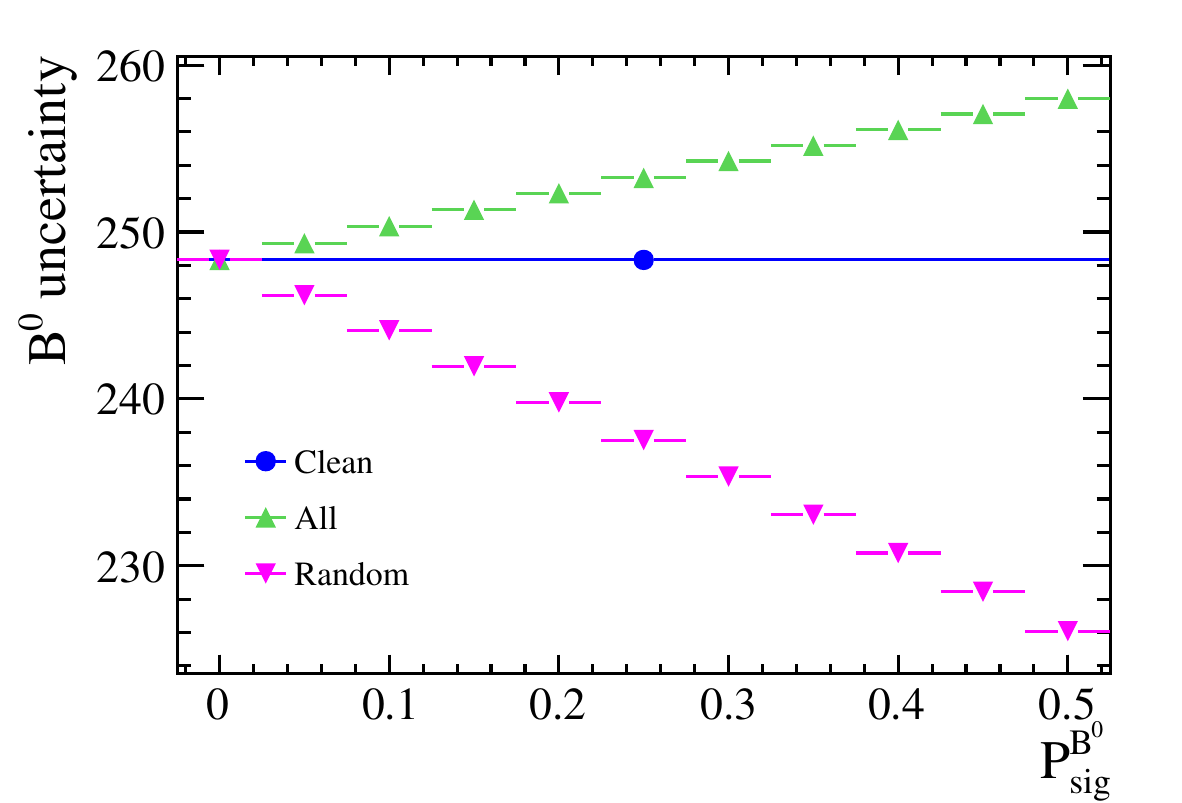}\\ 
\IG[0.48]{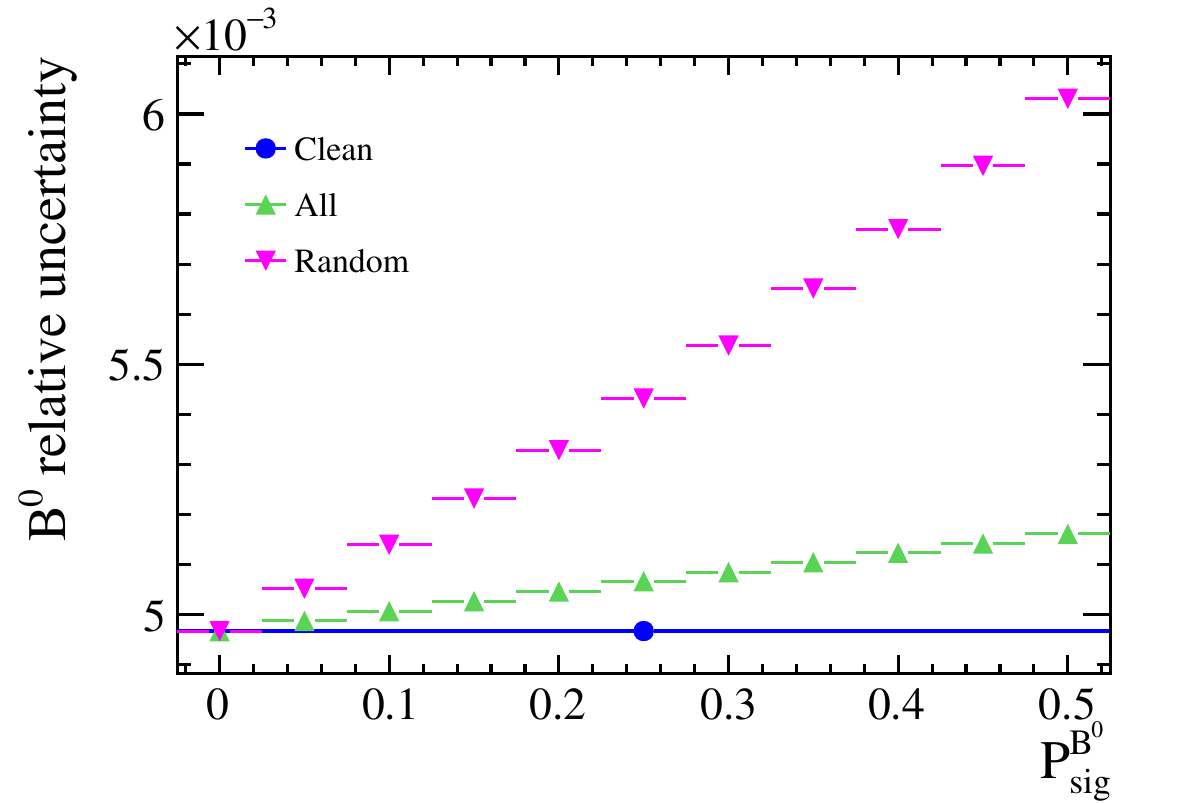}
\caption{Measured \Bz (top left) yield, (top right) uncertainty and (bottom) their ratio
versus \PBsig. All other probabilities of
Table~\ref{Tab:Symbols} are set to zero. Beware that all plots have a zero-suppressed vertical axis.
The yields vary
linearly with the probabilities while the uncertainties follow a square-root
law, which is almost linear on this scale.
}
\label{Fig:Toy:SigB0All_Random_Clean_1D}
\end{figure}
First the 
signal probabilities \PBsig and \PBbsig are varied
between $0$ and $0.5$ with $\PBbkg=\PBbbkg=0$. 
The resulting measured \Bz yields and uncertainties are reported in
Figure~\ref{Fig:Toy:SigB0All_Random_Clean_1D}.

The ideal clean case, in which always the correct candidate is picked,
is the benchmark scenario with which all other approaches are compared.
When using all candidates the correct yield is obtained, although with an increased
uncertainty.
With the technique consisting of randomly picking candidates,
the signal candidate is discarded in half of the events with multiple candidates.
The fit returns a biased \Bz yield, trivially depending on \PBsig as
\eq[eq:NBgen]{\left(\NBfit\right)^\text{random}=\NBgen\left(1-\frac{1}{2}\PBsig\right).}

The biases can be corrected\withSections{ (see Section~\ref{Sec:Efficiencies})}{}, but not
the loss of statistical precision.
The relative uncertainty on the \Bz yield~\cite{supplementary}
--- which is often the quantity for which the selection is optimised --- 
is shown in Fig.~\ref{Fig:Toy:SigB0All_Random_Clean_1D} (bottom).
The random selection technique performs worse than taking all
with respect to this figure of merit.

Next, an asymmetry between the \Bz and \Bzb yields is determined, defined by
\eq[eq:Araw]{\Araw=\frac{\NBfit-\NBbfit}{\NBfit+\NBbfit}.}

%
\begin{figure}[t]
\IG[0.48]{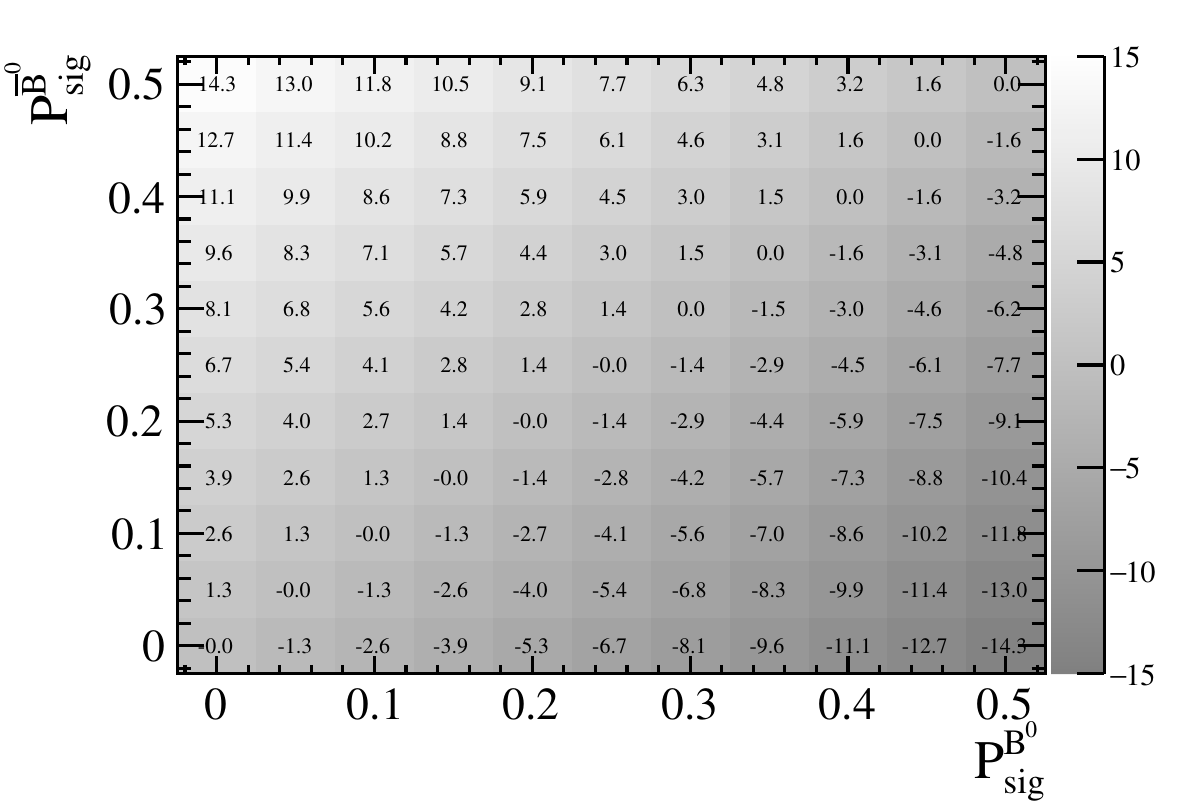}\hskip 0.04\textwidth
\IG[0.48]{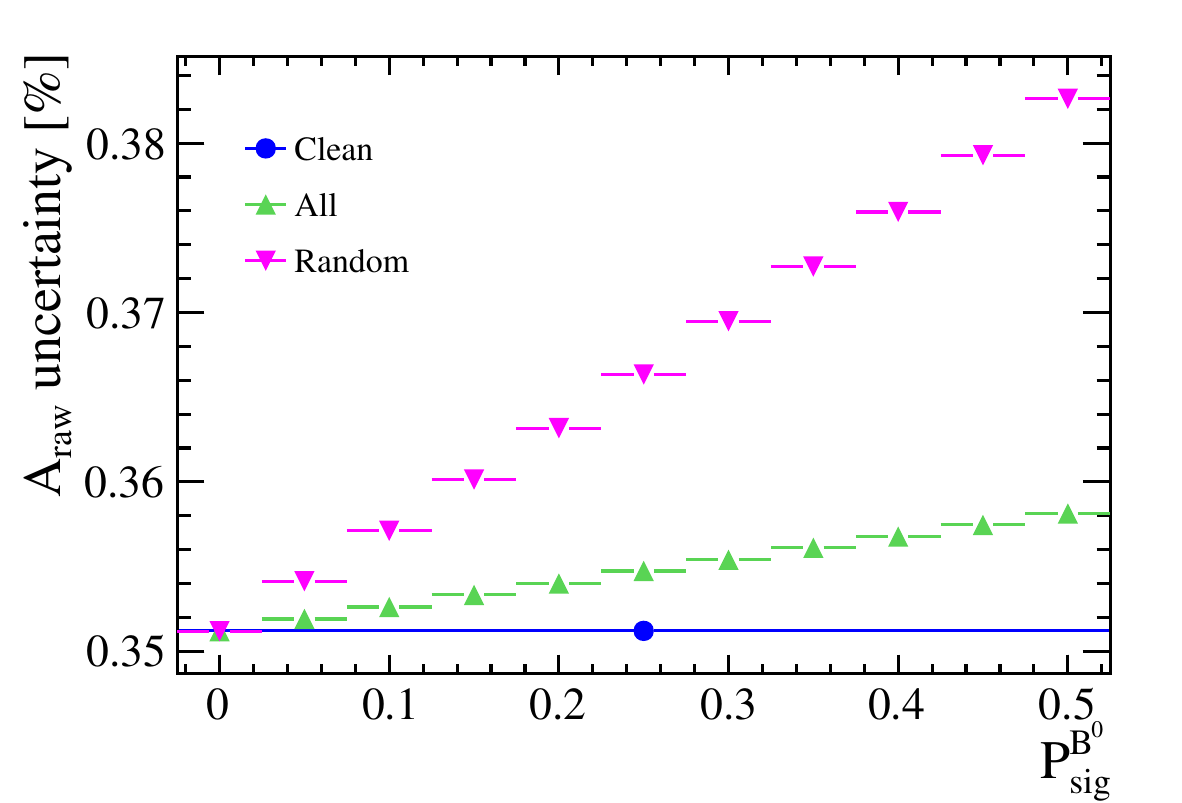}
\caption{(left) Measured \Araw (in percent)
versus \PBsig and \PBbsig for random picking. (right) Uncertainty 
on this quantity for the three techniques versus \PBsig, with $\PBbsig=0$. 
All other probabilities of
Table~\ref{Tab:Symbols} are set to zero.}\label{Fig:Toy:SigACP}
\end{figure}
In the following pseudoexperiments no asymmetry is generated and any measurement of
this quantity should be consistent with zero. 
Figure~\ref{Fig:Toy:SigACP} show this asymmetry versus the \PBsig and \PBbsig probabilities
in the range $0$ to $0.5$.
Asymmetries of up to $14\%$ can be generated when performing random picking.
Using Eq.~\ref{eq:NBgen}, the expected asymmetry is easily calculable as
\eq[eq:Arnd]{
\left(A_\text{raw}\right)^\text{random}=
\frac{\NBgen\left(1-\frac{1}{2}\PBsig\right) - \NBbgen\left(1-\frac{1}{2}\PBbsig\right)  }{ \NBgen\left(1-\frac{1}{2}\PBsig\right) + \NBbgen\left(1-\frac{1}{2}\PBbsig\right)}
\stackrel{A_\CP=0}{=}
\frac{\PBbsig-\PBsig  }{4-\PBsig - \PBbsig},
}
where the expression considerably simplifies when no true \CP asymmetry is present.
On the other hand, using all candidates leaves the asymmetry unbiased.

The uncertainties on this quantity vary from $0.362\%$, at $\PBsig=\PBbsig=0$, to values
of $0.366\%$ for random picking and $0.382\%$ when keeping all candidates.
These changes in statistical uncertainties are very small ($+0.02\%$) in comparison with the 
potential bias introduced by the random picking technique (up to $14\%$).
Similarly, varying the companion probabilities in background, \PBbkg and \PBbbkg, only slightly
affects the statistical uncertainties~\cite{PBbkg}.

\withSections{\subsection{Arbitration}\label{Sec:Toy:Best}}{}
The drawbacks of random picking may be cured by arbitration.
The selection of a ``best'' candidate is done using an observable $O$ 
which is not correlated to that which is used to determine the signal yield 
(here the candidate mass). For the technique to be effective, the true signal and 
the other candidates must have
different distributions of this quantity. This feature is exploited
to pick the candidate that is most
signal-like. 
The arbitration procedure has an efficiency \PBbest of picking 
the true signal, which should be as high as possible.
For the present discussion, only this efficiency matters,
and not the actual $O$ distributions.

\begin{figure}[t]
  \IG[0.48]{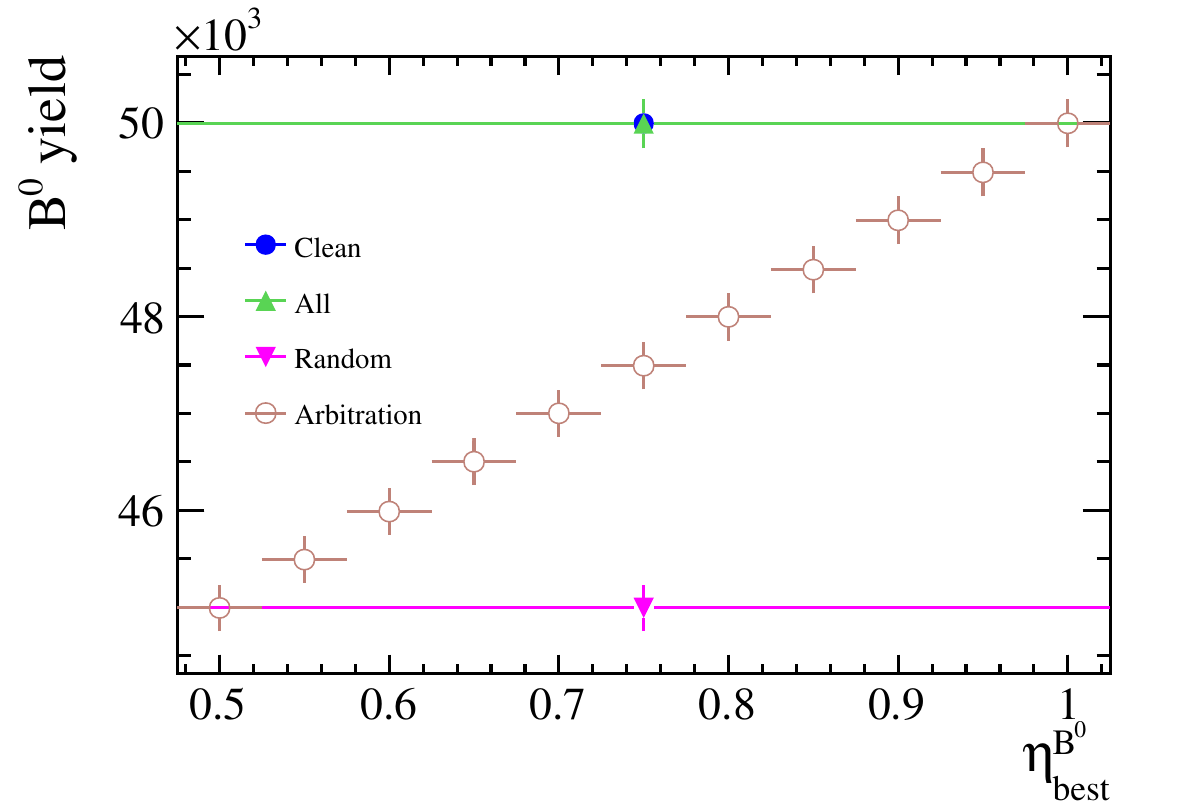}\hskip 0.04\textwidth
  \IG[0.48]{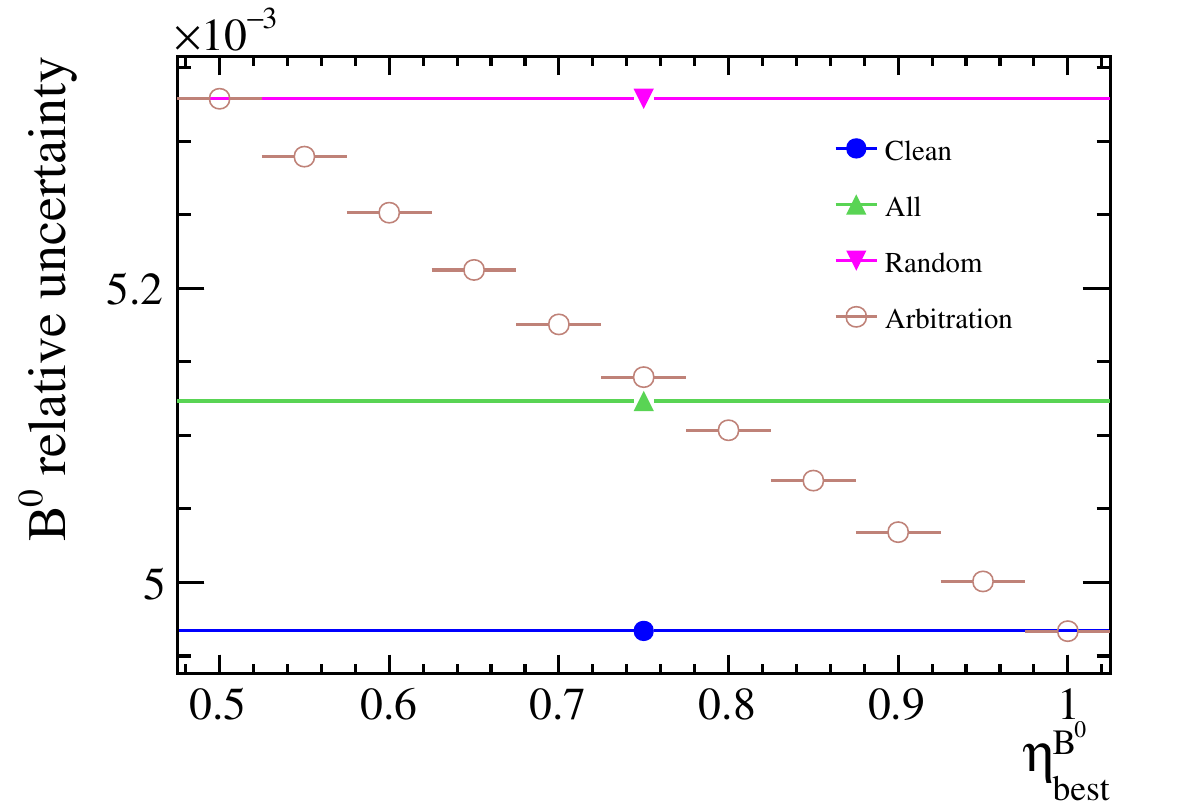}
  \caption{(left) \Bz yield and (right) relative uncertainty 
     versus efficiency of picking the signal candidate \PBbest.
    The value at $\PBbest=0.5$ corresponds to random picking.
    The other parameters are set to
    $\PBsig=\PBbkg=0.2$, $\PBswapB=\PBswapBkg=0$. 
  }\label{Fig:Toy:BestSelection_1D}
\end{figure}
The aim is illustrated in Figure~\ref{Fig:Toy:BestSelection_1D}~(left).
As the efficiency of picking the right candidate is improved, 
the measured \Bz yield increases linearly from the value
obtained with random picking to that of a perfect selection. 
The biases caused by random picking (Eq.~\ref{eq:NBgen}) are mitigated by 
arbitration as \PBbest approaches unity,
\eq[eq:NBgen2]{\left(\NBfit\right)^\text{arbitration}=\NBgen\left(1-\PBbest\PBsig\right).}
Similarly, the effect of the arbitration procedure on the biases of \Araw can be obtained by multiplying
the values of Figure~\ref{Fig:Toy:SigACP}~(left) by $2(1-\PBbest)$.

With increasing values of \PBbest, the relative uncertainty on the \Bz yield decreases and becomes better than
that obtained using all candidates.
The value of the crossing point in Fig.~\ref{Fig:Toy:BestSelection_1D} (right)
depends on \PBbest, \PBsig and \PBbkg~\cite{supplementary}.
It is to be noted again that the relative uncertainty varies very little.
As for random picking, the gain in statistical sensitivity is small
compared to the potential biases due to potentially different values of \PBsig and \PBbsig.

\begin{figure}[t]
  \IG[0.48]{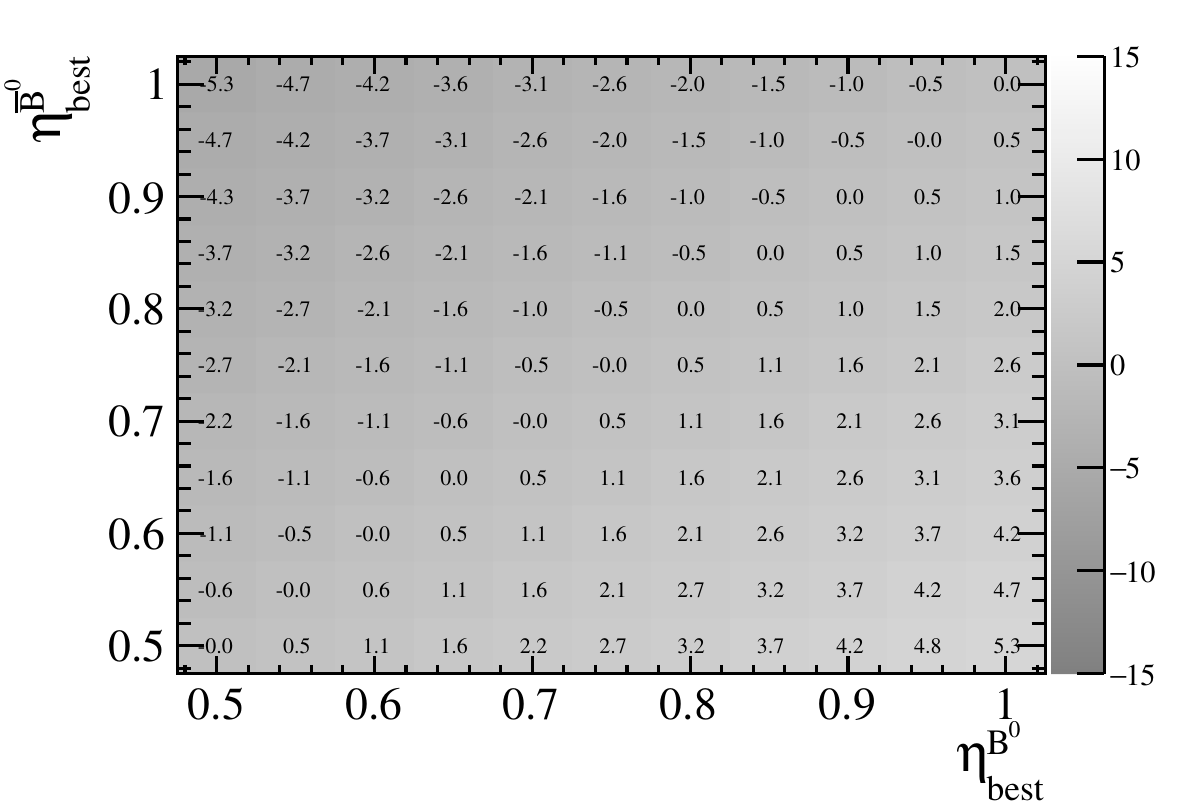}\hskip 0.04\textwidth
  \IG[0.48]{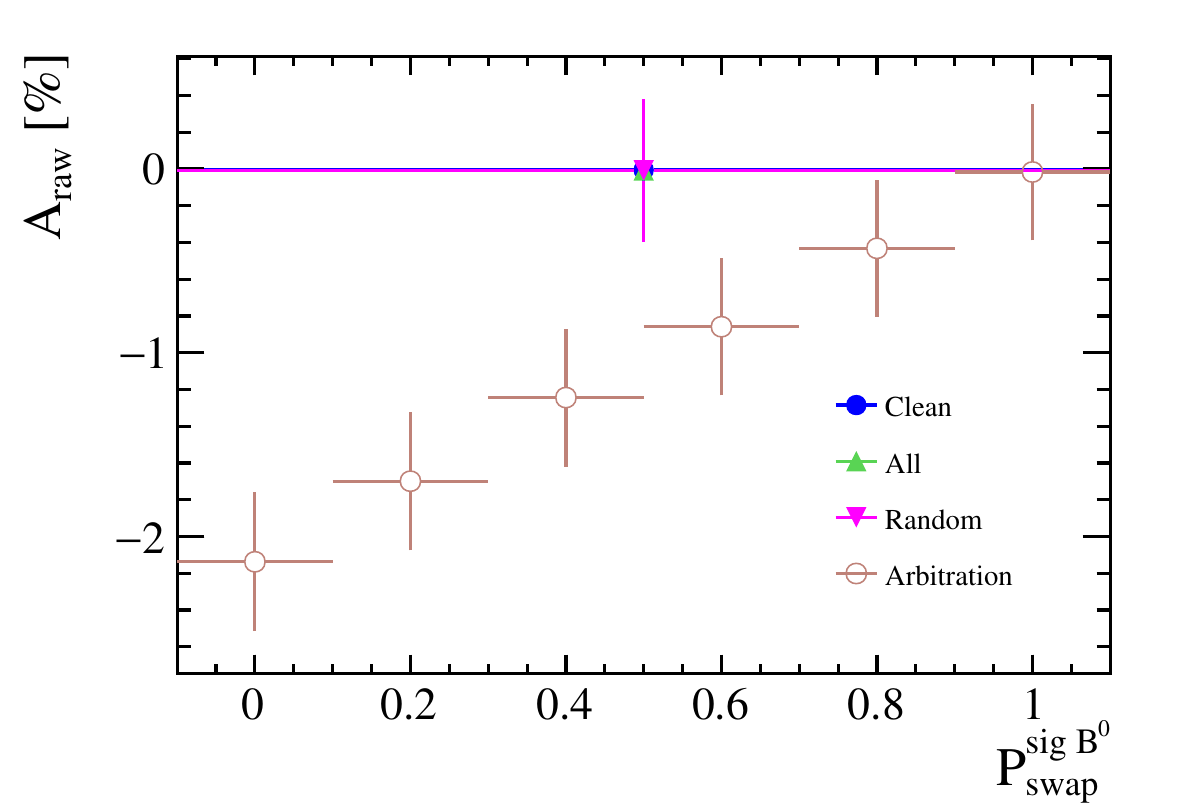}
  \caption{\Araw (in percent)
    versus (left) efficiencies of picking the signal candidate \PBbest and \PBbbest,
    and (right) versus flavour swapping probability \PBswapB.
    Other parameters are set to
    $\PBsig=\PBbsig=\PBbkg=\PBbbkg=0.2$.
    Additionally in the right plot, $\PBbest=\PBbbestTo=0.6$,
    $\PBbbest=\PBbestTo=0.8$ and
    $\PBbswapB=\PBswapBkg=0$.
    $\PBswapB=\PBswapBkg$ is imposed. 
    The vertical error bars represent the average uncertainty on the fitted asymmetry.
  }\label{Fig:Toy:BestSelection}\label{Fig:Toy:Bswap}
\end{figure}

A best-candidate selection only helps if the efficiency for selecting the correct
candidate is very large.
Otherwise, additional asymmetries can be generated, even if none are present in the data
prior to this operation.
The best candidate selection efficiency can be different for \Bz and \Bzb mesons, $\PBbbest\ne\PBbest$,
if the distributions of the $O$ observable
are different for \Bz and \Bzb signal,
background, or both.
Such an asymmetry of selection efficiencies can induce biases of the \Bz and \Bzb yields, generating
raw asymmetries, as shown in Figure~\ref{Fig:Toy:BestSelection}.
In the (unlikely) extreme cases, raw asymmetries of $5\%$ can be reached.

In a real measurement, analysts would not pick an observable which can generate
large asymmetries. However, even if checked, this fact is hardly ever reported
in publications. Also, the asymmetry of the observable is only known 
to a given precision, which in principle should be determined and
assigned as a systematic uncertainty~\cite{Negligible}.

\withSections{\subsection{Flavour swaps}\label{Sec:Toy:Swap}}{}
In the above, it is assumed that all companions have the
same flavour as the original candidate,
or that it is irrelevant. This is not necessarily the case. 
For overlapping candidates, the flavour of the
signal \B meson and that of the other candidate may be correlated, or anti-correlated~\cite{swaps2}.
The consequence are non-identical probabilities of being accompanied by a
candidate of different flavour \PBswapB, \PBbswapB for \Bz and \Bzb signal, and
\PBswapBkg, \PBbswapBkg for \Bz and \Bzb background.

In Figure~\ref{Fig:Toy:Bswap} (right)
the swapping probabilities \PBswapB and \PBbswapB are varied
(synchronously with \PBswapBkg and \PBbswapBkg, which have no effect),
for $\PBbest=0.6$ and $\PBbbest=0.8$.
Values of $-2.1\%$ for \Araw are obtained when the swapping
probabilities are identical. This corresponds to the
point at $\PBbest=0.6,\PBbbest=0.8$ in Fig.~\ref{Fig:Toy:BestSelection}~(left).
Different swapping probabilities can mitigate the problem
up to hiding it completely, as for the extreme case $\PBswapB=1$~\cite{NotGood}.
\withSections{\subsection{Continuous observables: lifetime and mass measurements}\label{Sec:Toy:Time}}{}
Next, the data described above is used to measure the lifetime of \Bz mesons. 
It is assumed that the decay-time distributions of the 
signal and background follow a falling exponential of constants $-\tauB$ and $-\tauB/3$,
respectively.
For companion candidates overlapping with the signal this constant is taken to be
$-\tauB/2$~\cite{TauMotivation}.

\begin{figure}[t]
\IG[0.48]{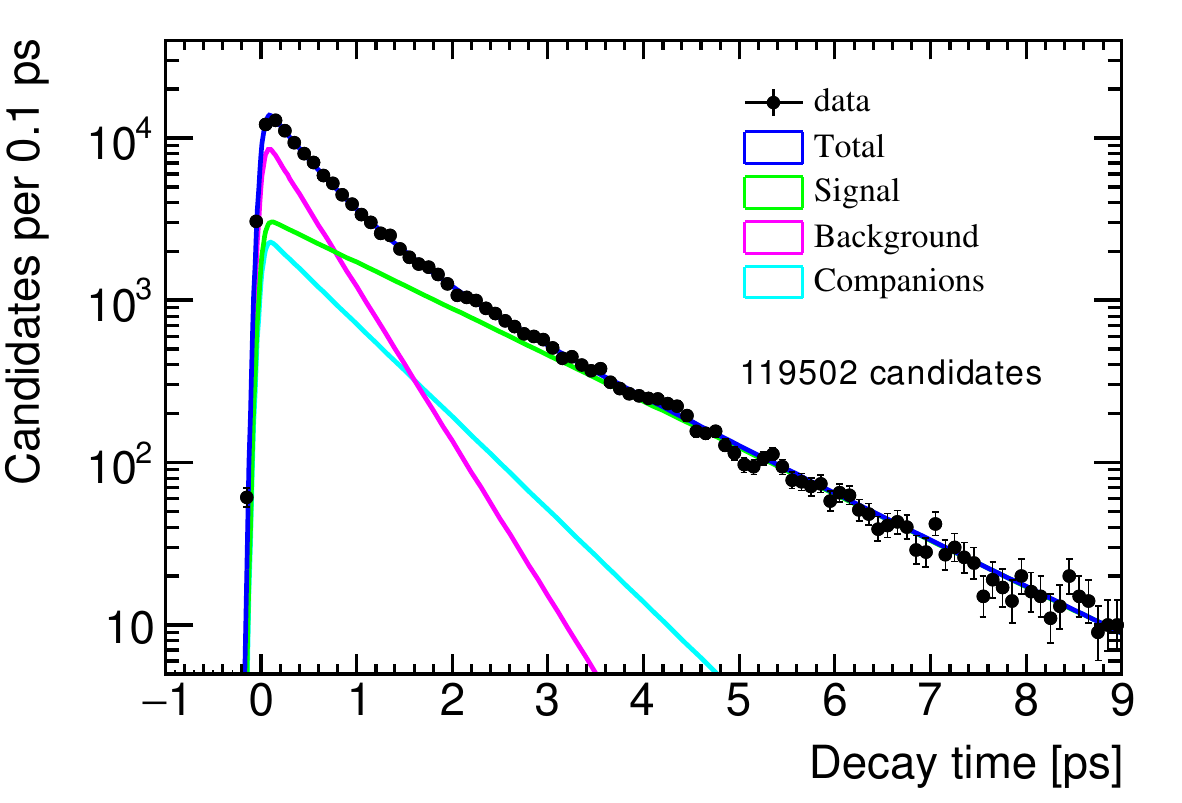}\hskip 0.04\textwidth
\IG[0.48]{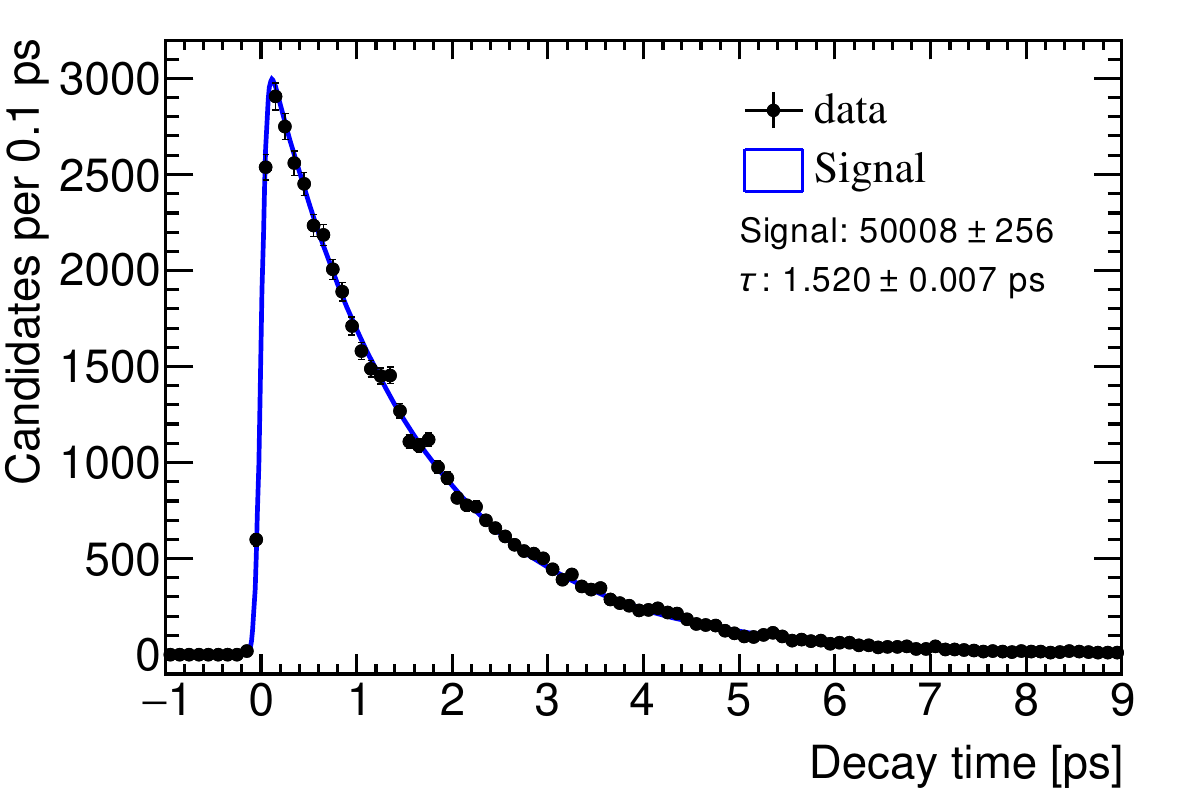}
\caption{Typical decay-time distribution after companion addition 
  for (left) all candidates and (right) background-subtracted signal.
  As in Fig.~\ref{Fig:Peaks} (left),
  the companion probability is
  $\PBsig=0.4$. All other probabilities of
  Table~\ref{Tab:Symbols} are set to zero.}\label{Fig:DecayTime}
\end{figure}

The three exponential distributions are convolved with a Gaussian-shaped resolution function~\cite{Resolution}.
No time-depended selection effects are considered. 
This situation leads to a distribution as shown in Fig.~\ref{Fig:DecayTime}~(left),
where the three contributions are shown. 

\begin{figure}[t]
  \IG[0.48]{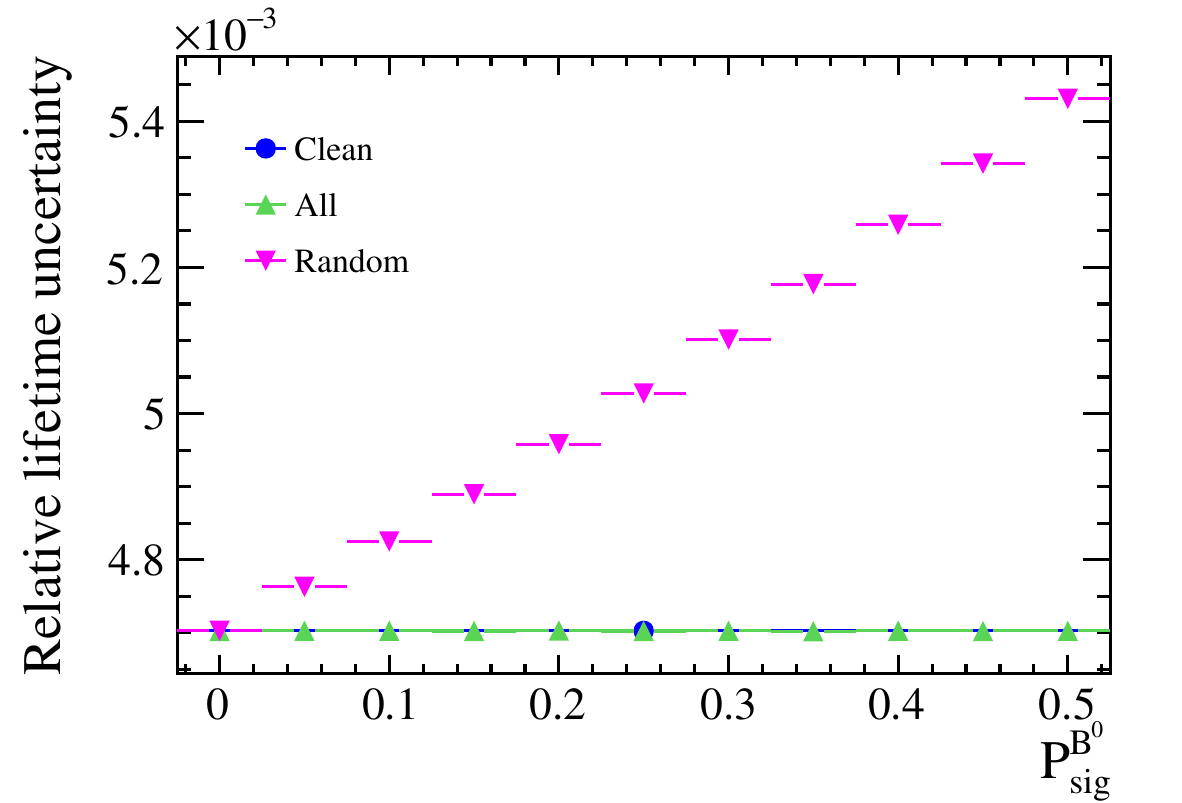}\hskip 0.04\textwidth
  \IG[0.48]{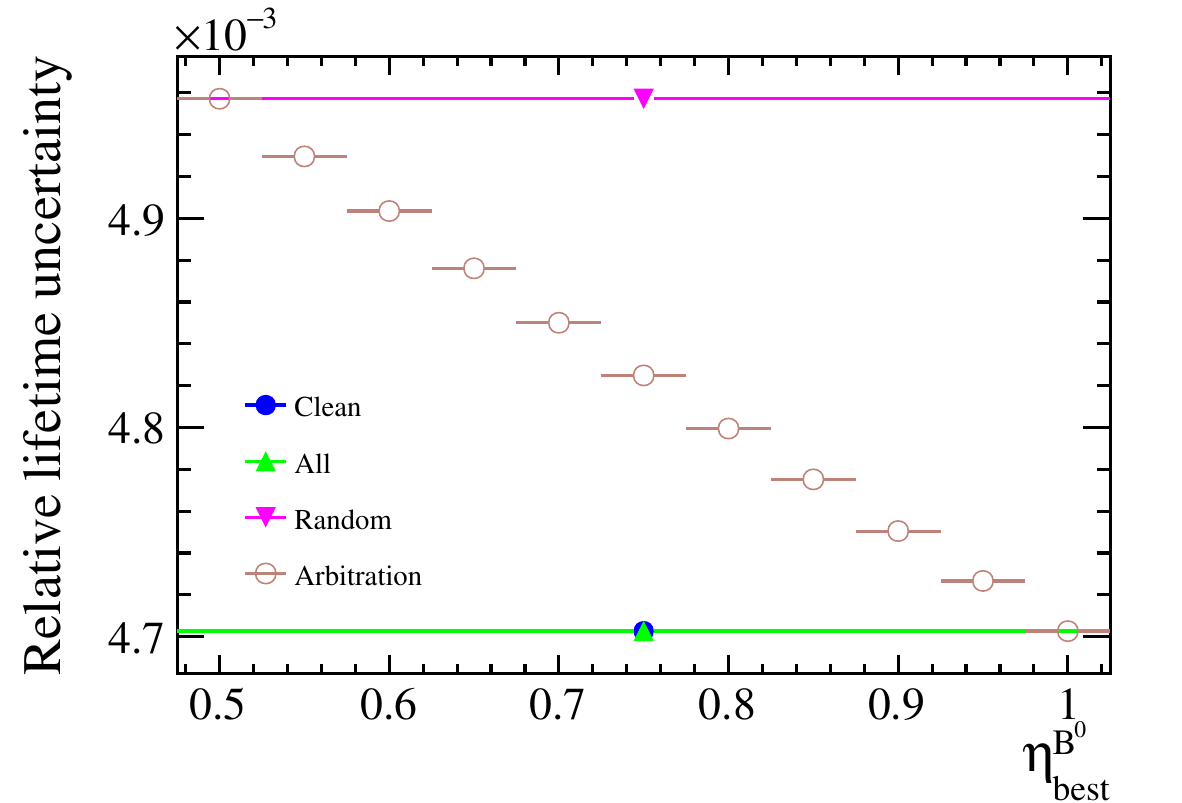}
  \caption{Relative uncertainty on lifetime
           versus (left) \PBsig and (right) \PBbest.
    All other probabilities are set to zero in the former case,
    while $\PBsig=\PBbkg=0.2$ in the latter case.
  }\label{Fig:Toy:Lifetime_1D}
\end{figure}

If the mass and decay time are not correlated for the three above-defined
species, the {\it sPlot} technique~\cite{Pivk:2004ty,sFit} can be used to statistically 
subtract the background and companion candidates using their distinct mass distributions.
The decay-time distribution for signal is thus obtained
with an unbinned maximum-likelihood fit of a single exponential to the
background-subtracted data, as
shown in Fig.~\ref{Fig:DecayTime} (right).
As companion candidates do not peak in mass they do not affect the signal
decay-time distribution and no biases can be caused~\cite{Peaking}.
The additional
background causes an increased relative uncertainty on the fitted
lifetime value, as shown in Fig.~\ref{Fig:Toy:Lifetime_1D}.
Taking all candidates does not bias the measurement,
and has a negligible effect on the uncertainty.
Random picking and arbitration degrade the resolution
due to the loss of signal.

The lifetime can also be determined
by fitting all data with a signal and one or several
background components~\cite{LHCb-PAPER-2014-038}.
If the specific features of overlapping companions
are ignored, only one component for the background is used.
\begin{figure}[t]
  \IG[0.48]{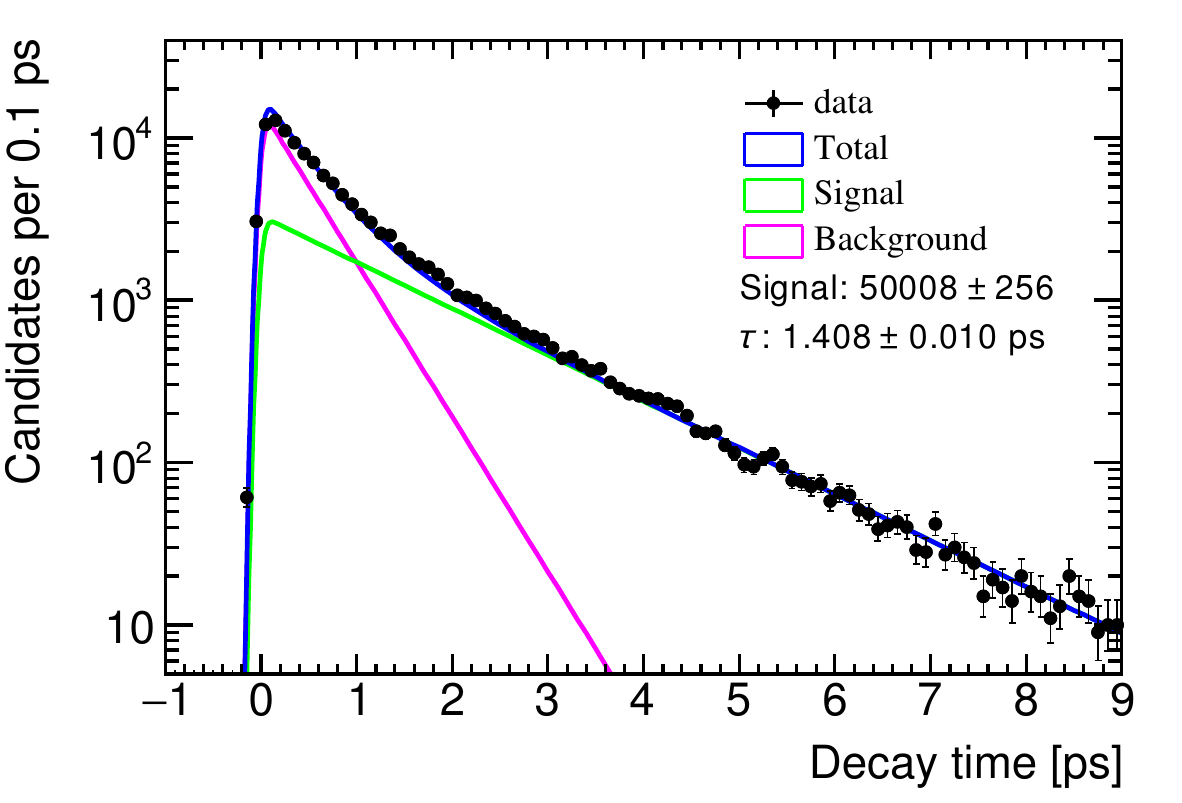}\hskip 0.04\textwidth
  \IG[0.48]{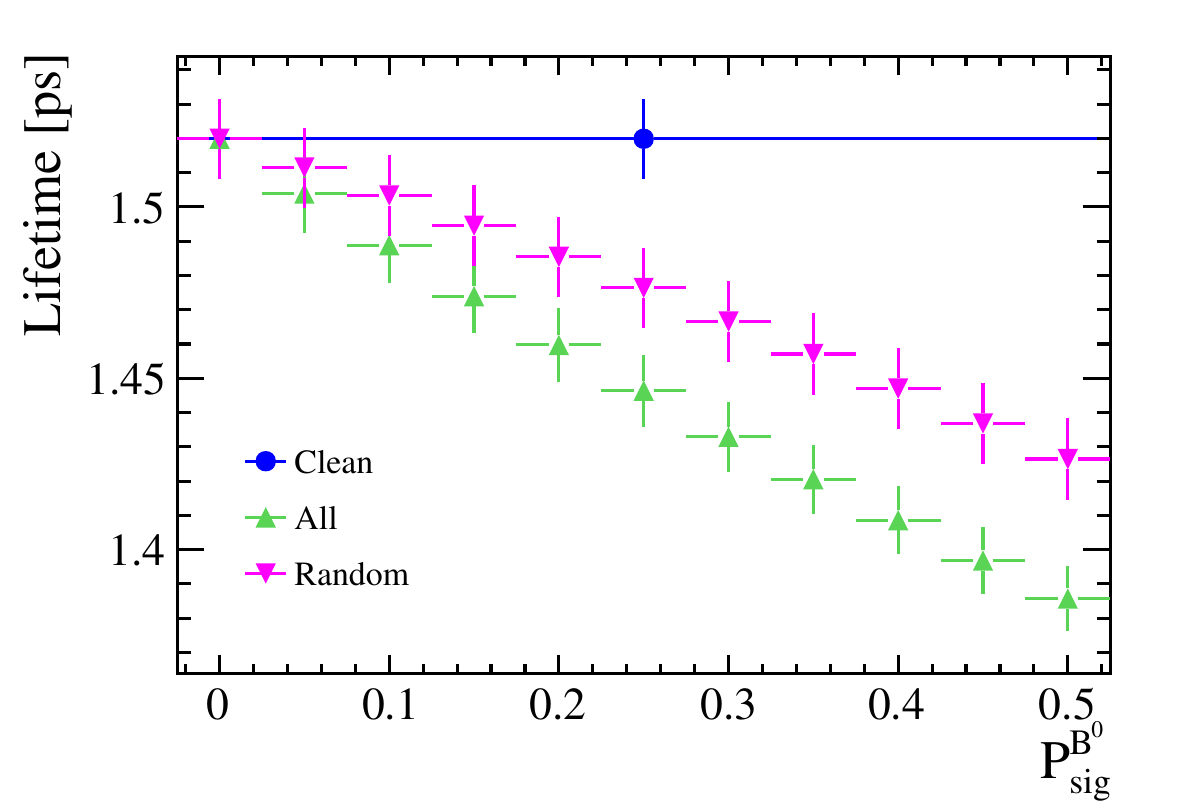}
  \caption{(left) Two-component fit to the lifetime data which was generated
  with three exponential functions.
  The companion probability is $\PBsig=0.4$. All other probabilities of
  Table~\ref{Tab:Symbols} are set to zero. (right) Measured lifetime
  versus \PBsig. The vertical error bars represent the average uncertainty on the
  fitted lifetime.
  }\label{Fig:Toy:BiasLifetime_1D}
\end{figure}
Such a fit is shown in Fig.~\ref{Fig:Toy:BiasLifetime_1D} (left).
The measured lifetime is biased, as seen in Fig.~\ref{Fig:Toy:BiasLifetime_1D} (right).
For large data samples, the missing component would lead to poor
fits, as in this example. However, it could easily be missed in case of 
smaller samples. Sums of similar exponential distributions
can be well fitted by a single exponential~\cite{DeBruyn:2012wj},
resulting in biased lifetime results. 

This issue is cured by the use of the {\it sPlot} technique, as shown above. 
But this technique does not apply to quantities which are correlated with the 
discriminating variable. 
For instance the measurement of the mass of a resonance requires a good
understanding of the background shape. If overlapping companions tend to accumulate at
masses close to but, somewhat lower than, the signal, the fit could 
return a biased value~\cite{HiggsMass,Chatrchyan:2013mxa,Aad:2015zhl,Aad:2014aba}.


\withSections{\subsection{Efficiency corrections}\label{Sec:Efficiencies}}{\vskip 1em}%

Biases can be caused by the presence of companion candidates,
or by the technique used to remove them. 
These biases can in principle be corrected for, but this correction requires
a good knowledge of their sizes.
The efficiency of the arbitration technique is $1-(1-\PBbest)\Rall$,
where the fraction of events with multiple candidates $\Rall$ is given
in Eq.~\ref{eq:MCrate}.
For a random selection $\PBbest=\sfrac{1}{2}$.
The fraction $\Rall$ is often reported in publications, 
giving the reader an estimate of the scale of the problem.
For a quantitative estimate of the effect of the multiple-candidate
handling procedure, \PBbest is needed, as well as 
the fraction of signal events with multiple candidates, $\Rsignal$
(defined in analogy with \Rall, setting $\NBKgen=0$ in Eq.~\ref{eq:MCrate}).
Only rarely are both numbers given.

Simulated signal events can be used to determine these numbers, provided that 
the simulation properly describes the companion candidate
rate and distribution. 
At \B or charm factories, where the underlying event is the other heavy 
meson~\cite{Adam:2007pv,Choi:2015lpc}, the simulation is often reported 
to correctly model multiple candidates.
The description of the
underlying event at hadron colliders is less reliable~\cite{Khachatryan:2010nk,LHCb-PAPER-2013-070}.
The analyst may have to deal with a different fraction
and composition of companion candidates in the simulation and the data.
Different candidate removal efficiencies in data and simulation lead to a potentially
large systematic uncertainty, which is rarely reported in publications.
This uncertainty depends on the properties of the 
underlying event, while prior to candidate removal only the signal  
efficiency is relevant.

\begin{figure}[t]
  \IG[0.48]{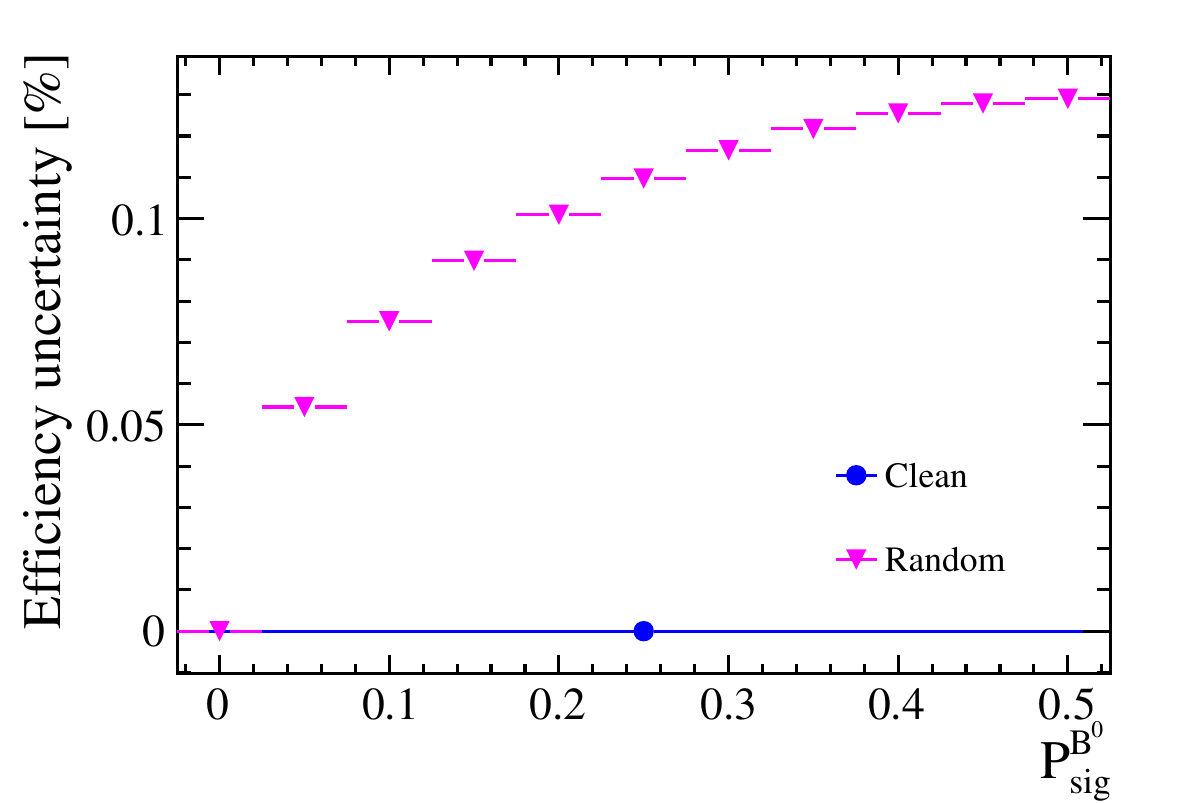}\hskip 0.04\textwidth
  \IG[0.48]{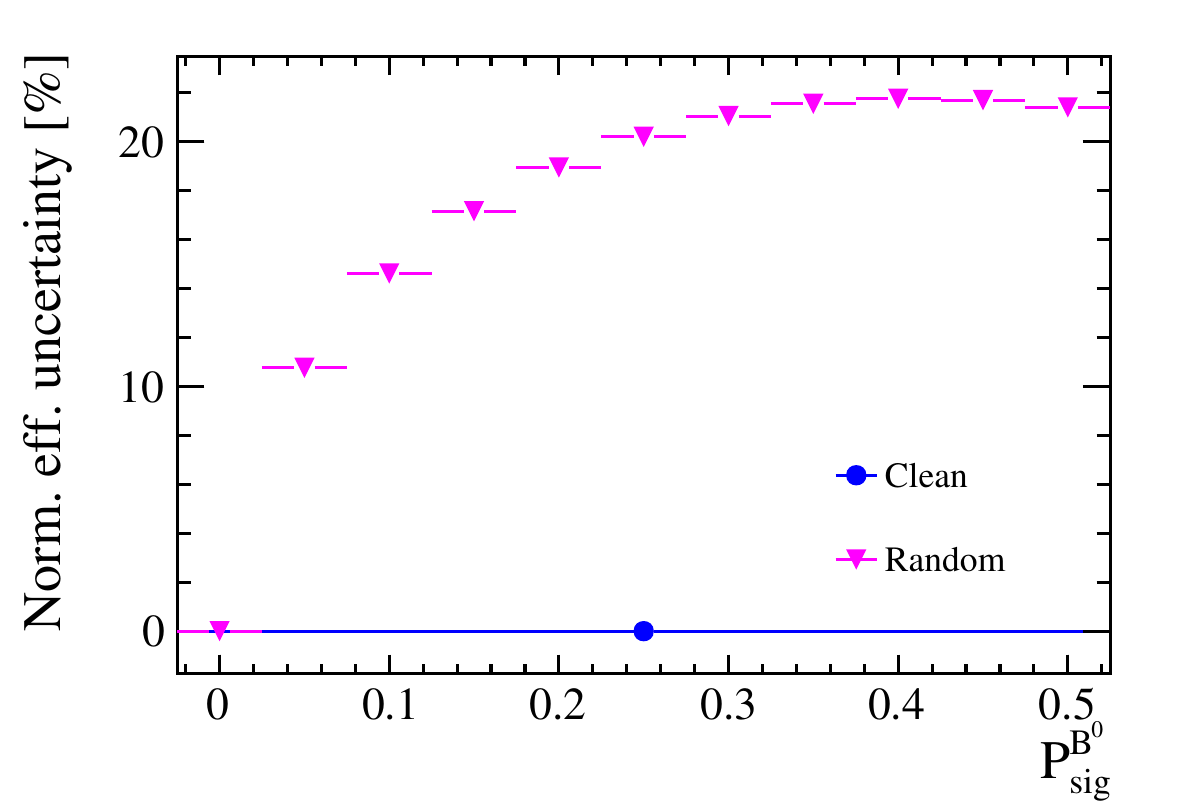}
  \caption{
  (left) Uncertainty on the  efficiency of randomly picking the true signal \B versus \PBsig.
  (right) Ratio of this uncertainty to the relative uncertainty on the signal yield
  (Fig.~\ref{Fig:Toy:SigB0All_Random_Clean_1D}, bottom).
  All other probabilities of %
  Table~\ref{Tab:Symbols} are set to zero.  
  }\label{Fig:Toy:EffCorrection}
\end{figure}

The efficiency can also be assessed using data, for instance 
applying the {\it sPlot} technique used in Sec.~\ref{Sec:Toy:Time}, or similar 
background-subtraction techniques. 
In the present set of pseudoexperiments,
the efficiency of random picking is determined without bias~\cite{sPlot}
from the candidate multiplicity in events with signal.
The uncertainty on this efficiency is in the
per-mille range, as shown in Fig.~\ref{Fig:Toy:EffCorrection} (left),
which is not negligible compared to the relative statistical
uncertainty on the signal yield (shown in Fig.~\ref{Fig:Toy:SigB0All_Random_Clean_1D}, bottom).
These two uncertainties are compared in Fig.~\ref{Fig:Toy:EffCorrection} (right). 
The systematic uncertainty related to the random picking efficiency
reaches up to 20\% of the statistical uncertainty in these pseudoexperiments.
It was already shown in Fig.~\ref{Fig:Toy:SigB0All_Random_Clean_1D} that
the statistical uncertainty of the randomly picked sample 
is larger than that of the sample with all candidates. 
The systematic uncertainty is an additional penalty.

Absolute efficiencies are difficult to determine at hadron colliders and 
thus potentially involve large systematic uncertainties.
It is therefore common practice to measure relative branching fractions
or cross-sections by normalising the signal to 
a well-known high-yield normalisation mode~\cite{LHCb-PAPER-2017-001,LHCb-PAPER-2011-041,LHCb-PAPER-2014-020,Chatrchyan:2013cda,LHCb-PAPER-2016-012}.
In this case, the ratio of candidate removal
efficiencies for the signal and normalisation modes matters. 
These efficiencies do not only depend on the signal and calibration modes,
but also on their respective backgrounds.
If the rate of companion candidates, as well as their properties, 
are not identical for both modes
(which is likely), a systematic uncertainty must be determined. 
In order to profit from the statistical power of the larger calibration sample,
analysts should demonstrate that the behaviour of companion candidates
is the same for the signal and the calibration mode. 
Otherwise, the systematic uncertainty described above applies.

\withSections{\section{Recommendations}\label{Sec:Recommendations}}{\vskip 1em}%
{In the following recommendations are given.}%
\begin{enumerate}\setlength\itemsep{-0.2em}%
\item Analysts should report the candidate multiplicity and 
   to which category the candidates belong\withSections{ (see in Section~\ref{Sec:Defs})}{}.
   It should be specified if they contribute to
   (and potentially bias) the
   signal component or only to the background.
\item The handling of multiple candidates should be performed after the full selection has been applied.
\item Analysts should study the available techniques
      and select that which minimises the biases
  and the total uncertainty.
\item Not doing anything usually avoids biases at the price of a slightly higher background level.
      The effect on the statistical uncertainty is usually negligible.
      In presence of many events with overlapping candidates, the uncertainty coverage
      should be checked using pseudoexperiments.
\item If overlaps with signal are present, it should be checked whether they behave signal-
      or background-like. If neither, a dedicated fit component is required
      to avoid biases.
\item A candidate-removal technique is a selection requirement. It has an associated signal efficiency which should be reported.
\item Any candidate-removal technique can cause biases which need to be studied and may
     require a correction.       
      Differences between the signal and simulation or a control samples 
      should be assessed and a systematic uncertainty assigned.
\item In case of arbitration using a best-candidate selection, additional biases can be due to the 
  choice of the discriminating variable. This technique is discouraged. 
  \item If arbitration is nevertheless chosen, analysts should demonstrate that it improves
  the precision of the measurement and assign a systematic uncertainty.
\item The most important recommendation is that analysts should have a strategy and
      describe it in the publication. If there is a possibility for any bias 
      a systematic uncertainty should be assessed.
\end{enumerate}

\withSections{\section{Conclusion}\label{Sec:Conclusion}}{\vskip 1em}%
\withSections{This}{In conclusion this}
paper shows that the presence of multiple candidates can
cause biases in measurements of rates, asymmetries or particle properties which
are much larger than the statistical precision. These biases
may be difficult to estimate using simulation or control samples.

In most cases the least biasing technique
is to keep all candidates, at the price of a slightly higher statistical uncertainty.
It is recommended not to select a single best candidate per event, as this procedure may generate biases.

In publications analysts should report the rate and nature of multiple candidates, outline the
strategy how to deal with them and assess systematic uncertainties.
Detailed recommendations are given in Sec.~\ref{Sec:Recommendations}.

\withSections{\section*{Acknowledgements}}{\vskip 1em}%
The author thanks 
I.~Belyaev,
U.~Egede, 
V.~Gligorov,
N.~de~Groot,
T.~du~Pree,
T.~Gershon,
D.~Mart\'{\i}nez Santos, 
G.~Raven, 
T.~Ruf,
P.~Seyfert,
S.~Stahl,
S.~Stone,
N.~Tuning
and 
I.~van Vulpen
for useful discussions. In particular
M.~Merk,
L.~Dufour,
and 
W.~Hulsbergen are thanked for carefully reading the manuscript.

This work is part of the NWO Institute Organisation (NWO-I), 
which is financed by the Netherlands Organisation for Scientific Research (NWO).


\addcontentsline{toc}{section}{References}
\setboolean{inbibliography}{true}
\bibliographystyle{LHCb}
\bibliography{footnotes,main,LHCb-PAPER,LHCb-CONF,LHCb-DP,LHCb-TDR,exp,theory,standard}

\ifx\mcitethebibliography\mciteundefinedmacro
\PackageError{LHCb.bst}{mciteplus.sty has not been loaded}
{This bibstyle requires the use of the mciteplus package.}\fi
\providecommand{\href}[2]{#2}
\begin{mcitethebibliography}{100}
\mciteSetBstSublistMode{n}
\mciteSetBstMaxWidthForm{subitem}{\alph{mcitesubitemcount})}
\mciteSetBstSublistLabelBeginEnd{\mcitemaxwidthsubitemform\space}
{\relax}{\relax}

\bibitem{LHCb-PAPER-2019-006}
LHCb collaboration, R.~Aaij {\em et~al.},
  \ifthenelse{\boolean{articletitles}}{\emph{{Observation of \CP violation in
  charm decays}},
  }{}\href{https://doi.org/10.1103/PhysRevLett.117.211803}{Phys.\ Rev.\ Lett.\
  \textbf{122} (2019) 211803},
  \href{http://arxiv.org/abs/1903.08726}{{\normalfont\ttfamily
  arXiv:1903.08726}}\relax
\mciteBstWouldAddEndPuncttrue
\mciteSetBstMidEndSepPunct{\mcitedefaultmidpunct}
{\mcitedefaultendpunct}{\mcitedefaultseppunct}\relax
\EndOfBibitem
\bibitem{LHCb-TDR-012}
LHCb collaboration, \ifthenelse{\boolean{articletitles}}{\emph{{Framework TDR
  for the LHCb Upgrade: Technical Design Report}}, }{}
  \href{http://cdsweb.cern.ch/search?p=CERN-LHCC-2012-007&f=reportnumber&action_search=Search&c=LHCb+Reports}
  {CERN-LHCC-2012-007}, 2012\relax
\mciteBstWouldAddEndPuncttrue
\mciteSetBstMidEndSepPunct{\mcitedefaultmidpunct}
{\mcitedefaultendpunct}{\mcitedefaultseppunct}\relax
\EndOfBibitem
\bibitem{Aushev:2010bq}
T.~Aushev {\em et~al.}, \ifthenelse{\boolean{articletitles}}{\emph{{Physics at
  Super B Factory}},
  }{}\href{http://arxiv.org/abs/1002.5012}{{\normalfont\ttfamily
  arXiv:1002.5012}}\relax
\mciteBstWouldAddEndPuncttrue
\mciteSetBstMidEndSepPunct{\mcitedefaultmidpunct}
{\mcitedefaultendpunct}{\mcitedefaultseppunct}\relax
\EndOfBibitem
\bibitem{Azzi:2019yne}
HL-LHC, HE-LHC Working Group 1, P.~Azzi {\em et~al.},
  \ifthenelse{\boolean{articletitles}}{\emph{{Standard Model Physics at the
  HL-LHC and HE-LHC}},
  }{}\href{http://arxiv.org/abs/1902.04070}{{\normalfont\ttfamily
  arXiv:1902.04070}}\relax
\mciteBstWouldAddEndPuncttrue
\mciteSetBstMidEndSepPunct{\mcitedefaultmidpunct}
{\mcitedefaultendpunct}{\mcitedefaultseppunct}\relax
\EndOfBibitem
\bibitem{Cepeda:2019klc}
HL-LHC, HE-LHC Working Group 2, M.~Cepeda {\em et~al.},
  \ifthenelse{\boolean{articletitles}}{\emph{{Higgs Physics at the HL-LHC and
  HE-LHC}}, }{}\href{http://arxiv.org/abs/1902.00134}{{\normalfont\ttfamily
  arXiv:1902.00134}}\relax
\mciteBstWouldAddEndPuncttrue
\mciteSetBstMidEndSepPunct{\mcitedefaultmidpunct}
{\mcitedefaultendpunct}{\mcitedefaultseppunct}\relax
\EndOfBibitem
\bibitem{Cerri:2018ypt}
A.~Cerri {\em et~al.},
  \ifthenelse{\boolean{articletitles}}{\emph{{Opportunities in Flavour Physics
  at the HL-LHC and HE-LHC}},
  }{}\href{http://arxiv.org/abs/1812.07638}{{\normalfont\ttfamily
  arXiv:1812.07638}}\relax
\mciteBstWouldAddEndPuncttrue
\mciteSetBstMidEndSepPunct{\mcitedefaultmidpunct}
{\mcitedefaultendpunct}{\mcitedefaultseppunct}\relax
\EndOfBibitem
\bibitem{Abada:2019lih}
A.~Abada {\em et~al.}, \ifthenelse{\boolean{articletitles}}{\emph{{FCC Physics
  Opportunities}},
  }{}\href{https://doi.org/10.1140/epjc/s10052-019-6904-3}{Eur.\ Phys.\ J.\
  \textbf{C79} (2019) 474}\relax
\mciteBstWouldAddEndPuncttrue
\mciteSetBstMidEndSepPunct{\mcitedefaultmidpunct}
{\mcitedefaultendpunct}{\mcitedefaultseppunct}\relax
\EndOfBibitem
\bibitem{Aad:2012tfa}
ATLAS collaboration, G.~Aad {\em et~al.},
  \ifthenelse{\boolean{articletitles}}{\emph{{Observation of a new particle in
  the search for the Standard Model Higgs boson with the ATLAS detector at the
  LHC}}, }{}\href{https://doi.org/10.1016/j.physletb.2012.08.020}{Phys.\ Lett.\
   \textbf{B716} (2012) 1},
  \href{http://arxiv.org/abs/1207.7214}{{\normalfont\ttfamily
  arXiv:1207.7214}}\relax
\mciteBstWouldAddEndPuncttrue
\mciteSetBstMidEndSepPunct{\mcitedefaultmidpunct}
{\mcitedefaultendpunct}{\mcitedefaultseppunct}\relax
\EndOfBibitem
\bibitem{Chatrchyan:2012ufa}
CMS collaboration, S.~Chatrchyan {\em et~al.},
  \ifthenelse{\boolean{articletitles}}{\emph{{Observation of a new boson at a
  mass of 125 GeV with the CMS experiment at the LHC}},
  }{}\href{https://doi.org/10.1016/j.physletb.2012.08.021}{Phys.\ Lett.\
  \textbf{B716} (2012) 30},
  \href{http://arxiv.org/abs/1207.7235}{{\normalfont\ttfamily
  arXiv:1207.7235}}\relax
\mciteBstWouldAddEndPuncttrue
\mciteSetBstMidEndSepPunct{\mcitedefaultmidpunct}
{\mcitedefaultendpunct}{\mcitedefaultseppunct}\relax
\EndOfBibitem
\bibitem{LHCb-PAPER-2017-001}
LHCb collaboration, R.~Aaij {\em et~al.},
  \ifthenelse{\boolean{articletitles}}{\emph{{Measurement of the
  \mbox{\decay{\Bs}{\mumu}} branching fraction and effective lifetime and
  search for \mbox{\decay{\Bz}{\mumu}} decays}},
  }{}\href{https://doi.org/10.1103/PhysRevLett.118.191801}{Phys.\ Rev.\ Lett.\
  \textbf{118} (2017) 191801},
  \href{http://arxiv.org/abs/1703.05747}{{\normalfont\ttfamily
  arXiv:1703.05747}}\relax
\mciteBstWouldAddEndPuncttrue
\mciteSetBstMidEndSepPunct{\mcitedefaultmidpunct}
{\mcitedefaultendpunct}{\mcitedefaultseppunct}\relax
\EndOfBibitem
\bibitem{Aaboud:2016ire}
ATLAS collaboration, M.~Aaboud {\em et~al.},
  \ifthenelse{\boolean{articletitles}}{\emph{{Study of the rare decays of
  $B^0_s$ and $B^0$ into muon pairs from data collected during the LHC Run 1
  with the ATLAS detector}},
  }{}\href{https://doi.org/10.1140/epjc/s10052-016-4338-8}{Eur.\ Phys.\ J.\
  \textbf{C76} (2016) 513},
  \href{http://arxiv.org/abs/1604.04263}{{\normalfont\ttfamily
  arXiv:1604.04263}}\relax
\mciteBstWouldAddEndPuncttrue
\mciteSetBstMidEndSepPunct{\mcitedefaultmidpunct}
{\mcitedefaultendpunct}{\mcitedefaultseppunct}\relax
\EndOfBibitem
\bibitem{LHCbINT}
Internal recommendations exist in LHCb and probably other collaborations. But
  these are not available for external readers. The present document is based
  on the LHCb recommendations.\relax
\mciteBstWouldAddEndPunctfalse
\mciteSetBstMidEndSepPunct{\mcitedefaultmidpunct}
{}{\mcitedefaultseppunct}\relax
\EndOfBibitem
\bibitem{expectation}
As an example, the likelihood of multiple signal candidate events can be
  determined in the case of \decay{\Dstarpm}{\DorDbar\pipm} with
  \decay{\DorDbar}{\Kp\Km} from Ref.~\cite{LHCb-PAPER-2019-006}. The
  probability of seeing another signal \Dstarpm meson in the same event (which
  thus contains a charm quark pair) is given by the \Dstarpm hadronisation
  fraction, multiplied by the branching fractions and the experimental
  efficiency. Using the \cquark{}\cquarkbar cross-section and hadronisation
  fractions from Ref.~\cite{LHCb-PAPER-2015-041}, one can infer this
  probability to be less than $10^{-6}$, while the multiple candidate rate is
  reported as about $10\%$.\relax
\mciteBstWouldAddEndPunctfalse
\mciteSetBstMidEndSepPunct{\mcitedefaultmidpunct}
{}{\mcitedefaultseppunct}\relax
\EndOfBibitem
\bibitem{LHCb-PAPER-2012-041}
LHCb collaboration, R.~Aaij {\em et~al.},
  \ifthenelse{\boolean{articletitles}}{\emph{{Prompt charm production in
  \proton\proton collisions at \mbox{$\sqs=$7\tev}}},
  }{}\href{https://doi.org/10.1016/j.nuclphysb.2013.02.010}{Nucl.\ Phys.\
  \textbf{B871} (2013) 1},
  \href{http://arxiv.org/abs/1302.2864}{{\normalfont\ttfamily
  arXiv:1302.2864}}\relax
\mciteBstWouldAddEndPuncttrue
\mciteSetBstMidEndSepPunct{\mcitedefaultmidpunct}
{\mcitedefaultendpunct}{\mcitedefaultseppunct}\relax
\EndOfBibitem
\bibitem{Proceedings}
This assumption may occasionally be wrong, as shown in conference proceedings
  which sometimes give more information than the corresponding
  publication.\relax
\mciteBstWouldAddEndPunctfalse
\mciteSetBstMidEndSepPunct{\mcitedefaultmidpunct}
{}{\mcitedefaultseppunct}\relax
\EndOfBibitem
\bibitem{LHCb-PAPER-2011-023}
LHCb collaboration, R.~Aaij {\em et~al.},
  \ifthenelse{\boolean{articletitles}}{\emph{{Evidence for \CP violation in
  time-integrated \mbox{\decay{\Dz}{h^-h^+}} decay rates}},
  }{}\href{https://doi.org/10.1103/PhysRevLett.108.111602}{Phys.\ Rev.\ Lett.\
  \textbf{108} (2012) 111602},
  \href{http://arxiv.org/abs/1112.0938}{{\normalfont\ttfamily
  arXiv:1112.0938}}\relax
\mciteBstWouldAddEndPuncttrue
\mciteSetBstMidEndSepPunct{\mcitedefaultmidpunct}
{\mcitedefaultendpunct}{\mcitedefaultseppunct}\relax
\EndOfBibitem
\bibitem{Aad:2012oxa}
ATLAS collaboration, G.~Aad {\em et~al.},
  \ifthenelse{\boolean{articletitles}}{\emph{{Search for a standard model Higgs
  boson in the mass range 200-600 GeV in the $H \to ZZ \to \ell^+ \ell^- q
  \bar{q}$ decay channel with the ATLAS detector}},
  }{}\href{https://doi.org/10.1016/j.physletb.2012.09.020}{Phys.\ Lett.\
  \textbf{B717} (2012) 70},
  \href{http://arxiv.org/abs/1206.2443}{{\normalfont\ttfamily
  arXiv:1206.2443}}\relax
\mciteBstWouldAddEndPuncttrue
\mciteSetBstMidEndSepPunct{\mcitedefaultmidpunct}
{\mcitedefaultendpunct}{\mcitedefaultseppunct}\relax
\EndOfBibitem
\bibitem{LHCb-PAPER-2014-061}
LHCb collaboration, R.~Aaij {\em et~al.},
  \ifthenelse{\boolean{articletitles}}{\emph{{Observation of two new \Xibm
  baryon resonances}},
  }{}\href{https://doi.org/10.1103/PhysRevLett.114.062004}{Phys.\ Rev.\ Lett.\
  \textbf{114} (2015) 062004},
  \href{http://arxiv.org/abs/1411.4849}{{\normalfont\ttfamily
  arXiv:1411.4849}}\relax
\mciteBstWouldAddEndPuncttrue
\mciteSetBstMidEndSepPunct{\mcitedefaultmidpunct}
{\mcitedefaultendpunct}{\mcitedefaultseppunct}\relax
\EndOfBibitem
\bibitem{Lees:2015jwa}
BaBar collaboration, {Lees, J.\  P.\ } {\em et~al.},
  \ifthenelse{\boolean{articletitles}}{\emph{{Search for a light Higgs
  resonance in radiative decays of the $\PUpsilon(1S)$ with a charm tag}},
  }{}\href{https://doi.org/10.1103/PhysRevD.91.071102}{Phys.\ Rev.\
  \textbf{D91} (2015) 071102},
  \href{http://arxiv.org/abs/1502.06019}{{\normalfont\ttfamily
  arXiv:1502.06019}}\relax
\mciteBstWouldAddEndPuncttrue
\mciteSetBstMidEndSepPunct{\mcitedefaultmidpunct}
{\mcitedefaultendpunct}{\mcitedefaultseppunct}\relax
\EndOfBibitem
\bibitem{LHCb-PAPER-2015-055}
LHCb collaboration, R.~Aaij {\em et~al.},
  \ifthenelse{\boolean{articletitles}}{\emph{{Measurement of the difference of
  time-integrated \CP asymmetries in \mbox{\decay{\Dz}{\Km\Kp}} and
  \mbox{\decay{\Dz}{\pim\pip}} decays}},
  }{}\href{https://doi.org/10.1103/PhysRevLett.116.191601}{Phys.\ Rev.\ Lett.\
  \textbf{116} (2016) 191601},
  \href{http://arxiv.org/abs/1602.03160}{{\normalfont\ttfamily
  arXiv:1602.03160}}\relax
\mciteBstWouldAddEndPuncttrue
\mciteSetBstMidEndSepPunct{\mcitedefaultmidpunct}
{\mcitedefaultendpunct}{\mcitedefaultseppunct}\relax
\EndOfBibitem
\bibitem{LHCb-PAPER-2016-026}
LHCb collaboration, R.~Aaij {\em et~al.},
  \ifthenelse{\boolean{articletitles}}{\emph{{Amplitude analysis of
  \mbox{\decay{\Bm}{\Dp \pim \pim}} decays}},
  }{}\href{https://doi.org/10.1103/PhysRevD.94.072001}{Phys.\ Rev.\
  \textbf{D94} (2016) 072001},
  \href{http://arxiv.org/abs/1608.01289}{{\normalfont\ttfamily
  arXiv:1608.01289}}\relax
\mciteBstWouldAddEndPuncttrue
\mciteSetBstMidEndSepPunct{\mcitedefaultmidpunct}
{\mcitedefaultendpunct}{\mcitedefaultseppunct}\relax
\EndOfBibitem
\bibitem{LHCb-PAPER-2016-029}
LHCb collaboration, R.~Aaij {\em et~al.},
  \ifthenelse{\boolean{articletitles}}{\emph{{Search for structure in the
  $\Bs\pipm$ invariant mass spectrum}},
  }{}\href{https://doi.org/10.1103/PhysRevLett.117.152003}{Phys.\ Rev.\ Lett.\
  \textbf{117} (2016) 152003},
  \href{http://arxiv.org/abs/1608.00435}{{\normalfont\ttfamily
  arXiv:1608.00435}}\relax
\mciteBstWouldAddEndPuncttrue
\mciteSetBstMidEndSepPunct{\mcitedefaultmidpunct}
{\mcitedefaultendpunct}{\mcitedefaultseppunct}\relax
\EndOfBibitem
\bibitem{LHCb-PAPER-2016-031}
LHCb collaboration, R.~Aaij {\em et~al.},
  \ifthenelse{\boolean{articletitles}}{\emph{{Measurement of the \bquark-quark
  production cross-section in 7 and 13\tev \proton\proton collisions}},
  }{}\href{https://doi.org/10.1103/PhysRevLett.118.052002}{Phys.\ Rev.\ Lett.\
  \textbf{118} (2017) 052002}, Erratum
  \href{https://doi.org/10.1103/PhysRevLett.119.169901}{ibid.\   \textbf{119}
  (2017) 169901}, \href{http://arxiv.org/abs/1612.05140}{{\normalfont\ttfamily
  arXiv:1612.05140}}\relax
\mciteBstWouldAddEndPuncttrue
\mciteSetBstMidEndSepPunct{\mcitedefaultmidpunct}
{\mcitedefaultendpunct}{\mcitedefaultseppunct}\relax
\EndOfBibitem
\bibitem{LHCb-PAPER-2016-061}
LHCb collaboration, R.~Aaij {\em et~al.},
  \ifthenelse{\boolean{articletitles}}{\emph{{Study of the $\Dz\proton$
  amplitude in \mbox{\decay{\Lb}{\Dz\proton\pim}} decays}},
  }{}\href{https://doi.org/10.1007/JHEP05(2017)030}{JHEP \textbf{05} (2017)
  030}, \href{http://arxiv.org/abs/1701.07873}{{\normalfont\ttfamily
  arXiv:1701.07873}}\relax
\mciteBstWouldAddEndPuncttrue
\mciteSetBstMidEndSepPunct{\mcitedefaultmidpunct}
{\mcitedefaultendpunct}{\mcitedefaultseppunct}\relax
\EndOfBibitem
\bibitem{LHCb-PAPER-2017-044}
LHCb collaboration, R.~Aaij {\em et~al.},
  \ifthenelse{\boolean{articletitles}}{\emph{{Search for \CP violation in
  \mbox{\decay{\Lc}{p \Km \Kp}} and \mbox{\decay{\Lc}{p\pim\pip}} decays}},
  }{}\href{https://doi.org/10.1007/JHEP03(2018)182}{JHEP \textbf{03} (2018)
  182}, \href{http://arxiv.org/abs/1712.07051}{{\normalfont\ttfamily
  arXiv:1712.07051}}\relax
\mciteBstWouldAddEndPuncttrue
\mciteSetBstMidEndSepPunct{\mcitedefaultmidpunct}
{\mcitedefaultendpunct}{\mcitedefaultseppunct}\relax
\EndOfBibitem
\bibitem{LHCb-PAPER-2018-005}
LHCb collaboration, R.~Aaij {\em et~al.},
  \ifthenelse{\boolean{articletitles}}{\emph{{Observation of the decay
  \mbox{\decay{\Lb}{\Lc p \antiproton \pim}}}},
  }{}\href{https://doi.org/10.1016/j.physletb.2018.07.033}{Phys.\ Lett.\
  \textbf{B784} (2018) 101},
  \href{http://arxiv.org/abs/1804.09617}{{\normalfont\ttfamily
  arXiv:1804.09617}}\relax
\mciteBstWouldAddEndPuncttrue
\mciteSetBstMidEndSepPunct{\mcitedefaultmidpunct}
{\mcitedefaultendpunct}{\mcitedefaultseppunct}\relax
\EndOfBibitem
\bibitem{LHCb-PAPER-2018-021}
LHCb collaboration, R.~Aaij {\em et~al.},
  \ifthenelse{\boolean{articletitles}}{\emph{{Prompt \Lc production in
  \proton{}Pb collisions at $\sqsnn = 5.02$\tev}},
  }{}\href{https://doi.org/10.1007/JHEP02(2019)102}{JHEP \textbf{02} (2019)
  102}, \href{http://arxiv.org/abs/1809.01404}{{\normalfont\ttfamily
  arXiv:1809.01404}}\relax
\mciteBstWouldAddEndPuncttrue
\mciteSetBstMidEndSepPunct{\mcitedefaultmidpunct}
{\mcitedefaultendpunct}{\mcitedefaultseppunct}\relax
\EndOfBibitem
\bibitem{LHCb-PAPER-2018-047}
LHCb collaboration, R.~Aaij {\em et~al.},
  \ifthenelse{\boolean{articletitles}}{\emph{{Measurement of the mass and
  production rate of \Xibm baryons}},
  }{}\href{https://doi.org/10.1103/PhysRevD.99.052006}{Phys.\ Rev.\
  \textbf{D99} (2019) 052006},
  \href{http://arxiv.org/abs/1901.07075}{{\normalfont\ttfamily
  arXiv:1901.07075}}\relax
\mciteBstWouldAddEndPuncttrue
\mciteSetBstMidEndSepPunct{\mcitedefaultmidpunct}
{\mcitedefaultendpunct}{\mcitedefaultseppunct}\relax
\EndOfBibitem
\bibitem{Sirunyan:2018grk}
CMS collaboration, A.~M. Sirunyan {\em et~al.},
  \ifthenelse{\boolean{articletitles}}{\emph{{Studies of $B^*_{s2}(5840)^0$ and
  $B_{s1}(5830)^0$ mesons including the observation of the
  $B^*_{s2}(5840)^0\to$ $B^0K_{S}^0$ decay in proton-proton collisions at
  $\sqrt{s}=$8 TeV}},
  }{}\href{https://doi.org/10.1140/epjc/s10052-018-6390-z}{Eur.\ Phys.\ J.\
  \textbf{C78} (2018) 939},
  \href{http://arxiv.org/abs/1809.03578}{{\normalfont\ttfamily
  arXiv:1809.03578}}\relax
\mciteBstWouldAddEndPuncttrue
\mciteSetBstMidEndSepPunct{\mcitedefaultmidpunct}
{\mcitedefaultendpunct}{\mcitedefaultseppunct}\relax
\EndOfBibitem
\bibitem{Aaboud:2018hgx}
ATLAS collaboration, M.~Aaboud {\em et~al.},
  \ifthenelse{\boolean{articletitles}}{\emph{{Search for a Structure in the
  $B^0_s \pi^\pm$ Invariant Mass Spectrum with the ATLAS Experiment}},
  }{}\href{https://doi.org/10.1103/PhysRevLett.120.202007}{Phys.\ Rev.\ Lett.\
  \textbf{120} (2018) 202007},
  \href{http://arxiv.org/abs/1802.01840}{{\normalfont\ttfamily
  arXiv:1802.01840}}\relax
\mciteBstWouldAddEndPuncttrue
\mciteSetBstMidEndSepPunct{\mcitedefaultmidpunct}
{\mcitedefaultendpunct}{\mcitedefaultseppunct}\relax
\EndOfBibitem
\bibitem{LHCb-PAPER-2019-003}
LHCb collaboration, R.~Aaij {\em et~al.},
  \ifthenelse{\boolean{articletitles}}{\emph{{Measurement of the \CP-violating
  phase \phis from \mbox{\decay{\Bs}{\jpsi\pip\pim}} decays in 13\tev\
  \proton\proton collisions}},
  }{}\href{http://arxiv.org/abs/1903.05530}{{\normalfont\ttfamily
  arXiv:1903.05530}}, {submitted to Phys. Lett. B}\relax
\mciteBstWouldAddEndPuncttrue
\mciteSetBstMidEndSepPunct{\mcitedefaultmidpunct}
{\mcitedefaultendpunct}{\mcitedefaultseppunct}\relax
\EndOfBibitem
\bibitem{LHCb-PAPER-2019-013}
LHCb collaboration, R.~Aaij {\em et~al.},
  \ifthenelse{\boolean{articletitles}}{\emph{{Updated measurement of
  time-dependent \CP-violating observables in \mbox{\decay{\Bs}{\jpsi \Kp\Km}}
  decays}}, }{}\href{http://arxiv.org/abs/1906.08356}{{\normalfont\ttfamily
  arXiv:1906.08356}}, {submitted to EPJC}\relax
\mciteBstWouldAddEndPuncttrue
\mciteSetBstMidEndSepPunct{\mcitedefaultmidpunct}
{\mcitedefaultendpunct}{\mcitedefaultseppunct}\relax
\EndOfBibitem
\bibitem{Abbott:1998sb}
\dzero collaboration, B.~Abbott {\em et~al.},
  \ifthenelse{\boolean{articletitles}}{\emph{{Small angle $\jpsi$ production in
  $p\bar{p}$ collisions at $\sqrt{s} = 1.8$ TeV}},
  }{}\href{https://doi.org/10.1103/PhysRevLett.82.35}{Phys.\ Rev.\ Lett.\
  \textbf{82} (1999) 35},
  \href{http://arxiv.org/abs/hep-ex/9807029}{{\normalfont\ttfamily
  arXiv:hep-ex/9807029}}\relax
\mciteBstWouldAddEndPuncttrue
\mciteSetBstMidEndSepPunct{\mcitedefaultmidpunct}
{\mcitedefaultendpunct}{\mcitedefaultseppunct}\relax
\EndOfBibitem
\bibitem{Acosta:2004yw}
CDF collaboration, D.~Acosta {\em et~al.},
  \ifthenelse{\boolean{articletitles}}{\emph{{Measurement of the $\jpsi$ meson
  and $b-$hadron production cross sections in $p\bar{p}$ collisions at
  $\sqrt{s} = 1960$ GeV}},
  }{}\href{https://doi.org/10.1103/PhysRevD.71.032001}{Phys.\ Rev.\
  \textbf{D71} (2005) 032001},
  \href{http://arxiv.org/abs/hep-ex/0412071}{{\normalfont\ttfamily
  arXiv:hep-ex/0412071}}\relax
\mciteBstWouldAddEndPuncttrue
\mciteSetBstMidEndSepPunct{\mcitedefaultmidpunct}
{\mcitedefaultendpunct}{\mcitedefaultseppunct}\relax
\EndOfBibitem
\bibitem{Adare:2006kf}
PHENIX collaboration, A.~Adare {\em et~al.},
  \ifthenelse{\boolean{articletitles}}{\emph{{$\jpsi$ production versus
  transverse momentum and rapidity in $p^+ p$ collisions at $\sqrt{s}$ =
  200~GeV}}, }{}\href{https://doi.org/10.1103/PhysRevLett.98.232002}{Phys.\
  Rev.\ Lett.\  \textbf{98} (2007) 232002},
  \href{http://arxiv.org/abs/hep-ex/0611020}{{\normalfont\ttfamily
  arXiv:hep-ex/0611020}}\relax
\mciteBstWouldAddEndPuncttrue
\mciteSetBstMidEndSepPunct{\mcitedefaultmidpunct}
{\mcitedefaultendpunct}{\mcitedefaultseppunct}\relax
\EndOfBibitem
\bibitem{Khachatryan:2010yr}
CMS collaboration, V.~Khachatryan {\em et~al.},
  \ifthenelse{\boolean{articletitles}}{\emph{{Prompt and non-prompt $\jpsi$
  production in $pp$ collisions at $\sqrt{s}=7$ TeV}},
  }{}\href{https://doi.org/10.1140/epjc/s10052-011-1575-8}{Eur.\ Phys.\ J.\
  \textbf{C71} (2011) 1575},
  \href{http://arxiv.org/abs/1011.4193}{{\normalfont\ttfamily
  arXiv:1011.4193}}\relax
\mciteBstWouldAddEndPuncttrue
\mciteSetBstMidEndSepPunct{\mcitedefaultmidpunct}
{\mcitedefaultendpunct}{\mcitedefaultseppunct}\relax
\EndOfBibitem
\bibitem{LHCb-PAPER-2011-003}
LHCb collaboration, R.~Aaij {\em et~al.},
  \ifthenelse{\boolean{articletitles}}{\emph{{Measurement of \jpsi production
  in \proton\proton collisions at \mbox{$\sqs=$7\tev}}},
  }{}\href{https://doi.org/10.1140/epjc/s10052-011-1645-y}{Eur.\ Phys.\ J.\
  \textbf{C71} (2011) 1645},
  \href{http://arxiv.org/abs/1103.0423}{{\normalfont\ttfamily
  arXiv:1103.0423}}\relax
\mciteBstWouldAddEndPuncttrue
\mciteSetBstMidEndSepPunct{\mcitedefaultmidpunct}
{\mcitedefaultendpunct}{\mcitedefaultseppunct}\relax
\EndOfBibitem
\bibitem{LHCb-PAPER-2011-013}
LHCb collaboration, R.~Aaij {\em et~al.},
  \ifthenelse{\boolean{articletitles}}{\emph{{Observation of \jpsi-pair
  production in \proton\proton collisions at \mbox{$\sqs=$7\tev}}},
  }{}\href{https://doi.org/10.1016/j.physletb.2011.12.015}{Phys.\ Lett.\
  \textbf{B707} (2012) 52},
  \href{http://arxiv.org/abs/1109.0963}{{\normalfont\ttfamily
  arXiv:1109.0963}}\relax
\mciteBstWouldAddEndPuncttrue
\mciteSetBstMidEndSepPunct{\mcitedefaultmidpunct}
{\mcitedefaultendpunct}{\mcitedefaultseppunct}\relax
\EndOfBibitem
\bibitem{Aamodt:2011gj}
ALICE collaboration, K.~Aamodt {\em et~al.},
  \ifthenelse{\boolean{articletitles}}{\emph{{Rapidity and transverse momentum
  dependence of inclusive J$/\psi$ production in $pp$ collisions at $\sqrt{s} =
  7$ TeV}}, }{}\href{https://doi.org/10.1016/j.physletb.2011.09.054}{Phys.\
  Lett.\  \textbf{B704} (2011) 442}, Erratum
  \href{https://doi.org/10.1016/j.physletb.2012.10.060}{ibid.\   \textbf{B718}
  (2012) 692}, \href{http://arxiv.org/abs/1105.0380}{{\normalfont\ttfamily
  arXiv:1105.0380}}\relax
\mciteBstWouldAddEndPuncttrue
\mciteSetBstMidEndSepPunct{\mcitedefaultmidpunct}
{\mcitedefaultendpunct}{\mcitedefaultseppunct}\relax
\EndOfBibitem
\bibitem{Aad:2011sp}
ATLAS collaboration, G.~Aad {\em et~al.},
  \ifthenelse{\boolean{articletitles}}{\emph{{Measurement of the differential
  cross-sections of inclusive, prompt and non-prompt $\jpsi$ production in
  proton-proton collisions at $\sqrt{s}=7$ TeV}},
  }{}\href{https://doi.org/10.1016/j.nuclphysb.2011.05.015}{Nucl.\ Phys.\
  \textbf{B850} (2011) 387},
  \href{http://arxiv.org/abs/1104.3038}{{\normalfont\ttfamily
  arXiv:1104.3038}}\relax
\mciteBstWouldAddEndPuncttrue
\mciteSetBstMidEndSepPunct{\mcitedefaultmidpunct}
{\mcitedefaultendpunct}{\mcitedefaultseppunct}\relax
\EndOfBibitem
\bibitem{Chatrchyan:2011kc}
CMS collaboration, S.~Chatrchyan {\em et~al.},
  \ifthenelse{\boolean{articletitles}}{\emph{{$\jpsi$ and $\psi(2S)$ production
  in $pp$ collisions at $\sqrt{s}=7$ TeV}},
  }{}\href{https://doi.org/10.1007/JHEP02(2012)011}{JHEP \textbf{02} (2012)
  011}, \href{http://arxiv.org/abs/1111.1557}{{\normalfont\ttfamily
  arXiv:1111.1557}}\relax
\mciteBstWouldAddEndPuncttrue
\mciteSetBstMidEndSepPunct{\mcitedefaultmidpunct}
{\mcitedefaultendpunct}{\mcitedefaultseppunct}\relax
\EndOfBibitem
\bibitem{Khachatryan:2014iia}
CMS collaboration, V.~Khachatryan {\em et~al.},
  \ifthenelse{\boolean{articletitles}}{\emph{{Measurement of prompt $\jpsi$
  pair production in pp collisions at $ \sqrt{s} $ = 7 TeV}},
  }{}\href{https://doi.org/10.1007/JHEP09(2014)094}{JHEP \textbf{09} (2014)
  094}, \href{http://arxiv.org/abs/1406.0484}{{\normalfont\ttfamily
  arXiv:1406.0484}}\relax
\mciteBstWouldAddEndPuncttrue
\mciteSetBstMidEndSepPunct{\mcitedefaultmidpunct}
{\mcitedefaultendpunct}{\mcitedefaultseppunct}\relax
\EndOfBibitem
\bibitem{Aaltonen:2013atp}
CDF collaboration, T.~A. Aaltonen {\em et~al.},
  \ifthenelse{\boolean{articletitles}}{\emph{{Study of orbitally excited $B$
  mesons and evidence for a new $B\pi$ resonance}},
  }{}\href{https://doi.org/10.1103/PhysRevD.90.012013}{Phys.\ Rev.\
  \textbf{D90} (2014) 012013},
  \href{http://arxiv.org/abs/1309.5961}{{\normalfont\ttfamily
  arXiv:1309.5961}}\relax
\mciteBstWouldAddEndPuncttrue
\mciteSetBstMidEndSepPunct{\mcitedefaultmidpunct}
{\mcitedefaultendpunct}{\mcitedefaultseppunct}\relax
\EndOfBibitem
\bibitem{Khachatryan:2016hje}
CMS collaboration, V.~Khachatryan {\em et~al.},
  \ifthenelse{\boolean{articletitles}}{\emph{{Search for Resonant Production of
  High-Mass Photon Pairs in Proton-Proton Collisions at $\sqrt{s} =8$ and $13$
  TeV}}, }{}\href{https://doi.org/10.1103/PhysRevLett.117.051802}{Phys.\ Rev.\
  Lett.\  \textbf{117} (2016) 051802},
  \href{http://arxiv.org/abs/1606.04093}{{\normalfont\ttfamily
  arXiv:1606.04093}}\relax
\mciteBstWouldAddEndPuncttrue
\mciteSetBstMidEndSepPunct{\mcitedefaultmidpunct}
{\mcitedefaultendpunct}{\mcitedefaultseppunct}\relax
\EndOfBibitem
\bibitem{Khachatryan:2014ira}
CMS collaboration, V.~Khachatryan {\em et~al.},
  \ifthenelse{\boolean{articletitles}}{\emph{{Observation of the diphoton decay
  of the Higgs boson and measurement of its properties}},
  }{}\href{https://doi.org/10.1140/epjc/s10052-014-3076-z}{Eur.\ Phys.\ J.\
  \textbf{C74} (2014) 3076},
  \href{http://arxiv.org/abs/1407.0558}{{\normalfont\ttfamily
  arXiv:1407.0558}}\relax
\mciteBstWouldAddEndPuncttrue
\mciteSetBstMidEndSepPunct{\mcitedefaultmidpunct}
{\mcitedefaultendpunct}{\mcitedefaultseppunct}\relax
\EndOfBibitem
\bibitem{TheBABAR:2016lja}
BaBar collaboration, J.~P. Lees {\em et~al.},
  \ifthenelse{\boolean{articletitles}}{\emph{{Measurement of the inclusive
  electron spectrum from B meson decays and determination of $|V_{ub}|$}},
  }{}\href{https://doi.org/10.1103/PhysRevD.95.072001}{Phys.\ Rev.\
  \textbf{D95} (2016) 072001},
  \href{http://arxiv.org/abs/1611.05624}{{\normalfont\ttfamily
  arXiv:1611.05624}}\relax
\mciteBstWouldAddEndPuncttrue
\mciteSetBstMidEndSepPunct{\mcitedefaultmidpunct}
{\mcitedefaultendpunct}{\mcitedefaultseppunct}\relax
\EndOfBibitem
\bibitem{PhysRevD.93.052016}
Belle collaboration, V.~Bhardwaj {\em et~al.},
  \ifthenelse{\boolean{articletitles}}{\emph{{Inclusive and exclusive
  measurements of $B$ decays to ${\ensuremath{\chi}}_{c1}$ and
  ${\ensuremath{\chi}}_{c2}$ at Belle}},
  }{}\href{https://doi.org/10.1103/PhysRevD.93.052016}{Phys.\ Rev.\ D
  \textbf{93} (2016) 052016},
  \href{http://arxiv.org/abs/1512.02672}{{\normalfont\ttfamily
  arXiv:1512.02672}}\relax
\mciteBstWouldAddEndPuncttrue
\mciteSetBstMidEndSepPunct{\mcitedefaultmidpunct}
{\mcitedefaultendpunct}{\mcitedefaultseppunct}\relax
\EndOfBibitem
\bibitem{LHCb-PAPER-2015-027}
LHCb collaboration, R.~Aaij {\em et~al.},
  \ifthenelse{\boolean{articletitles}}{\emph{{\B flavour tagging using charm
  decays at the LHCb experiment}},
  }{}\href{https://doi.org/10.1088/1748-0221/10/10/P10005}{JINST \textbf{10}
  (2015) P10005}, \href{http://arxiv.org/abs/1507.07892}{{\normalfont\ttfamily
  arXiv:1507.07892}}\relax
\mciteBstWouldAddEndPuncttrue
\mciteSetBstMidEndSepPunct{\mcitedefaultmidpunct}
{\mcitedefaultendpunct}{\mcitedefaultseppunct}\relax
\EndOfBibitem
\bibitem{LHCb-PAPER-2012-001}
LHCb collaboration, R.~Aaij {\em et~al.},
  \ifthenelse{\boolean{articletitles}}{\emph{{Observation of \CP violation in
  \mbox{\decay{\Bpm}{\D\Kpm}} decays}},
  }{}\href{https://doi.org/10.1016/j.physletb.2012.04.060}{Phys.\ Lett.\
  \textbf{B712} (2012) 203}, Erratum
  \href{https://doi.org/10.1016/j.physletb.2012.05.060}{ibid.\   \textbf{B713}
  (2012) 351}, \href{http://arxiv.org/abs/1203.3662}{{\normalfont\ttfamily
  arXiv:1203.3662}}\relax
\mciteBstWouldAddEndPuncttrue
\mciteSetBstMidEndSepPunct{\mcitedefaultmidpunct}
{\mcitedefaultendpunct}{\mcitedefaultseppunct}\relax
\EndOfBibitem
\bibitem{LHCb-PAPER-2018-015}
LHCb collaboration, R.~Aaij {\em et~al.},
  \ifthenelse{\boolean{articletitles}}{\emph{{Observation of the decay
  \mbox{\decay{\Bs}{\Dbar^{\ast 0} \phi}} and search for the mode
  \mbox{\decay{\Bz}{\Dbar^0 \phi}}}},
  }{}\href{https://doi.org/10.1103/PhysRevD.98.071103}{Phys.\ Rev.\
  \textbf{D98} (2018) 071103(R)},
  \href{http://arxiv.org/abs/1807.01892}{{\normalfont\ttfamily
  arXiv:1807.01892}}\relax
\mciteBstWouldAddEndPuncttrue
\mciteSetBstMidEndSepPunct{\mcitedefaultmidpunct}
{\mcitedefaultendpunct}{\mcitedefaultseppunct}\relax
\EndOfBibitem
\bibitem{LHCb-PAPER-2017-021}
LHCb collaboration, R.~Aaij {\em et~al.},
  \ifthenelse{\boolean{articletitles}}{\emph{{Measurement of \CP observables in
  \mbox{\decay{\Bpm}{D^{(\ast)}\Kpm}} and \mbox{\decay{\Bpm}{D^{(\ast)}\pipm}}
  decays}}, }{}\href{https://doi.org/10.1016/j.physletb.2017.11.070}{Phys.\
  Lett.\  \textbf{B777} (2017) 16},
  \href{http://arxiv.org/abs/1708.06370}{{\normalfont\ttfamily
  arXiv:1708.06370}}\relax
\mciteBstWouldAddEndPuncttrue
\mciteSetBstMidEndSepPunct{\mcitedefaultmidpunct}
{\mcitedefaultendpunct}{\mcitedefaultseppunct}\relax
\EndOfBibitem
\bibitem{Ablikim:2015swa}
BES III collaboration, M.~Ablikim {\em et~al.},
  \ifthenelse{\boolean{articletitles}}{\emph{{Confirmation of a charged
  charmoniumlike state $Z_c(3885)^{\mp}$ in
  $e^+e^-\to\pi^{\pm}(D\bar{D}^*)^\mp$ with double $D$ tag}},
  }{}\href{https://doi.org/10.1103/PhysRevD.92.092006}{Phys.\ Rev.\
  \textbf{D92} (2015) 092006},
  \href{http://arxiv.org/abs/1509.01398}{{\normalfont\ttfamily
  arXiv:1509.01398}}\relax
\mciteBstWouldAddEndPuncttrue
\mciteSetBstMidEndSepPunct{\mcitedefaultmidpunct}
{\mcitedefaultendpunct}{\mcitedefaultseppunct}\relax
\EndOfBibitem
\bibitem{LHCb-PAPER-2013-065}
LHCb collaboration, R.~Aaij {\em et~al.},
  \ifthenelse{\boolean{articletitles}}{\emph{{Measurements of the \Bp, \Bz, \Bs
  meson and \Lb baryon lifetimes}},
  }{}\href{https://doi.org/10.1007/JHEP04(2014)114}{JHEP \textbf{04} (2014)
  114}, \href{http://arxiv.org/abs/1402.2554}{{\normalfont\ttfamily
  arXiv:1402.2554}}\relax
\mciteBstWouldAddEndPuncttrue
\mciteSetBstMidEndSepPunct{\mcitedefaultmidpunct}
{\mcitedefaultendpunct}{\mcitedefaultseppunct}\relax
\EndOfBibitem
\bibitem{Sandilya:2013rhy}
Belle collaboration, S.~Sandilya {\em et~al.},
  \ifthenelse{\boolean{articletitles}}{\emph{{Search for Bottomonium States in
  Exclusive Radiative $\Upsilon(2S)$ Decays}},
  }{}\href{https://doi.org/10.1103/PhysRevLett.111.112001}{Phys.\ Rev.\ Lett.\
  \textbf{111} (2013) 112001},
  \href{http://arxiv.org/abs/1306.6212}{{\normalfont\ttfamily
  arXiv:1306.6212}}\relax
\mciteBstWouldAddEndPuncttrue
\mciteSetBstMidEndSepPunct{\mcitedefaultmidpunct}
{\mcitedefaultendpunct}{\mcitedefaultseppunct}\relax
\EndOfBibitem
\bibitem{Abdallah:2005cx}
DELPHI collaboration, J.~Abdallah {\em et~al.},
  \ifthenelse{\boolean{articletitles}}{\emph{{Determination of heavy quark
  non-perturbative parameters from spectral moments in semileptonic \B
  decays}}, }{}\href{https://doi.org/10.1140/epjc/s2005-02406-7}{Eur.\ Phys.\
  J.\  \textbf{C45} (2006) 35},
  \href{http://arxiv.org/abs/hep-ex/0510024}{{\normalfont\ttfamily
  arXiv:hep-ex/0510024}}\relax
\mciteBstWouldAddEndPuncttrue
\mciteSetBstMidEndSepPunct{\mcitedefaultmidpunct}
{\mcitedefaultendpunct}{\mcitedefaultseppunct}\relax
\EndOfBibitem
\bibitem{Abdallah:2004rz}
DELPHI collaboration, J.~Abdallah {\em et~al.},
  \ifthenelse{\boolean{articletitles}}{\emph{{Measurement of $|V_{cb}|$ using
  the semileptonic decay \decay{\Bz}{\Dstarp\ellm\neub_\ell}}},
  }{}\href{https://doi.org/10.1140/epjc/s2004-01598-6}{Eur.\ Phys.\ J.\
  \textbf{C33} (2004) 213},
  \href{http://arxiv.org/abs/hep-ex/0401023}{{\normalfont\ttfamily
  arXiv:hep-ex/0401023}}\relax
\mciteBstWouldAddEndPuncttrue
\mciteSetBstMidEndSepPunct{\mcitedefaultmidpunct}
{\mcitedefaultendpunct}{\mcitedefaultseppunct}\relax
\EndOfBibitem
\bibitem{Barate:1998yi}
ALEPH collaboration, R.~Barate {\em et~al.},
  \ifthenelse{\boolean{articletitles}}{\emph{{The Forward - backward asymmetry
  for charm quarks at the Z}},
  }{}\href{https://doi.org/10.1016/S0370-2693(98)00818-1}{Phys.\ Lett.\
  \textbf{B434} (1998) 415},
  \href{http://arxiv.org/abs/hep-ex/9811015}{{\normalfont\ttfamily
  arXiv:hep-ex/9811015}}\relax
\mciteBstWouldAddEndPuncttrue
\mciteSetBstMidEndSepPunct{\mcitedefaultmidpunct}
{\mcitedefaultendpunct}{\mcitedefaultseppunct}\relax
\EndOfBibitem
\bibitem{Schumann:2005ej}
Belle collaboration, J.~Schumann {\em et~al.},
  \ifthenelse{\boolean{articletitles}}{\emph{{Observation of
  \decay{\Bzb}{\Dz\eta'} and \decay{\Bzb}{\Dstarz\eta'}}},
  }{}\href{https://doi.org/10.1103/PhysRevD.72.011103}{Phys.\ Rev.\
  \textbf{D72} (2005) 011103},
  \href{http://arxiv.org/abs/hep-ex/0501013}{{\normalfont\ttfamily
  arXiv:hep-ex/0501013}}\relax
\mciteBstWouldAddEndPuncttrue
\mciteSetBstMidEndSepPunct{\mcitedefaultmidpunct}
{\mcitedefaultendpunct}{\mcitedefaultseppunct}\relax
\EndOfBibitem
\bibitem{Nishida:2004fk}
Belle collaboration, S.~Nishida {\em et~al.},
  \ifthenelse{\boolean{articletitles}}{\emph{{Observation of $\Bp
  \to\Kp\Peta\Pgamma$}},
  }{}\href{https://doi.org/10.1016/j.physletb.2005.01.097}{Phys.\ Lett.\
  \textbf{B610} (2005) 23},
  \href{http://arxiv.org/abs/hep-ex/0411065}{{\normalfont\ttfamily
  arXiv:hep-ex/0411065}}\relax
\mciteBstWouldAddEndPuncttrue
\mciteSetBstMidEndSepPunct{\mcitedefaultmidpunct}
{\mcitedefaultendpunct}{\mcitedefaultseppunct}\relax
\EndOfBibitem
\bibitem{Aaboud:2016zpr}
ATLAS collaboration, M.~Aaboud {\em et~al.},
  \ifthenelse{\boolean{articletitles}}{\emph{{Search for squarks and gluinos in
  events with hadronically decaying tau leptons, jets and missing transverse
  momentum in proton-proton collisions at $\sqrt{s}=13$ TeV recorded with the
  ATLAS detector}},
  }{}\href{https://doi.org/10.1140/epjc/s10052-016-4481-2}{Eur.\ Phys.\ J.\
  \textbf{C76} (2016) 683},
  \href{http://arxiv.org/abs/1607.05979}{{\normalfont\ttfamily
  arXiv:1607.05979}}\relax
\mciteBstWouldAddEndPuncttrue
\mciteSetBstMidEndSepPunct{\mcitedefaultmidpunct}
{\mcitedefaultendpunct}{\mcitedefaultseppunct}\relax
\EndOfBibitem
\bibitem{Khachatryan:2015isa}
CMS collaboration, V.~Khachatryan {\em et~al.},
  \ifthenelse{\boolean{articletitles}}{\emph{{Angular analysis of the decay $
  B^0 \to K^{*0} \mu^{+} \mu^{-}$ from pp collisions at $\sqrt{s}=8$ TeV}},
  }{}\href{https://doi.org/10.1016/j.physletb.2015.12.020}{Phys.\ Lett.\
  \textbf{B753} (2016) 424},
  \href{http://arxiv.org/abs/1507.08126}{{\normalfont\ttfamily
  arXiv:1507.08126}}\relax
\mciteBstWouldAddEndPuncttrue
\mciteSetBstMidEndSepPunct{\mcitedefaultmidpunct}
{\mcitedefaultendpunct}{\mcitedefaultseppunct}\relax
\EndOfBibitem
\bibitem{Chatrchyan:2013mxa}
CMS collaboration, S.~Chatrchyan {\em et~al.},
  \ifthenelse{\boolean{articletitles}}{\emph{{Measurement of the properties of
  a Higgs boson in the four-lepton final state}},
  }{}\href{https://doi.org/10.1103/PhysRevD.89.092007}{Phys.\ Rev.\
  \textbf{D89} (2014) 092007},
  \href{http://arxiv.org/abs/1312.5353}{{\normalfont\ttfamily
  arXiv:1312.5353}}\relax
\mciteBstWouldAddEndPuncttrue
\mciteSetBstMidEndSepPunct{\mcitedefaultmidpunct}
{\mcitedefaultendpunct}{\mcitedefaultseppunct}\relax
\EndOfBibitem
\bibitem{Aad:2014aba}
ATLAS collaboration, G.~Aad {\em et~al.},
  \ifthenelse{\boolean{articletitles}}{\emph{{Measurement of the Higgs boson
  mass from the $H\rightarrow \gamma\gamma$ and $H \rightarrow ZZ^{*}
  \rightarrow 4\ell$ channels with the ATLAS detector using 25 fb$^{-1}$ of
  $pp$ collision data}},
  }{}\href{https://doi.org/10.1103/PhysRevD.90.052004}{Phys.\ Rev.\
  \textbf{D90} (2014) 052004},
  \href{http://arxiv.org/abs/1406.3827}{{\normalfont\ttfamily
  arXiv:1406.3827}}\relax
\mciteBstWouldAddEndPuncttrue
\mciteSetBstMidEndSepPunct{\mcitedefaultmidpunct}
{\mcitedefaultendpunct}{\mcitedefaultseppunct}\relax
\EndOfBibitem
\bibitem{Aaboud:2018krd}
ATLAS, M.~Aaboud {\em et~al.},
  \ifthenelse{\boolean{articletitles}}{\emph{{Angular analysis of $B^0_d
  \rightarrow K^{*}\mu^+\mu^-$ decays in $pp$ collisions at $\sqrt{s}= 8$ TeV
  with the ATLAS detector}},
  }{}\href{https://doi.org/10.1007/JHEP10(2018)047}{JHEP \textbf{10} (2018)
  047}, \href{http://arxiv.org/abs/1805.04000}{{\normalfont\ttfamily
  arXiv:1805.04000}}\relax
\mciteBstWouldAddEndPuncttrue
\mciteSetBstMidEndSepPunct{\mcitedefaultmidpunct}
{\mcitedefaultendpunct}{\mcitedefaultseppunct}\relax
\EndOfBibitem
\bibitem{LHCb-PAPER-2018-014}
LHCb collaboration, R.~Aaij {\em et~al.},
  \ifthenelse{\boolean{articletitles}}{\emph{{Observation of the decay
  \mbox{\decay{\Bs}{\Dbar^0 \Kp\Km}}}},
  }{}\href{https://doi.org/10.1103/PhysRevD.98.072006}{Phys.\ Rev.\
  \textbf{D98} (2018) 072006},
  \href{http://arxiv.org/abs/1807.01891}{{\normalfont\ttfamily
  arXiv:1807.01891}}\relax
\mciteBstWouldAddEndPuncttrue
\mciteSetBstMidEndSepPunct{\mcitedefaultmidpunct}
{\mcitedefaultendpunct}{\mcitedefaultseppunct}\relax
\EndOfBibitem
\bibitem{LHCb-PAPER-2018-042}
LHCb collaboration, R.~Aaij {\em et~al.},
  \ifthenelse{\boolean{articletitles}}{\emph{{Study of the
  \mbox{\decay{\Bz}{\rho(770)^0 K^*(892)^0}} decay with an amplitude analysis
  of \mbox{\decay{\Bz}{(\pip\pim) (\Kp\pim)}} decays}},
  }{}\href{https://doi.org/10.1007/JHEP05(2019)026}{JHEP \textbf{05} (2019)
  026}, \href{http://arxiv.org/abs/1812.07008}{{\normalfont\ttfamily
  arXiv:1812.07008}}\relax
\mciteBstWouldAddEndPuncttrue
\mciteSetBstMidEndSepPunct{\mcitedefaultmidpunct}
{\mcitedefaultendpunct}{\mcitedefaultseppunct}\relax
\EndOfBibitem
\bibitem{Bevan:2014iga}
Belle and BaBar collaborations, A.~J. Bevan {\em et~al.},
  \ifthenelse{\boolean{articletitles}}{\emph{{The Physics of the $B$
  Factories}}, }{}\href{https://doi.org/10.1140/epjc/s10052-014-3026-9}{Eur.\
  Phys.\ J.\  \textbf{C74} (2014) 3026},
  \href{http://arxiv.org/abs/1406.6311}{{\normalfont\ttfamily
  arXiv:1406.6311}}\relax
\mciteBstWouldAddEndPuncttrue
\mciteSetBstMidEndSepPunct{\mcitedefaultmidpunct}
{\mcitedefaultendpunct}{\mcitedefaultseppunct}\relax
\EndOfBibitem
\bibitem{Aubert:2003jq}
BaBar collaboration, B.~Aubert {\em et~al.},
  \ifthenelse{\boolean{articletitles}}{\emph{{Measurement of the branching
  fractions for the exclusive decays of $B^0$ and $B^+$ to $\bar{D}^{(*)}
  D^{(*)} K$}}, }{}\href{https://doi.org/10.1103/PhysRevD.68.092001}{Phys.\
  Rev.\  \textbf{D68} (2003) 092001},
  \href{http://arxiv.org/abs/hep-ex/0305003}{{\normalfont\ttfamily
  arXiv:hep-ex/0305003}}\relax
\mciteBstWouldAddEndPuncttrue
\mciteSetBstMidEndSepPunct{\mcitedefaultmidpunct}
{\mcitedefaultendpunct}{\mcitedefaultseppunct}\relax
\EndOfBibitem
\bibitem{Lees:2014lra}
BaBar collaboration, J.~P. Lees {\em et~al.},
  \ifthenelse{\boolean{articletitles}}{\emph{{Study of $B^{\pm,0} \to \jpsi K^+
  K^- K^{\pm,0}$ and search for $B^0 \to \jpsi\phi$ at BABAR}},
  }{}\href{https://doi.org/10.1103/PhysRevD.91.012003}{Phys.\ Rev.\
  \textbf{D91} (2015) 012003},
  \href{http://arxiv.org/abs/1407.7244}{{\normalfont\ttfamily
  arXiv:1407.7244}}\relax
\mciteBstWouldAddEndPuncttrue
\mciteSetBstMidEndSepPunct{\mcitedefaultmidpunct}
{\mcitedefaultendpunct}{\mcitedefaultseppunct}\relax
\EndOfBibitem
\bibitem{Ishikawa:2006fh}
Belle collaboration, A.~Ishikawa {\em et~al.},
  \ifthenelse{\boolean{articletitles}}{\emph{{Measurement of Forward-Backward
  Asymmetry and Wilson Coefficients in $B\to K^*\ellp\ellm$}},
  }{}\href{https://doi.org/10.1103/PhysRevLett.96.251801}{Phys.\ Rev.\ Lett.\
  \textbf{96} (2006) 251801},
  \href{http://arxiv.org/abs/hep-ex/0603018}{{\normalfont\ttfamily
  arXiv:hep-ex/0603018}}\relax
\mciteBstWouldAddEndPuncttrue
\mciteSetBstMidEndSepPunct{\mcitedefaultmidpunct}
{\mcitedefaultendpunct}{\mcitedefaultseppunct}\relax
\EndOfBibitem
\bibitem{Ablikim:2015hih}
BES III collaboration, M.~Ablikim {\em et~al.},
  \ifthenelse{\boolean{articletitles}}{\emph{{Measurement of $y_{\CP}$ in
  $D^0-\overline{D}^0$ oscillation using quantum correlations in $e^+e^-\to
  D^0\overline{D}^0$ at $\sqrt{s}$ = 3.773\,GeV}},
  }{}\href{https://doi.org/10.1016/j.physletb.2015.04.008}{Phys.\ Lett.\
  \textbf{B744} (2015) 339},
  \href{http://arxiv.org/abs/1501.01378}{{\normalfont\ttfamily
  arXiv:1501.01378}}\relax
\mciteBstWouldAddEndPuncttrue
\mciteSetBstMidEndSepPunct{\mcitedefaultmidpunct}
{\mcitedefaultendpunct}{\mcitedefaultseppunct}\relax
\EndOfBibitem
\bibitem{Adam:2007pv}
CLEO collaboration, N.~E. Adam {\em et~al.},
  \ifthenelse{\boolean{articletitles}}{\emph{{A Study of Exclusive Charmless
  Semileptonic B Decay and $|V_{ub}|$}},
  }{}\href{https://doi.org/10.1103/PhysRevLett.99.041802}{Phys.\ Rev.\ Lett.\
  \textbf{99} (2007) 041802},
  \href{http://arxiv.org/abs/hep-ex/0703041}{{\normalfont\ttfamily
  arXiv:hep-ex/0703041}}\relax
\mciteBstWouldAddEndPuncttrue
\mciteSetBstMidEndSepPunct{\mcitedefaultmidpunct}
{\mcitedefaultendpunct}{\mcitedefaultseppunct}\relax
\EndOfBibitem
\bibitem{Choi:2015lpc}
Belle collaboration, S.-K. Choi {\em et~al.},
  \ifthenelse{\boolean{articletitles}}{\emph{{Measurements of $B\rightarrow
  \bar{D} D_{s0} ^{*+}(2317)$ decay rates and a search for isospin partners of
  the $D_{s0}^{*+} (2317)$}},
  }{}\href{https://doi.org/10.1103/PhysRevD.91.092011}{Phys.\ Rev.\
  \textbf{D91} (2015) 092011}, Addendum
  \href{https://doi.org/10.1103/PhysRevD.92.039905}{ibid.\   \textbf{D92}
  (2015) 039905}, \href{http://arxiv.org/abs/1504.02637}{{\normalfont\ttfamily
  arXiv:1504.02637}}\relax
\mciteBstWouldAddEndPuncttrue
\mciteSetBstMidEndSepPunct{\mcitedefaultmidpunct}
{\mcitedefaultendpunct}{\mcitedefaultseppunct}\relax
\EndOfBibitem
\bibitem{Pattern}
A candidate arbitration is sometimes part of the design of the selection. For
  instance the highest \pt lepton, or the most energetic jet may be used as
  seed to further reconstruction. 
  between data and simulation, or signal and control channels, 
  such cases. More generally, the particle trajectory reconstruction software
  --- known as pattern recognition --- applies arbitration by
  design~\cite{Fruhwirth:1987fm}. The pattern recognition decides which
  detector signal (''hit'') to add to the trajectory of a particle based its
  the extrapolation to the detector. The closest hit is usually selected. The
  tracking efficiency is calibrated using
  data~\cite{ATLAS:2012jma,Chatrchyan:2014fea,LHCb-DP-2013-002}, thus removing
  potential biases.\relax
\mciteBstWouldAddEndPunctfalse
\mciteSetBstMidEndSepPunct{\mcitedefaultmidpunct}
{}{\mcitedefaultseppunct}\relax
\EndOfBibitem
\bibitem{PV}
In many analyses the correct determination of the creation point of the
  particle of interest is equally important as identifying the signal. Examples
  are measurements of decay-time dependent observables, or analyses of
  candidates which do not allow the determination of a decay vertex as
  diphotons. At colliders the origin is usually the primary vertex, for which
  ambiguities remaining after the selection are addressed by
  arbitration~\cite{Aad:2011zb,Aad:2011sp,Aad:2011bw,LHCb-DP-2012-004,LHCb-PAPER-2013-040,LHCb-PAPER-2014-004,LHCb-PAPER-2014-026,LHCb-PAPER-2014-059,Aad:2014vma,Aad:2014yka,LHCb-PAPER-2015-037,Aaboud:2016ire}
  or random
  picking~\cite{LHCb-PAPER-2011-041,LHCb-PAPER-2013-015,LHCb-PAPER-2014-020,LHCb-PAPER-2015-004,LHCb-PAPER-2015-005}.
  Wrong assignments should be accounted for by a resolution function or by an
  additional background contribution.\relax
\mciteBstWouldAddEndPunctfalse
\mciteSetBstMidEndSepPunct{\mcitedefaultmidpunct}
{}{\mcitedefaultseppunct}\relax
\EndOfBibitem
\bibitem{LHCb-PAPER-2011-041}
LHCb collaboration, R.~Aaij {\em et~al.},
  \ifthenelse{\boolean{articletitles}}{\emph{{Measurement of the
  \mbox{\decay{\Bs}{\jpsi \KS}} branching fraction}},
  }{}\href{https://doi.org/10.1016/j.physletb.2012.05.062}{Phys.\ Lett.\
  \textbf{B713} (2012) 172},
  \href{http://arxiv.org/abs/1205.0934}{{\normalfont\ttfamily
  arXiv:1205.0934}}\relax
\mciteBstWouldAddEndPuncttrue
\mciteSetBstMidEndSepPunct{\mcitedefaultmidpunct}
{\mcitedefaultendpunct}{\mcitedefaultseppunct}\relax
\EndOfBibitem
\bibitem{LHCb-PAPER-2013-015}
LHCb collaboration, R.~Aaij {\em et~al.},
  \ifthenelse{\boolean{articletitles}}{\emph{{Measurement of the effective
  \mbox{\decay{\Bs}{\jpsi\KS}} lifetime}},
  }{}\href{https://doi.org/10.1016/j.nuclphysb.2013.04.021}{Nucl.\ Phys.\
  \textbf{B873} (2013) 275},
  \href{http://arxiv.org/abs/1304.4500}{{\normalfont\ttfamily
  arXiv:1304.4500}}\relax
\mciteBstWouldAddEndPuncttrue
\mciteSetBstMidEndSepPunct{\mcitedefaultmidpunct}
{\mcitedefaultendpunct}{\mcitedefaultseppunct}\relax
\EndOfBibitem
\bibitem{LHCb-PAPER-2014-020}
LHCb collaboration, R.~Aaij {\em et~al.},
  \ifthenelse{\boolean{articletitles}}{\emph{{Observation of the
  \mbox{\decay{\Lb}{\jpsi\proton\pim}} decay}},
  }{}\href{https://doi.org/10.1007/JHEP07(2014)103}{JHEP \textbf{07} (2014)
  103}, \href{http://arxiv.org/abs/1406.0755}{{\normalfont\ttfamily
  arXiv:1406.0755}}\relax
\mciteBstWouldAddEndPuncttrue
\mciteSetBstMidEndSepPunct{\mcitedefaultmidpunct}
{\mcitedefaultendpunct}{\mcitedefaultseppunct}\relax
\EndOfBibitem
\bibitem{ATLAS:2014fka}
ATLAS collaboration, G.~Aad {\em et~al.},
  \ifthenelse{\boolean{articletitles}}{\emph{{Searches for heavy long-lived
  charged particles with the ATLAS detector in proton-proton collisions at $
  \sqrt{s}=8 $ TeV}}, }{}\href{https://doi.org/10.1007/JHEP01(2015)068}{JHEP
  \textbf{01} (2015) 068},
  \href{http://arxiv.org/abs/1411.6795}{{\normalfont\ttfamily
  arXiv:1411.6795}}\relax
\mciteBstWouldAddEndPuncttrue
\mciteSetBstMidEndSepPunct{\mcitedefaultmidpunct}
{\mcitedefaultendpunct}{\mcitedefaultseppunct}\relax
\EndOfBibitem
\bibitem{LHCb-PAPER-2015-005}
LHCb collaboration, R.~Aaij {\em et~al.},
  \ifthenelse{\boolean{articletitles}}{\emph{{Measurement of the time-dependent
  \CP asymmetries in \mbox{\decay{\Bs}{\jpsi\KS}}}},
  }{}\href{https://doi.org/10.1007/JHEP06(2015)131}{JHEP \textbf{06} (2015)
  131}, \href{http://arxiv.org/abs/1503.07055}{{\normalfont\ttfamily
  arXiv:1503.07055}}\relax
\mciteBstWouldAddEndPuncttrue
\mciteSetBstMidEndSepPunct{\mcitedefaultmidpunct}
{\mcitedefaultendpunct}{\mcitedefaultseppunct}\relax
\EndOfBibitem
\bibitem{LHCb-PAPER-2015-050}
LHCb collaboration, R.~Aaij {\em et~al.},
  \ifthenelse{\boolean{articletitles}}{\emph{{Observation of
  \mbox{\decay{\Bs}{\Dzb\KS}} and evidence for \mbox{\decay{\Bs}{\Dstarzb\KS}}
  decays}}, }{}\href{https://doi.org/10.1103/PhysRevLett.116.161802}{Phys.\
  Rev.\ Lett.\  \textbf{116} (2016) 161802},
  \href{http://arxiv.org/abs/1603.02408}{{\normalfont\ttfamily
  arXiv:1603.02408}}\relax
\mciteBstWouldAddEndPuncttrue
\mciteSetBstMidEndSepPunct{\mcitedefaultmidpunct}
{\mcitedefaultendpunct}{\mcitedefaultseppunct}\relax
\EndOfBibitem
\bibitem{LHCb-PAPER-2016-017}
LHCb collaboration, R.~Aaij {\em et~al.},
  \ifthenelse{\boolean{articletitles}}{\emph{{Measurement of the
  \mbox{\decay{\Bs}{\jpsi\eta}} lifetime}},
  }{}\href{https://doi.org/10.1016/j.physletb.2016.10.006}{Phys.\ Lett.\
  \textbf{B762} (2016) 484},
  \href{http://arxiv.org/abs/1607.06314}{{\normalfont\ttfamily
  arXiv:1607.06314}}\relax
\mciteBstWouldAddEndPuncttrue
\mciteSetBstMidEndSepPunct{\mcitedefaultmidpunct}
{\mcitedefaultendpunct}{\mcitedefaultseppunct}\relax
\EndOfBibitem
\bibitem{LHCb-PAPER-2016-030}
LHCb collaboration, R.~Aaij {\em et~al.},
  \ifthenelse{\boolean{articletitles}}{\emph{{Measurement of matter-antimatter
  differences in beauty baryon decays}},
  }{}\href{https://doi.org/10.1038/nphys4021}{Nature Physics \textbf{13} (2017)
  391}, \href{http://arxiv.org/abs/1609.05216}{{\normalfont\ttfamily
  arXiv:1609.05216}}\relax
\mciteBstWouldAddEndPuncttrue
\mciteSetBstMidEndSepPunct{\mcitedefaultmidpunct}
{\mcitedefaultendpunct}{\mcitedefaultseppunct}\relax
\EndOfBibitem
\bibitem{LHCb-PAPER-2016-035}
LHCb collaboration, R.~Aaij {\em et~al.},
  \ifthenelse{\boolean{articletitles}}{\emph{{Measurement of \CP asymmetry in
  \mbox{\decay{\Dz}{\Kp\Km}} decays}},
  }{}\href{https://doi.org/10.1016/j.physletb.2017.01.061}{Phys.\ Lett.\
  \textbf{B767} (2017) 177},
  \href{http://arxiv.org/abs/1610.09476}{{\normalfont\ttfamily
  arXiv:1610.09476}}\relax
\mciteBstWouldAddEndPuncttrue
\mciteSetBstMidEndSepPunct{\mcitedefaultmidpunct}
{\mcitedefaultendpunct}{\mcitedefaultseppunct}\relax
\EndOfBibitem
\bibitem{LHCb-PAPER-2016-057}
LHCb collaboration, R.~Aaij {\em et~al.},
  \ifthenelse{\boolean{articletitles}}{\emph{{Measurement of the \jpsi pair
  production cross-section in \proton\proton collisions at
  \mbox{$\sqs=$13\tev}}},
  }{}\href{https://doi.org/10.1007/JHEP06(2017)047}{JHEP \textbf{06} (2017)
  047}, Erratum \href{https://doi.org/10.1007/JHEP10(2017)068}{ibid.\
  \textbf{10} (2017) 068},
  \href{http://arxiv.org/abs/1612.07451}{{\normalfont\ttfamily
  arXiv:1612.07451}}\relax
\mciteBstWouldAddEndPuncttrue
\mciteSetBstMidEndSepPunct{\mcitedefaultmidpunct}
{\mcitedefaultendpunct}{\mcitedefaultseppunct}\relax
\EndOfBibitem
\bibitem{LHCb-PAPER-2017-013}
LHCb collaboration, R.~Aaij {\em et~al.},
  \ifthenelse{\boolean{articletitles}}{\emph{{Test of lepton universality with
  \mbox{\decay{\Bz}{\Kstarz\ellell}} decays}},
  }{}\href{https://doi.org/10.1007/JHEP08(2017)055}{JHEP \textbf{08} (2017)
  055}, \href{http://arxiv.org/abs/1705.05802}{{\normalfont\ttfamily
  arXiv:1705.05802}}\relax
\mciteBstWouldAddEndPuncttrue
\mciteSetBstMidEndSepPunct{\mcitedefaultmidpunct}
{\mcitedefaultendpunct}{\mcitedefaultseppunct}\relax
\EndOfBibitem
\bibitem{LHCb-PAPER-2017-018}
LHCb collaboration, R.~Aaij {\em et~al.},
  \ifthenelse{\boolean{articletitles}}{\emph{{Observation of the doubly charmed
  baryon \Xiccpp}},
  }{}\href{https://doi.org/10.1103/PhysRevLett.119.112001}{Phys.\ Rev.\ Lett.\
  \textbf{119} (2017) 112001},
  \href{http://arxiv.org/abs/1707.01621}{{\normalfont\ttfamily
  arXiv:1707.01621}}\relax
\mciteBstWouldAddEndPuncttrue
\mciteSetBstMidEndSepPunct{\mcitedefaultmidpunct}
{\mcitedefaultendpunct}{\mcitedefaultseppunct}\relax
\EndOfBibitem
\bibitem{LHCb-PAPER-2017-029}
LHCb collaboration, R.~Aaij {\em et~al.},
  \ifthenelse{\boolean{articletitles}}{\emph{{Measurement of \CP violation in
  \mbox{\decay{\Bz}{\jpsi \KS}} and \mbox{\decay{\Bz}{\psitwos \KS}} decays}},
  }{}\href{https://doi.org/10.1007/JHEP11(2017)170}{JHEP \textbf{11} (2017)
  170}, \href{http://arxiv.org/abs/1709.03944}{{\normalfont\ttfamily
  arXiv:1709.03944}}\relax
\mciteBstWouldAddEndPuncttrue
\mciteSetBstMidEndSepPunct{\mcitedefaultmidpunct}
{\mcitedefaultendpunct}{\mcitedefaultseppunct}\relax
\EndOfBibitem
\bibitem{LHCb-PAPER-2017-041}
LHCb collaboration, R.~Aaij {\em et~al.},
  \ifthenelse{\boolean{articletitles}}{\emph{{Measurement of the ratio of
  branching fractions of the decays \mbox{\decay{\Lb}{\psitwos\Lz}} and
  \mbox{\decay{\Lb}{\jpsi\Lz}}}},
  }{}\href{https://doi.org/10.1007/JHEP03(2019)126}{JHEP \textbf{03} (2019)
  126}, \href{http://arxiv.org/abs/1902.02092}{{\normalfont\ttfamily
  arXiv:1902.02092}}\relax
\mciteBstWouldAddEndPuncttrue
\mciteSetBstMidEndSepPunct{\mcitedefaultmidpunct}
{\mcitedefaultendpunct}{\mcitedefaultseppunct}\relax
\EndOfBibitem
\bibitem{LHCb-PAPER-2017-048}
LHCb collaboration, R.~Aaij {\em et~al.},
  \ifthenelse{\boolean{articletitles}}{\emph{{First measurement of the
  \CP-violating phase $\phis^{d\dquarkbar}$ in
  \mbox{\decay{\Bs}{(\Kp\pim)(\Km\pip)}} decays}},
  }{}\href{https://doi.org/10.1007/JHEP03(2018)140}{JHEP \textbf{03} (2018)
  140}, \href{http://arxiv.org/abs/1712.08683}{{\normalfont\ttfamily
  arXiv:1712.08683}}\relax
\mciteBstWouldAddEndPuncttrue
\mciteSetBstMidEndSepPunct{\mcitedefaultmidpunct}
{\mcitedefaultendpunct}{\mcitedefaultseppunct}\relax
\EndOfBibitem
\bibitem{LHCb-PAPER-2018-001}
LHCb collaboration, R.~Aaij {\em et~al.},
  \ifthenelse{\boolean{articletitles}}{\emph{{Search for \CP violation using
  triple product asymmetries in \mbox{\decay{\Lb}{p\Km\pip\pim}},
  \mbox{\decay{\Lb}{p\Km\Kp\Km}}, and \mbox{\decay{\Xires^0_b}{p\Km\Km\pip}}
  decays}}, }{}\href{https://doi.org/10.1007/JHEP08(2018)039}{JHEP \textbf{08}
  (2018) 039}, \href{http://arxiv.org/abs/1805.03941}{{\normalfont\ttfamily
  arXiv:1805.03941}}\relax
\mciteBstWouldAddEndPuncttrue
\mciteSetBstMidEndSepPunct{\mcitedefaultmidpunct}
{\mcitedefaultendpunct}{\mcitedefaultseppunct}\relax
\EndOfBibitem
\bibitem{LHCb-PAPER-2018-025}
LHCb collaboration, R.~Aaij {\em et~al.},
  \ifthenelse{\boolean{articletitles}}{\emph{{Search for \CP violation in
  \mbox{\decay{\Lb}{p\Km} and \mbox{\decay{\Lb}{p\pim}}} decays}},
  }{}\href{https://doi.org/10.1016/j.physletb.2018.10.039}{Phys.\ Lett.\
  \textbf{B784} (2018) 101},
  \href{http://arxiv.org/abs/1807.06544}{{\normalfont\ttfamily
  arXiv:1807.06544}}\relax
\mciteBstWouldAddEndPuncttrue
\mciteSetBstMidEndSepPunct{\mcitedefaultmidpunct}
{\mcitedefaultendpunct}{\mcitedefaultseppunct}\relax
\EndOfBibitem
\bibitem{LHCb-PAPER-2018-006}
LHCb collaboration, R.~Aaij {\em et~al.},
  \ifthenelse{\boolean{articletitles}}{\emph{{Measurement of \CP asymmetries in
  two-body \BdorBs-meson decays to charged pions and kaons}},
  }{}\href{https://doi.org/10.1103/PhysRevD.98.032004}{Phys.\ Rev.\
  \textbf{D98} (2018) 032004},
  \href{http://arxiv.org/abs/1805.06759}{{\normalfont\ttfamily
  arXiv:1805.06759}}\relax
\mciteBstWouldAddEndPuncttrue
\mciteSetBstMidEndSepPunct{\mcitedefaultmidpunct}
{\mcitedefaultendpunct}{\mcitedefaultseppunct}\relax
\EndOfBibitem
\bibitem{LHCb-PAPER-2018-020}
LHCb collaboration, R.~Aaij {\em et~al.},
  \ifthenelse{\boolean{articletitles}}{\emph{{Measurement of angular and \CP
  asymmetries in \mbox{\decay{\Dz}{\pip\pim\mumu}} and
  \mbox{\decay{\Dz}{\Kp\Km\mumu}} decays}},
  }{}\href{https://doi.org/10.1103/PhysRevLett.121.091801}{Phys.\ Rev.\ Lett.\
  \textbf{121} (2018) 091801},
  \href{http://arxiv.org/abs/1806.10793}{{\normalfont\ttfamily
  arXiv:1806.10793}}\relax
\mciteBstWouldAddEndPuncttrue
\mciteSetBstMidEndSepPunct{\mcitedefaultmidpunct}
{\mcitedefaultendpunct}{\mcitedefaultseppunct}\relax
\EndOfBibitem
\bibitem{Artuso:1999dd}
CLEO collaboration, M.~Artuso {\em et~al.},
  \ifthenelse{\boolean{articletitles}}{\emph{{Search for the decay
  \decay{\Bzb}{\Dstarz\gamma}}},
  }{}\href{https://doi.org/10.1103/PhysRevLett.84.4292}{Phys.\ Rev.\ Lett.\
  \textbf{84} (2000) 4292},
  \href{http://arxiv.org/abs/hep-ex/0001002}{{\normalfont\ttfamily
  arXiv:hep-ex/0001002}}\relax
\mciteBstWouldAddEndPuncttrue
\mciteSetBstMidEndSepPunct{\mcitedefaultmidpunct}
{\mcitedefaultendpunct}{\mcitedefaultseppunct}\relax
\EndOfBibitem
\bibitem{Adams:2011sq}
CLEO collaboration, G.~S. Adams {\em et~al.},
  \ifthenelse{\boolean{articletitles}}{\emph{{Amplitude analyses of the decays
  $\chi_{c1} \to \eta \pi^+ \pi^-$ and $\chi_{c1} \to \eta' \pi^+ \pi^-$}},
  }{}\href{https://doi.org/10.1103/PhysRevD.84.112009}{Phys.\ Rev.\
  \textbf{D84} (2011) 112009},
  \href{http://arxiv.org/abs/1109.5843}{{\normalfont\ttfamily
  arXiv:1109.5843}}\relax
\mciteBstWouldAddEndPuncttrue
\mciteSetBstMidEndSepPunct{\mcitedefaultmidpunct}
{\mcitedefaultendpunct}{\mcitedefaultseppunct}\relax
\EndOfBibitem
\bibitem{Aad:2011dm}
ATLAS collaboration, G.~Aad {\em et~al.},
  \ifthenelse{\boolean{articletitles}}{\emph{{Measurement of the inclusive
  $W^\pm$ and $Z/\gamma^\ast$ cross sections in the electron and muon decay
  channels in $pp$ collisions at $\sqrt{s}=7$ TeV with the ATLAS detector}},
  }{}\href{https://doi.org/10.1103/PhysRevD.85.072004}{Phys.\ Rev.\
  \textbf{D85} (2012) 072004},
  \href{http://arxiv.org/abs/1109.5141}{{\normalfont\ttfamily
  arXiv:1109.5141}}\relax
\mciteBstWouldAddEndPuncttrue
\mciteSetBstMidEndSepPunct{\mcitedefaultmidpunct}
{\mcitedefaultendpunct}{\mcitedefaultseppunct}\relax
\EndOfBibitem
\bibitem{LHCb-PAPER-2015-049}
LHCb collaboration, R.~Aaij {\em et~al.},
  \ifthenelse{\boolean{articletitles}}{\emph{{Measurement of forward \W and \Z
  boson production in \proton\proton collisions at \mbox{$\sqs=$8\tev} }},
  }{}\href{https://doi.org/10.1007/JHEP01(2016)155}{JHEP \textbf{01} (2016)
  155}, \href{http://arxiv.org/abs/1511.08039}{{\normalfont\ttfamily
  arXiv:1511.08039}}\relax
\mciteBstWouldAddEndPuncttrue
\mciteSetBstMidEndSepPunct{\mcitedefaultmidpunct}
{\mcitedefaultendpunct}{\mcitedefaultseppunct}\relax
\EndOfBibitem
\bibitem{Aghasyan:2017utv}
COMPASS collaboration, M.~Aghasyan {\em et~al.},
  \ifthenelse{\boolean{articletitles}}{\emph{{Observation of $X(3872)$
  muoproduction at COMPASS}},
  }{}\href{https://doi.org/10.1016/j.physletb.2018.07.008}{Phys.\ Lett.\
  \textbf{B783} (2018) 334},
  \href{http://arxiv.org/abs/1707.01796}{{\normalfont\ttfamily
  arXiv:1707.01796}}\relax
\mciteBstWouldAddEndPuncttrue
\mciteSetBstMidEndSepPunct{\mcitedefaultmidpunct}
{\mcitedefaultendpunct}{\mcitedefaultseppunct}\relax
\EndOfBibitem
\bibitem{LHCb-PAPER-2018-036}
LHCb collaboration, R.~Aaij {\em et~al.},
  \ifthenelse{\boolean{articletitles}}{\emph{{Measurement of the branching
  fraction and \CP asymmetry in \mbox{\decay{\Bp}{\jpsi\rhop}} decays}},
  }{}\href{https://doi.org/10.1140/epjc/s10052-019-6698-3}{Eur.\ Phys.\ J.\
  \textbf{C79} (2019) 537},
  \href{http://arxiv.org/abs/1812.07041}{{\normalfont\ttfamily
  arXiv:1812.07041}}\relax
\mciteBstWouldAddEndPuncttrue
\mciteSetBstMidEndSepPunct{\mcitedefaultmidpunct}
{\mcitedefaultendpunct}{\mcitedefaultseppunct}\relax
\EndOfBibitem
\bibitem{Uniform}
An exponential shape would be more physical, but would be an unnecessary
  complication of the model.\relax
\mciteBstWouldAddEndPunctfalse
\mciteSetBstMidEndSepPunct{\mcitedefaultmidpunct}
{}{\mcitedefaultseppunct}\relax
\EndOfBibitem
\bibitem{FlatMass}
In the general case one would expect a different mass distribution for
  overlapping candidates. This different distribution could then be taken into
  account in the fit. The author is not aware of any publication where this was
  done. Therefore it is assumed here that overlaps behave
  background-like.\relax
\mciteBstWouldAddEndPunctfalse
\mciteSetBstMidEndSepPunct{\mcitedefaultmidpunct}
{}{\mcitedefaultseppunct}\relax
\EndOfBibitem
\bibitem{swaps}
For example a \decay{\Bd}{\jpsi\Kstarz} signal decay with \decay{\jpsi}{\mumu}
  and \decay{\Kstarz}{\Kp\pim} could have a companion
  \decay{\Bdb}{\jpsi\Kstarzb} candidate which shares the same \jpsi but adds an
  unrelated \decay{\Kstarzb}{\Km\pip} candidate. Such a situation produces a
  correlation between the \Bd signal yield and the \Bdb background yield.\relax
\mciteBstWouldAddEndPunctfalse
\mciteSetBstMidEndSepPunct{\mcitedefaultmidpunct}
{}{\mcitedefaultseppunct}\relax
\EndOfBibitem
\bibitem{infer}
Weighting leads to the same results as the random selection provided the
  weights are properly taken into account in the fit. Event removal is
  equivalent to random picking with probability $0$ instead of $\sfrac{1}{2}$.
  The resulting biases have twice the size of those obtained with this
  method.\relax
\mciteBstWouldAddEndPunctfalse
\mciteSetBstMidEndSepPunct{\mcitedefaultmidpunct}
{}{\mcitedefaultseppunct}\relax
\EndOfBibitem
\bibitem{supplementary}
Mathematical expressions can be found in the \arxiv{ancillary}{supplementary}
  material\arxiv{ at
  \url{https://arxiv.org/src/1703.01128/anc/Ancillary.pdf}.}{.}\relax
\mciteBstWouldAddEndPunctfalse
\mciteSetBstMidEndSepPunct{\mcitedefaultmidpunct}
{}{\mcitedefaultseppunct}\relax
\EndOfBibitem
\bibitem{PBbkg}
Variation of the probability of a background candidate to have a companion is
  tested with $\PBsig=\PBbsig=0$. When taking all candidates, the effect of
  larger \PBbkg and \PBbbkg values is to increase the background level, which
  does not bias any measurement but degrades the uncertainty in the same way as
  \PBsig and \PBbsig do. However, when a candidate selection technique is
  applied, only background candidates get removed resulting in a situation
  equivalent to the clean case.\relax
\mciteBstWouldAddEndPunctfalse
\mciteSetBstMidEndSepPunct{\mcitedefaultmidpunct}
{}{\mcitedefaultseppunct}\relax
\EndOfBibitem
\bibitem{Negligible}
This systematic uncertainty may be negligible and thus not be reported.\relax
\mciteBstWouldAddEndPunctfalse
\mciteSetBstMidEndSepPunct{\mcitedefaultmidpunct}
{}{\mcitedefaultseppunct}\relax
\EndOfBibitem
\bibitem{swaps2}
In the decay \decay{\Bp}{\Dp\piz} companion candidates can be generated
  starting from the signal \Dp and adding a fake \piz. By construction, the two
  candidates have the same flavour and the swapping probabilities are zero. In
  the decay \decay{\Bp}{\D_\CP\mup\neu} where $\D_\CP$ is a \Dz meson decaying
  to a \CP eigenstate, if a true $\D_\CP$ is combined with a muon from the
  other \B, the latter is likely to be of the opposite flavour and hence the
  flavours of signal and overlapping candidate are anti-correlated. The
  swapping probabilities are close to unity. For the decay
  \decay{\Bz}{\jpsi\Kstarz}~\cite{swaps}, the flavour of the fake \Bd candidate
  depends on the \Kstarz to \Kstarzb production ratio, which may be slightly
  (anti-)correlated with that of \Bd versus \Bdb,
  $\PBswapB\simeq\PBbswapB\simeq0.5$.\relax
\mciteBstWouldAddEndPunctfalse
\mciteSetBstMidEndSepPunct{\mcitedefaultmidpunct}
{}{\mcitedefaultseppunct}\relax
\EndOfBibitem
\bibitem{NotGood}
Such situations are not necessarily desirable, as these probabilities may not
  be reproduced well in the simulation or in a control sample, leading to
  incorrect efficiency corrections.\relax
\mciteBstWouldAddEndPunctfalse
\mciteSetBstMidEndSepPunct{\mcitedefaultmidpunct}
{}{\mcitedefaultseppunct}\relax
\EndOfBibitem
\bibitem{TauMotivation}
It is common for \B background candidates to be formed from a combination of
  final-state particles from different \bquark hadrons, resulting in a visible
  lifetime that is lower than that of a true \bquark hadron. In the case of
  overlaps, some components of the candidate are from the true signal,
  resulting in a visible lifetime which is larger than that of background but
  lower than \tauB. In reality the actual factors may differ considerably from
  the values $\sfrac{1}{2}$ and $\sfrac{1}{3}$ used here and the distributions
  would not be pure exponential functions. However, the conclusions are not
  affected as long as the distributions are smooth curves.\relax
\mciteBstWouldAddEndPunctfalse
\mciteSetBstMidEndSepPunct{\mcitedefaultmidpunct}
{}{\mcitedefaultseppunct}\relax
\EndOfBibitem
\bibitem{Resolution}
This resolution function is irrelevant for this study, but helps making the fit
  more stable.\relax
\mciteBstWouldAddEndPunctfalse
\mciteSetBstMidEndSepPunct{\mcitedefaultmidpunct}
{}{\mcitedefaultseppunct}\relax
\EndOfBibitem
\bibitem{Pivk:2004ty}
M.~Pivk and F.~R. Le~Diberder,
  \ifthenelse{\boolean{articletitles}}{\emph{{sPlot: A statistical tool to
  unfold data distributions}},
  }{}\href{https://doi.org/10.1016/j.nima.2005.08.106}{Nucl.\ Instrum.\ Meth.\
  \textbf{A555} (2005) 356},
  \href{http://arxiv.org/abs/physics/0402083}{{\normalfont\ttfamily
  arXiv:physics/0402083}}\relax
\mciteBstWouldAddEndPuncttrue
\mciteSetBstMidEndSepPunct{\mcitedefaultmidpunct}
{\mcitedefaultendpunct}{\mcitedefaultseppunct}\relax
\EndOfBibitem
\bibitem{sFit}
Y.~Xie, \ifthenelse{\boolean{articletitles}}{\emph{{sFit: a method for
  background subtraction in maximum likelihood fit}},
  }{}\href{http://arxiv.org/abs/0905.0724}{{\normalfont\ttfamily
  arXiv:0905.0724}}\relax
\mciteBstWouldAddEndPuncttrue
\mciteSetBstMidEndSepPunct{\mcitedefaultmidpunct}
{\mcitedefaultendpunct}{\mcitedefaultseppunct}\relax
\EndOfBibitem
\bibitem{Peaking}
The situation may be very different if the overlaps are peaking in mass.\relax
\mciteBstWouldAddEndPunctfalse
\mciteSetBstMidEndSepPunct{\mcitedefaultmidpunct}
{}{\mcitedefaultseppunct}\relax
\EndOfBibitem
\bibitem{LHCb-PAPER-2014-038}
LHCb collaboration, R.~Aaij {\em et~al.},
  \ifthenelse{\boolean{articletitles}}{\emph{{Measurement of \CP asymmetry in
  \mbox{\decay{\Bs}{\Dsmp\Kpm}} decays}},
  }{}\href{https://doi.org/10.1007/JHEP11(2014)060}{JHEP \textbf{11} (2014)
  060}, \href{http://arxiv.org/abs/1407.6127}{{\normalfont\ttfamily
  arXiv:1407.6127}}\relax
\mciteBstWouldAddEndPuncttrue
\mciteSetBstMidEndSepPunct{\mcitedefaultmidpunct}
{\mcitedefaultendpunct}{\mcitedefaultseppunct}\relax
\EndOfBibitem
\bibitem{DeBruyn:2012wj}
K.~De~Bruyn {\em et~al.}, \ifthenelse{\boolean{articletitles}}{\emph{{Branching
  Ratio Measurements of $B_s$ Decays}},
  }{}\href{https://doi.org/10.1103/PhysRevD.86.014027}{Phys.\ Rev.\
  \textbf{D86} (2012) 014027},
  \href{http://arxiv.org/abs/1204.1735}{{\normalfont\ttfamily
  arXiv:1204.1735}}\relax
\mciteBstWouldAddEndPuncttrue
\mciteSetBstMidEndSepPunct{\mcitedefaultmidpunct}
{\mcitedefaultendpunct}{\mcitedefaultseppunct}\relax
\EndOfBibitem
\bibitem{HiggsMass}
This risk has been recognised by the CMS collaboration in the analysis of Higgs
  bosons decaying to four leptons~\cite{Chatrchyan:2013mxa} used in the Higgs
  boson mass measurement~\cite{Aad:2015zhl}. It is reported that the used
  arbitration procedure based on dilepton masses does not sculpt the background
  shape. The arbitration procedure applied by the ATLAS collaboration is
  different and there is no statement on potential
  biases~\cite{Aad:2014aba}.\relax
\mciteBstWouldAddEndPunctfalse
\mciteSetBstMidEndSepPunct{\mcitedefaultmidpunct}
{}{\mcitedefaultseppunct}\relax
\EndOfBibitem
\bibitem{Aad:2015zhl}
ATLAS and CMS collaborations, G.~Aad {\em et~al.},
  \ifthenelse{\boolean{articletitles}}{\emph{{Combined Measurement of the Higgs
  Boson Mass in $pp$ Collisions at $\sqrt{s}=7$ and $8$ TeV with the ATLAS and
  CMS Experiments}},
  }{}\href{https://doi.org/10.1103/PhysRevLett.114.191803}{Phys.\ Rev.\ Lett.\
  \textbf{114} (2015) 191803},
  \href{http://arxiv.org/abs/1503.07589}{{\normalfont\ttfamily
  arXiv:1503.07589}}\relax
\mciteBstWouldAddEndPuncttrue
\mciteSetBstMidEndSepPunct{\mcitedefaultmidpunct}
{\mcitedefaultendpunct}{\mcitedefaultseppunct}\relax
\EndOfBibitem
\bibitem{Khachatryan:2010nk}
CMS collaboration, V.~Khachatryan {\em et~al.},
  \ifthenelse{\boolean{articletitles}}{\emph{{Charged particle multiplicities
  in $pp$ interactions at $\sqrt{s}=0.9$, 2.36, and 7 TeV}},
  }{}\href{https://doi.org/10.1007/JHEP01(2011)079}{JHEP \textbf{01} (2011)
  079}, \href{http://arxiv.org/abs/1011.5531}{{\normalfont\ttfamily
  arXiv:1011.5531}}\relax
\mciteBstWouldAddEndPuncttrue
\mciteSetBstMidEndSepPunct{\mcitedefaultmidpunct}
{\mcitedefaultendpunct}{\mcitedefaultseppunct}\relax
\EndOfBibitem
\bibitem{LHCb-PAPER-2013-070}
LHCb collaboration, R.~Aaij {\em et~al.},
  \ifthenelse{\boolean{articletitles}}{\emph{{Measurement of charged particle
  multiplicities and densities in \proton\proton collisions at
  \mbox{$\sqs=$7\tev} in the forward region}},
  }{}\href{https://doi.org/10.1140/epjc/s10052-014-2888-1}{Eur.\ Phys.\ J.\
  \textbf{C74} (2014) 2888},
  \href{http://arxiv.org/abs/1402.4430}{{\normalfont\ttfamily
  arXiv:1402.4430}}\relax
\mciteBstWouldAddEndPuncttrue
\mciteSetBstMidEndSepPunct{\mcitedefaultmidpunct}
{\mcitedefaultendpunct}{\mcitedefaultseppunct}\relax
\EndOfBibitem
\bibitem{sPlot}
This may not be true in the general case. If the mass distribution of companion
  candidates and of the background are different, the sPlot
  technique~\cite{Pivk:2004ty} may not be able to disentangle all species
  correctly, unless all these distributions are known.\relax
\mciteBstWouldAddEndPunctfalse
\mciteSetBstMidEndSepPunct{\mcitedefaultmidpunct}
{}{\mcitedefaultseppunct}\relax
\EndOfBibitem
\bibitem{Chatrchyan:2013cda}
CMS collaboration, S.~Chatrchyan {\em et~al.},
  \ifthenelse{\boolean{articletitles}}{\emph{{Angular analysis and branching
  fraction measurement of the decay $B^0 \to K^{*0} \mu^+\mu^-$}},
  }{}\href{https://doi.org/10.1016/j.physletb.2013.10.017}{Phys.\ Lett.\
  \textbf{B727} (2013) 77},
  \href{http://arxiv.org/abs/1308.3409}{{\normalfont\ttfamily
  arXiv:1308.3409}}\relax
\mciteBstWouldAddEndPuncttrue
\mciteSetBstMidEndSepPunct{\mcitedefaultmidpunct}
{\mcitedefaultendpunct}{\mcitedefaultseppunct}\relax
\EndOfBibitem
\bibitem{LHCb-PAPER-2016-012}
LHCb collaboration, R.~Aaij {\em et~al.},
  \ifthenelse{\boolean{articletitles}}{\emph{{Measurement of the S-wave
  fraction in \mbox{\decay{\Bz}{\Kp\pim\mumu}} decays and the
  \mbox{\decay{\Bz}{\Kstar(892)^0\mumu}} differential branching fraction}},
  }{}\href{https://doi.org/10.1007/JHEP11(2016)047}{JHEP \textbf{11} (2016)
  047}, Erratum \href{https://doi.org/10.1007/JHEP04(2017)142}{ibid.\
  \textbf{04} (2017) 142},
  \href{http://arxiv.org/abs/1606.04731}{{\normalfont\ttfamily
  arXiv:1606.04731}}\relax
\mciteBstWouldAddEndPuncttrue
\mciteSetBstMidEndSepPunct{\mcitedefaultmidpunct}
{\mcitedefaultendpunct}{\mcitedefaultseppunct}\relax
\EndOfBibitem
\bibitem{LHCb-PAPER-2015-041}
LHCb collaboration, R.~Aaij {\em et~al.},
  \ifthenelse{\boolean{articletitles}}{\emph{{Measurements of prompt charm
  production cross-sections in \proton\proton collisions at $\sqs = $13\tev}},
  }{}\href{https://doi.org/10.1007/JHEP03(2016)159}{JHEP \textbf{03} (2016)
  159}, Erratum \href{https://doi.org/10.1007/JHEP09(2016)013}{ibid.\
  \textbf{09} (2016) 013}, Erratum
  \href{https://doi.org/10.1007/JHEP05(2017)074}{ibid.\   \textbf{05} (2017)
  074}, \href{http://arxiv.org/abs/1510.01707}{{\normalfont\ttfamily
  arXiv:1510.01707}}\relax
\mciteBstWouldAddEndPuncttrue
\mciteSetBstMidEndSepPunct{\mcitedefaultmidpunct}
{\mcitedefaultendpunct}{\mcitedefaultseppunct}\relax
\EndOfBibitem
\bibitem{Fruhwirth:1987fm}
R.~Fr{\"u}hwirth, \ifthenelse{\boolean{articletitles}}{\emph{{Application of
  Kalman filtering to track and vertex fitting}},
  }{}\href{https://doi.org/10.1016/0168-9002(87)90887-4}{Nucl.\ Instrum.\
  Meth.\  \textbf{A262} (1987) 444}\relax
\mciteBstWouldAddEndPuncttrue
\mciteSetBstMidEndSepPunct{\mcitedefaultmidpunct}
{\mcitedefaultendpunct}{\mcitedefaultseppunct}\relax
\EndOfBibitem
\bibitem{ATLAS:2012jma}
ATLAS collaboration, \ifthenelse{\boolean{articletitles}}{\emph{{Performance of
  the ATLAS Inner Detector Track and Vertex Reconstruction in the High Pile-Up
  LHC Environment}}, }{} Tech. Rep.
  \href{https://cds.cern.ch/record/1435196}{ATLAS-CONF-2012-042}, CERN, Geneva,
  2012\relax
\mciteBstWouldAddEndPuncttrue
\mciteSetBstMidEndSepPunct{\mcitedefaultmidpunct}
{\mcitedefaultendpunct}{\mcitedefaultseppunct}\relax
\EndOfBibitem
\bibitem{Chatrchyan:2014fea}
CMS collaboration, S.~Chatrchyan {\em et~al.},
  \ifthenelse{\boolean{articletitles}}{\emph{{Description and performance of
  track and primary-vertex reconstruction with the CMS tracker}},
  }{}\href{https://doi.org/10.1088/1748-0221/9/10/P10009}{JINST \textbf{9}
  (2014) P10009}, \href{http://arxiv.org/abs/1405.6569}{{\normalfont\ttfamily
  arXiv:1405.6569}}\relax
\mciteBstWouldAddEndPuncttrue
\mciteSetBstMidEndSepPunct{\mcitedefaultmidpunct}
{\mcitedefaultendpunct}{\mcitedefaultseppunct}\relax
\EndOfBibitem
\bibitem{LHCb-DP-2013-002}
LHCb collaboration, R.~Aaij {\em et~al.},
  \ifthenelse{\boolean{articletitles}}{\emph{{Measurement of the track
  reconstruction efficiency at LHCb}},
  }{}\href{https://doi.org/10.1088/1748-0221/10/02/P02007}{JINST \textbf{10}
  (2015) P02007}, \href{http://arxiv.org/abs/1408.1251}{{\normalfont\ttfamily
  arXiv:1408.1251}}\relax
\mciteBstWouldAddEndPuncttrue
\mciteSetBstMidEndSepPunct{\mcitedefaultmidpunct}
{\mcitedefaultendpunct}{\mcitedefaultseppunct}\relax
\EndOfBibitem
\bibitem{Aad:2011zb}
ATLAS collaboration, G.~Aad {\em et~al.},
  \ifthenelse{\boolean{articletitles}}{\emph{{Search for displaced vertices
  arising from decays of new heavy particles in 7 TeV pp collisions at ATLAS}},
  }{}\href{https://doi.org/10.1016/j.physletb.2011.12.057}{Phys.\ Lett.\
  \textbf{B707} (2012) 478},
  \href{http://arxiv.org/abs/1109.2242}{{\normalfont\ttfamily
  arXiv:1109.2242}}\relax
\mciteBstWouldAddEndPuncttrue
\mciteSetBstMidEndSepPunct{\mcitedefaultmidpunct}
{\mcitedefaultendpunct}{\mcitedefaultseppunct}\relax
\EndOfBibitem
\bibitem{Aad:2011bw}
ATLAS collaboration, G.~Aad {\em et~al.},
  \ifthenelse{\boolean{articletitles}}{\emph{{Search for strong gravity
  signatures in same-sign dimuon final states using the ATLAS detector at the
  LHC}}, }{}\href{https://doi.org/10.1016/j.physletb.2012.02.049}{Phys.\ Lett.\
   \textbf{B709} (2012) 322},
  \href{http://arxiv.org/abs/1111.0080}{{\normalfont\ttfamily
  arXiv:1111.0080}}\relax
\mciteBstWouldAddEndPuncttrue
\mciteSetBstMidEndSepPunct{\mcitedefaultmidpunct}
{\mcitedefaultendpunct}{\mcitedefaultseppunct}\relax
\EndOfBibitem
\bibitem{LHCb-DP-2012-004}
R.~Aaij {\em et~al.}, \ifthenelse{\boolean{articletitles}}{\emph{{The \lhcb
  trigger and its performance in 2011}},
  }{}\href{https://doi.org/10.1088/1748-0221/8/04/P04022}{JINST \textbf{8}
  (2013) P04022}, \href{http://arxiv.org/abs/1211.3055}{{\normalfont\ttfamily
  arXiv:1211.3055}}\relax
\mciteBstWouldAddEndPuncttrue
\mciteSetBstMidEndSepPunct{\mcitedefaultmidpunct}
{\mcitedefaultendpunct}{\mcitedefaultseppunct}\relax
\EndOfBibitem
\bibitem{LHCb-PAPER-2013-040}
LHCb collaboration, R.~Aaij {\em et~al.},
  \ifthenelse{\boolean{articletitles}}{\emph{{First measurement of
  time-dependent \CP violation in \mbox{\decay{\Bs}{\Kp\Km}} decays}},
  }{}\href{https://doi.org/10.1007/JHEP10(2013)183}{JHEP \textbf{10} (2013)
  183}, \href{http://arxiv.org/abs/1308.1428}{{\normalfont\ttfamily
  arXiv:1308.1428}}\relax
\mciteBstWouldAddEndPuncttrue
\mciteSetBstMidEndSepPunct{\mcitedefaultmidpunct}
{\mcitedefaultendpunct}{\mcitedefaultseppunct}\relax
\EndOfBibitem
\bibitem{LHCb-PAPER-2014-004}
LHCb collaboration, R.~Aaij {\em et~al.},
  \ifthenelse{\boolean{articletitles}}{\emph{{Study of the kinematic
  dependences of \Lb production in \proton\proton collisions and a measurement
  of the \mbox{\decay{\Lb}{\Lc\pim}} branching fraction}},
  }{}\href{https://doi.org/10.1007/JHEP08(2014)143}{JHEP \textbf{08} (2014)
  143}, \href{http://arxiv.org/abs/1405.6842}{{\normalfont\ttfamily
  arXiv:1405.6842}}\relax
\mciteBstWouldAddEndPuncttrue
\mciteSetBstMidEndSepPunct{\mcitedefaultmidpunct}
{\mcitedefaultendpunct}{\mcitedefaultseppunct}\relax
\EndOfBibitem
\bibitem{LHCb-PAPER-2014-026}
LHCb collaboration, R.~Aaij {\em et~al.},
  \ifthenelse{\boolean{articletitles}}{\emph{{Measurement of \CP violation in
  \mbox{\decay{\Bs}{\phiz\phiz}} decays}},
  }{}\href{https://doi.org/10.1103/PhysRevD.90.052011}{Phys.\ Rev.\
  \textbf{D90} (2014) 052011},
  \href{http://arxiv.org/abs/1407.2222}{{\normalfont\ttfamily
  arXiv:1407.2222}}\relax
\mciteBstWouldAddEndPuncttrue
\mciteSetBstMidEndSepPunct{\mcitedefaultmidpunct}
{\mcitedefaultendpunct}{\mcitedefaultseppunct}\relax
\EndOfBibitem
\bibitem{LHCb-PAPER-2014-059}
LHCb collaboration, R.~Aaij {\em et~al.},
  \ifthenelse{\boolean{articletitles}}{\emph{{Precision measurement of \CP
  violation in \mbox{\decay{\Bs}{\jpsi\Kp\Km}} decays}},
  }{}\href{https://doi.org/10.1103/PhysRevLett.114.041801}{Phys.\ Rev.\ Lett.\
  \textbf{114} (2015) 041801},
  \href{http://arxiv.org/abs/1411.3104}{{\normalfont\ttfamily
  arXiv:1411.3104}}\relax
\mciteBstWouldAddEndPuncttrue
\mciteSetBstMidEndSepPunct{\mcitedefaultmidpunct}
{\mcitedefaultendpunct}{\mcitedefaultseppunct}\relax
\EndOfBibitem
\bibitem{Aad:2014vma}
ATLAS collaboration, G.~Aad {\em et~al.},
  \ifthenelse{\boolean{articletitles}}{\emph{{Search for direct production of
  charginos, neutralinos and sleptons in final states with two leptons and
  missing transverse momentum in $pp$ collisions at $\sqrt{s} =$ 8 TeV with the
  ATLAS detector}}, }{}\href{https://doi.org/10.1007/JHEP05(2014)071}{JHEP
  \textbf{05} (2014) 071},
  \href{http://arxiv.org/abs/1403.5294}{{\normalfont\ttfamily
  arXiv:1403.5294}}\relax
\mciteBstWouldAddEndPuncttrue
\mciteSetBstMidEndSepPunct{\mcitedefaultmidpunct}
{\mcitedefaultendpunct}{\mcitedefaultseppunct}\relax
\EndOfBibitem
\bibitem{Aad:2014yka}
ATLAS collaboration, G.~Aad {\em et~al.},
  \ifthenelse{\boolean{articletitles}}{\emph{{Search for the direct production
  of charginos, neutralinos and staus in final states with at least two
  hadronically decaying taus and missing transverse momentum in $pp$ collisions
  at $\sqrt{s}$ = 8 TeV with the ATLAS detector}},
  }{}\href{https://doi.org/10.1007/JHEP10(2014)096}{JHEP \textbf{10} (2014)
  096}, \href{http://arxiv.org/abs/1407.0350}{{\normalfont\ttfamily
  arXiv:1407.0350}}\relax
\mciteBstWouldAddEndPuncttrue
\mciteSetBstMidEndSepPunct{\mcitedefaultmidpunct}
{\mcitedefaultendpunct}{\mcitedefaultseppunct}\relax
\EndOfBibitem
\bibitem{LHCb-PAPER-2015-037}
LHCb collaboration, R.~Aaij {\em et~al.},
  \ifthenelse{\boolean{articletitles}}{\emph{{Measurement of forward \jpsi
  production cross-sections in \proton\proton collisions at
  \mbox{$\sqs=$13\tev}}},
  }{}\href{https://doi.org/10.1007/JHEP10(2015)172}{JHEP \textbf{10} (2015)
  172}, Erratum \href{https://doi.org/10.1007/JHEP05(2017)063}{ibid.\
  \textbf{05} (2017) 063},
  \href{http://arxiv.org/abs/1509.00771}{{\normalfont\ttfamily
  arXiv:1509.00771}}\relax
\mciteBstWouldAddEndPuncttrue
\mciteSetBstMidEndSepPunct{\mcitedefaultmidpunct}
{\mcitedefaultendpunct}{\mcitedefaultseppunct}\relax
\EndOfBibitem
\bibitem{LHCb-PAPER-2015-004}
LHCb collaboration, R.~Aaij {\em et~al.},
  \ifthenelse{\boolean{articletitles}}{\emph{{Measurement of \CP violation in
  \mbox{\decay{\Bz}{\jpsi\KS}} decays}},
  }{}\href{https://doi.org/10.1103/PhysRevLett.115.031601}{Phys.\ Rev.\ Lett.\
  \textbf{115} (2015) 031601},
  \href{http://arxiv.org/abs/1503.07089}{{\normalfont\ttfamily
  arXiv:1503.07089}}\relax
\mciteBstWouldAddEndPuncttrue
\mciteSetBstMidEndSepPunct{\mcitedefaultmidpunct}
{\mcitedefaultendpunct}{\mcitedefaultseppunct}\relax
\EndOfBibitem
\end{mcitethebibliography}
\end{document}